\RequirePackage{fix-cm}
\documentclass[natbib,smallextended]{svjour3}       
\smartqed  
\usepackage{graphicx}
\usepackage{mathptmx}      
\usepackage{amssymb}

 \journalname{my journal}
\newcommand{\rhead}{r_{_{\rm h}}}
\newcommand{\rL}{r_{_{\rm L}}}

\newcommand{\rj}{r_{_{\rm j}}}

\newcommand{\pL}{p_{_{\rm L}}}
\newcommand{\pj}{p_{_{\rm j}}}
\newcommand{\pc}{p_{_{\rm c}}}

\newcommand{\bh}{\beta_{_{\rm h}}}
\newcommand{\Gh}{\Gamma_{_{\rm h}}}
\newcommand{\uh}{u_{_{\rm h}}}

\newcommand{\Gj}{\Gamma_{_{\rm j}}}

\newcommand{\Lj}{L_{_{\rm j}}}

\newcommand{\rrel}{R_{_{\rm rel}}}
\newcommand{\tb}{t_{_{\rm b}}}

\newcommand{\bhbar}{\bar{\beta}_{_{\rm h}}}
\newcommand{\Ghbar}{\bar{\Gamma}_{_{\rm h}}}

\newcommand{\mean}[1]{\langle{#1}\rangle}
\newcommand{\fracb}[2]{\left(\frac{#1}{#2}\right)}

\def \simless {\mathbin{\lower 3pt\hbox{$\rlap{\raise 5pt
              \hbox{$\char'074$}}\mathchar"7218$}}}
\def \simgreat {\mathbin{\lower 3pt\hbox{$\rlap{\raise 5pt
              \hbox{$\char'076$}}\mathchar"7218$}}}

\begin{document}

\title{Gamma-Ray Bursts as Sources of Strong Magnetic Fields
}

\titlerunning{GRBs as Sources of Strong Magnetic Fields}        

\author{Jonathan Granot, Tsvi Piran, Omer Bromberg, Judith L. Racusin, and Fr\'{e}d\'{e}ric Daigne}

\authorrunning{Granot et al.} 

\institute{Jonathan Granot \at
              Department of Natural Sciences, The Open University of Israel, 1 University Road, P.O. Box 808, Ra'anana 4353701, Israel \\
              \email{granot@openu.ac.il}   \\
           \and
           Tsvi Piran \at
           Racah Institute of Physics, The Hebrew University of Jerusalem, Jerusalem 91904, Israel  \\
           \email{tsvi@phys.huji.ac.il}   \\
           \and
           Omer Bromberg \at
           Department of Astrophysical Sciences, Princeton University, 4 Ivy Ln., Princeton NJ 08544, USA \\
           \email{omerb@astro.princeton.edu}   \\
           \and
           Judith L. Racusin \at
           Astrophysics Science Division, NASA Goddard Space Flight
           Center, Greenbelt MD, USA \\
           \email{judith.racusin@nasa.gov}   \\
           \and
           Fr\'{e}d\'{e}ric Daigne \at
           Institut d'Astrophysique de Paris, UMR 7095 Universit\'{e} Pierre et Marie Curie; CNRS, 
           98 bis, boulevard Arago, 75014 Paris, France \\
           \email{daigne@iap.fr}   \\
}

\date{Received: date / Accepted: date}

\maketitle

\begin{abstract}
  Gamma-Ray Bursts (GRBs) are the strongest explosions in the
  Universe, which due to their extreme character likely involve some of
  the strongest magnetic fields in nature.  This review discusses the
  possible roles of magnetic fields in GRBs, from their central
  engines, through the launching, acceleration and collimation of
  their ultra-relativistic jets, to the dissipation and particle
  acceleration that power their $\gamma$-ray emission, and the powerful
  blast wave they drive into the surrounding medium that generates
  their long-lived afterglow emission. An emphasis is put on
  particular areas in which there have been interesting developments in
  recent years.  
\end{abstract}

\section{Introduction}

Gamma-ray Bursts (GRBs) are among the most extreme objects in the
Universe.  They are the most luminous cosmic explosions, and therefore
serve as beacons at the edge of the visible Universe that can be used
as cosmic probes. GRBs provide short timescale insight into end-stage
stellar evolution, and serve as probes of extreme physics such as strong
gravity, very large densities and magnetic fields, extremely energetic
particles, and relativistic bulk motions. They are also promising
sources of high-energy neutrinos and gravitational waves.
 
GRBs can be roughly divided into two main sub-classes: (i) Long-duration
($\gtrsim 2\;$s) soft-spectrum bursts that are found in star-forming
regions and are associated with broad-lined Type Ic supernovae, implying a
massive star progenitor, which is most likely low-metalicity and
rapidly rotating near this cataclysmic end of its life, and lives in a
gas-rich environment not far from where it was formed. In order to
produce a GRB, the central engine must drive a strong relativistic jet
that bores its way through the stellar envelope and produces the GRB
well outside of the progenitor star; (ii) Short-duration ($\lesssim
2\;$s) hard-spectrum bursts that are thought to arise from the merger
of a binary neutron star system (or a neutron star and a stellar-mass
black hole) that emits gravitational waves as it inspirals and
coalesces, producing a central engine driven jet. Such systems live in
low density environments, possibly with a prior supernova kick that
pushed them into the outskirts of their host galaxies. 
A third sub-class,  whose importance was realized only recently
\citep{Soderberg06,Campana06,Liang07,Virgili09,Bromberg11,Bromberg12,NS12},  
involves low-luminosity GRBs, whose overall isotropic equivalent radiated energy is 
$E_{\gamma,{\rm iso}}\lesssim 10^{49}\;$erg. They also typically have a smooth, 
single-peaked lightcurve, and their $\nu F_\nu$ spectrum typically peaks at a lower 
than average photon energy (usually $E_{\rm p}\lesssim 100\;$keV). While observed 
rarely, because of their low luminosity,  they are the most numerous group in nature 
(in terms of their rate per unit volume). They most likely do not arise from the same 
emission mechanism as regular long GRBs \citep[e.g.,][]{Bromberg12,NS12}. 

The phenomenology of GRBs is generally separated into two
observational phases: the prompt emission and the afterglow.  These
two phases are traditionally differentiated largely based upon
instrumental measurement methods, but they do seem to also be
physically distinct -- they arise from different emission mechanisms
and occur at different distances from the central source.  However,
the dividing lines between the prompt emission and the afterglow have
blurred in recent years.  In the standard Fireball model
\citep[e.g.,][]{Piran99,Piran04,KZ15}, the prompt emission (i.e. the
burst of $\gamma$-rays) is due to dissipative internal shocks within
the outflow, while the long-lived broadband afterglow is the result of
the jet driving a strong relativistic forward shock into the
surrounding medium as it decelerates and transfers its energy to the
shocked external medium. Unless the outflow is highly magnetized when
it is decelerated by the external medium, this deceleration can occur
through a strong reverse shock that results in a bright optical flash, 
which is also sometimes detected
\citep[e.g.,][]{Akerlof99,SP99,Mundell07b,Racusin08,Vestrand14}
just after the onset of the prompt emission, and decays largely
independently of the forward shock emission.

The field of GRBs is relatively young, with several revolutions in our
understanding of these objects thanks to new observations over the
last two decades. The role of magnetic fields in GRBs is relevant to
many topics in this field.  They affect the properties of the compact
object (neutron star and/or black hole) that powers the central
engine, and how it launches the jet. Magnetic fields may also play an
important role in the acceleration and collimation of the relativistic
jets in GRBs, as well as in their composition. They can contribute to
the energy dissipation and particle acceleration that powers the
prompt GRB emission, and may play a key role in its emission
mechanism. A strong magnetic field can suppress the reverse shock and
its emission. During the afterglow, the amplification of the weak
magnetic field in the external medium by the afterglow shock and its
subsequent behavior in the shocked external medium downstream of the
shock play a key role in the particle acceleration by the shock and in
the shaping of the afterglow emission. In fact, it seems hard to find
any important part of GRB physics where magnetic fields might be
safely ignored.

In this review, we explore the evidence for extreme magnetic
fields in GRBs, and how magnetic fields are intertwined with our
understanding of the mechanisms that produce the relativistic jets
that power these objects. As we cannot cover here all of the relevant
topics in detail, we have instead chosen to focus on specific topics
in which there has been recent progress \citep[see e.g.][for an earlier review]{Piran05m}.  
First, in \S\ref{sec:progenitors} a brief overview is given on the progenitors
of both long and short GRBs, with the main thrust being devoted to the
possible role of millisecond magnetars -- newly born, very rapidly
rotating and highly magnetized neutron stars -- a topic that has
recently received a lot of attention in the literature and in the GRB
community. Next, \S\ref{jetcomp} discusses the dynamics of GRB
jets. It starts with long GRB jets as they bore their way out of their
massive star progenitors, and then moves on to discuss more generally
the possible role of magnetic fields in the acceleration and
collimation of GRB jets, both in steady-state and highly time-variable
outflows, as well as in the interaction of the jet with the external
medium and the reverse shock.  In \S\ref{sec:prompt} we discuss the
role of magnetic fields in the dissipation and radiation that power
the prompt $\gamma$-ray emission, and what GRB observations can tell us
about the conditions within the emitting region. Finally,
\S\ref{afterglow} is devoted to the role of magnetic fields in the
afterglow. It focuses on their effects on the afterglow and
reverse shock emission and their polarization, and how this can teach
us about the magnetic field structure in the GRB outflow and its
amplification in the afterglow shock as well as its structure and
possible decay further downstream of this shock. Our conclusions are
discussed in \S\ref{sec:conc}.

\section{GRB Progenitors, Central Engine, and the Role of Magnetars}
\label{sec:progenitors}

Long duration GRBs are associated with Type Ic supernovae
\citep[e.g.,][]{WB06}, which directly relate them to the death of
massive stars stripped of their hydrogen and helium. This supports the
popular Collapsar model \citep{Woosley93,MW99} in which 
a central engine lunches a relativistic jet that penetrates the stellar envelope 
and powers the GRB. Typically, within the Collapsar models the central engine 
is considered to be a an accreting newly-formed stellar-mass black hole at the 
center of the progenitor star. 
The most popular model for short duration GRBs features the merger of two
neutron stars in a tight binary system \citep{ELPS89}, which may again form a black
hole surrounded by an accretion disk as they coalesce. Therefore, in
both long and short duration GRBs, despite their very different
progenitors, the central engine that is formed during the explosion
and launches the relativistic jets might still be similar in nature --
accretion onto a newly formed black hole.

An attractive alternative possibility that has gained popularity in
recent years is that GRB central engines may involve magnetars
\citep{Usov92,DT92,Bucciantini08,DallOsso09} -- highly magnetized
neutron stars with surface magnetic fields of order $B \sim
10^{15}\;$G, which in this case are newly born and very rapidly
rotating, with $\sim 1\;$ms periods (and hence dubbed millisecond
magnetars). In this model the main energy source is the neutron star's
rotational energy, and a very strong magnetic field is needed for a
rapid extraction of this rotational energy and to channel it
into a relativistic outflow.  A rapidly rotating neutron star may
arise in the collapse of a rotating stellar core and the magnetic
field can be amplified in this collapse \citep[e.g.,][]{DT92}. As the
magnetar's energy is naturally extracted in the form of a Poynting
flux (though this flux is initially not significantly collimated) it naturally 
leads to a magnetically dominated outflow. Collimation of the outflow into a 
narrow jet may, however, be facilitated by the interaction of the outgoing 
strong magnetohydrodynamic (MHD) wind with the progenitor star's 
envelope \citep[e.g.,][]{Bucciantini07,Bucciantini08,Bucciantini09,BGLP14}.

After the first $\sim 10-100\;$s or so from the neutron star formation, the 
neutrino-driven winds subside and the baryon loading on the
MHD wind significantly decreases. As a result the initial wind magnetization
parameter $\sigma_0$ significantly increases and becomes $\gg 1$ (the
magnetization parameter $\sigma$ is the Poynting-to-matter energy flux
ratio, or proper enthalpy density ratio). At this stage, the neutron
star spin-down and its associate luminosity are approximately given by
the magnetic dipole in vacuum formula \citep[which also approximately
holds in the force-free regime;][]{Spitkovsky06},
\begin{equation}
\label{eq:magnetar}
L (t) = \frac{ B^2 R^6 \Omega^4_0 /(6 c^3) }{\left[1 + 2 B^2 R^6  \Omega^2_0 t /(6 I c^3)\right]^2} 
= \frac{E_0 }{t_0 (1+t/t_0)^2}\approx L_0\times\left\{\matrix{1 & (t<t_0)\ , \cr & \cr
(t/t_0)^{-2}  & (t>t_0)\ , }\right.
\end{equation} 
where $I$ is the neutron star's moment of inertia, $R$ is its radius, $B$
is the surface dipole magnetic field at the pole, $\Omega_0$ is the
initial angular velocity, $E_0\approx\frac{1}{2}I\Omega_0^2$ is the
initial rotational energy, $L_0 = E_0/t_0$ is the initial spin-down
luminosity and
\begin{equation}\label{eq:t0} 
t_0 = \frac{3 I c^3}{B^2 R^6 \Omega^2_0} 
\approx 2\times10^{3}\fracb{B}{10^{15}\,{\rm G}}^{-2}\fracb{P}{1\,{\rm ms}}^2\;{\rm s}
\end{equation}
is the initial spin-down time (using typical values of $R\approx
10\;$km, $I\approx10^{45}\;{\rm g\;cm^{2}}$).  This spindown
luminosity initially (at $t<t_0$) has a plateau at $L_0$, and then (at
$t>t_0$) falls off as $t^{-2}$.  Both $L_0$ and $t_0$ can be tuned
with the proper choice of the initial angular velocity $\Omega_0$ and the
magnetic field $B$.  With a choice of $B \sim 10^{15.5-16}\;$G one can
arrange $t_0$ fit the prompt duration in which case the magnetar is
invoked to power the prompt GRB.  With a lower magnetic field of order
$10^{14.5-15}\;$G, $t_0$ is of order several thousand seconds,
comparable to the duration of the plateau phase in some X-ray afterglows.

The magnetar model gained a lot of popularity with the discovery by
{\it Swift} of plateaus in the X-ray afterglow light curves of many GRBs
\citep{nousek06,zhang06}, whose shape resembles the overall shape of
Magnetar's spindown luminosity \citep{troja07,DallOsso09,DallOsso11,Rowlinson13,Rowlinson14}.  
Somewhat surprisingly, even though this tentative evidence\footnote{These plateaus
have several alternative explanations, which are at least as compelling as the magnetar
explanation, such as promptly ejected slow material that gradually catches up with the 
afterglow shock \citep{nousek06,GK06}, time-varying afterglow shock microphysical 
parameters \citep{GKP06}, viewing angle effects \citep{EG06}, or a two-component jet 
\citep{PKG05,GKP06}.} for magnetar-like activity was obtained for the afterglow phase, 
it was interpreted in the community as evidence for a magnetar operating as the main source 
of energy for the prompt emission as well. Both interpretations face some difficulties.

The magnetic field needed to produce the prompt emission is larger by
about one order of magnitude than the one observed even in the
strongest magnetars. This may not be that puzzling as there is ample
evidence of magnetic field decay in magnetars \citep{DallOsso12}, and
the observed magnetars are typically a few thousand years old. It is
possible that the magnetic fields of newborn magnetars are large
enough. 

A more serious problem concerns the energy budget.  The rotational
energy of a typical neutron star, even when rotating at breakup
velocity, is at best marginally sufficient to power the most powerful
GRBs \citep{Cenko10}. This is especially so if we also take into account 
the efficiency of converting this rotational energy into the prompt flux of
$\gamma$-rays.  Of course, magnetars could still power less energetic
GRBs. However, this would require one to invoke two kinds of central
engines, as a different energy source would be needed to power the
most energetic GRBs.

Even more perplexing is the situation concerning the longer duration
plateaus in the afterglow light curves. Here, the needed values of
magnetic fields are indeed typical for those arising in the observed
magnetars, and the overall energy budget is reasonable as well.
However, another question arises: if a low magnetic field magnetar has
powered the afterglow plateau, then what has powered the prompt GRB? 
Can a magnetar fire twice?  The simple answer, according to
Eq.~(\ref{eq:magnetar}), is no. This is as long as the magnetic field
remains constant during the slowdown time scale. However, one can
come up with a fine-tuned model in which the magnetic field decays on
a timescale shorter than $t_0$. In this case the duration of the
magnetar activity is not determined by $t_0$, but by the magnetic field
decay time.  Once the magnetic field has decayed, a second slower
magnetar phase appears with a new $t_0$. Overall such a model requires
a the magnetic field that is extremely large initially, leading to the
prompt emission and then it decreases, just at the right time (and
before all the rotational energy is exhausted) to a lower level in
which the weaker magnetar powers (using the remaining rotational
energy) the afterglow plateau.

An alternative option is as follows: in the first $\sim 10-100\;$s or
so after the formation of the neutron star, the strong neutrino-driven
winds cause a large baryon loading on the MHD wind that prevents the
formation of a very high initial magnetization ($\sigma_0\lesssim
10-100$). Therefore, during at least part of this time the spindown
luminosity can significantly deviate from the form
Eq.~(\ref{eq:magnetar}) -- the formula for a magnetic dipole in vacuum
-- and may in fact be significantly higher, and closer to the result
for a magnetic monopole in vacuum, since most of the magnetic field
lines are opened by the strong baryon loading
\citep[e.g.,][]{Metzger11}.  This can increase the spindown luminosity
by a factor of $\sim (R_L/R)^{2}\sim 10^{1.5}$, where
$R_L=c/\Omega_0$ is the initial value of the light-cylinder radius.
However, as in the early magnetic field decay scenario mentioned
above, also this solution would require fine tuning in order to
extract just the right amount of rotational energy over just the right
timescale. Moreover, in this case the bulk of the large amount of energy 
that is released on the timescale of the prompt GRB is given to relatively 
low-$\sigma_0$ baryon-rich material, which could not attain sufficiently 
large asymptotic Lorentz factors that are needed to power a GRB.

Another possible solution to this problem was suggested recently
\citep{Rezzolla15,Ciolfi15} -- the ``time reversal model", which
postulates magnetar activity for the plateau but an accretion disk for
the prompt phase.  According to this model, first a magnetar with $t_0
\sim 10^4\;$s is born and launches a fast MHD wind whose interaction
with slower matter that was ejected earlier produces the afterglow
plateau. In this scenario the magnetar is a supramassive neutron star,
i.e. supported against gravitational collapse by its very fast rotation, 
so once it spins down significantly it collapses to a black hole, and 
an accretion disk that forms during this collapse powers the prompt
emission.\footnote{This scenario is rather similar to the ``supranova
  model" that was suggested much earlier \citep{VS98,VS99}.}  In spite
of this reversed time sequence the plateau is observed after the
prompt emission because it involved the interaction of the winds and
this phase introduces a time delay (in the observer frame).  While it
is appealing, this model requires the formation of a disk during the
collapse of the supramassive neutron star.  However, \cite{Margalit15}
have recently argued that this is impossible.

A different possible solution is if the prompt GRB is powered by the
energy in a strong initial differential rotation \citep{KR98}. The strong
differential rotation winds-up strong toroidal magnetic field loops, which are 
buoyantly pushed out of the neutron star surface and power the prompt GRB.
This lasts until they exhaust all of the differential rotation energy on the 
timescale of the prompt GRB emission of long duration GRBs. 
The rotational energy of the remaining uniformly rotating neutron star 
could then power the plateaus on its longer magnetic dipole spindown time
$t_0$. This might possibly work for long duration GRBs that are not
too energetic (as the energy in differential rotation is somewhat
lower than the total rotational energy).

Recently, it was suggested that millisecond magnetars might also be at
work in short duration GRBs
\citep[e.g.,][]{FX06,Rosswog07,Metzger08,Rowlinson10,Bucciantini12,Rowlinson13,Fan13b,GOW14}. 
Newly formed millisecond magnetars were suggested to produce the extended
emission seen on a timescale of $\sim 10^2\;$s in some short GRBs
\citep[e.g.][]{Metzger08,Bucciantini12}. In this picture the initial
short GRB spike may arise from short-lived accretion following the
merger. The extended emission is driven by the spindown power that is
released over $\sim 10^2\;$s, and takes several seconds to break out
of the surrounding mildly relativistic material that is ejected
quasi-isotropically during the merger. In this scenario, however, the
jets that power the short GRB itself (the initial hard spike) are
launched within the first second or so after the formation of a
newly-born millisecond magnetar, when the neutrino-driven wind is very
vigorous, the magnetization is low ($\sigma_0\lesssim 1$), and the
baryon loading is very significant near the star and the inner
accretion disc where the jet may be launched.  It is therefore unclear
whether in this case the jets could eventually reach a high enough
Lorentz factor to produce the GRB.  Moreover, in a binary merger the
neutron star is formed extremely rapidly rotating (near breakup), and
its rotational energy of a few $\times 10^{52}\;$erg is eventually
injected into the afterglow shock. This should naturally produce a
bright afterglow emission while the observed afterglows of short GRBs
(either with or without an extended emission) are typically much
dimmer than those of long GRBs. While this might in part be attributed
to a lower external density on average, this cannot fully account for
the dimmer afterglows of short GRBs essentially over the entire
broad-band spectrum, from radio to GeV energies.

\section{Jet Propagation and Dynamics \label{jetcomp}}

The question of the jet composition is still a major issue in our
understanding of GRBs.  It affects the location of the emission site,
the mechanism of the emission and the particle acceleration.  There are
two main possibilities that are commonly discussed in the literature:
a hydrodynamic jet and a Poynting flux dominated jet (for the jet dynamics
discussed here we do not make the distinction between baryonic and
$e^\pm$ pairs particle content).  The main advantage of a hydrodynamic
jet is fast and robust acceleration, which allows the jet to reach
very high Lorentz factors relatively close to the central
source. Magnetic acceleration, on the other hand, is slower and less
robust.  However, hydromagnetic jet launching implies dynamically
strong magnetic fields near the central source, which can naturally
avoid excessive baryonic loading into the central part of the jet, and
thus allow it to reach large asymptotic Lorentz factors far from the
source. The required very small baryon loading is hard to naturally
achieve in a purely hydrodynamic jet. Here we focus on the possible
role of magnetic fields in the jet dynamics and propagation, while
keeping in mind these two main options for the jet composition.

\subsection{Jet Propagation in the Stellar Envelope\label{subsec:collapsar}}

In order to produce a regular GRB, a collapsar jet needs to
successfully break out of its progenitor star.  After it breaks out,
the jet can accelerate freely and eventually generate the observed
$\gamma$-ray photons far from the star in a region where they can
escape (see \S\ref{sec:prompt}). Before it emerges from the stellar
surface, the jet propagates inside the star by pushing the stellar
material in front of it, forming a bow shock ahead of the jet.  The
stellar material that crosses this shock is heated and forms a cocoon
around the jet, which in turn applies pressure on the jet and
collimates it. The collimated jet propagates at a different velocity
than a freely expanding jet. It continuously injects energy into the
cocoon through a slower moving {\it head} that forms at the front of
the jet. The head dissipates the jet's energy and channels it into the
cocoon. Therefore, the continuous propagation of the jet through the
star depends on the supply of fresh energy from the source. If the
engine stops injecting energy, the head will essentially stop
propagating once the information about the energy cutoff will reach
it, and the jet will fail. {The {\it breakout time}, $\tb$, is defined
  as the time of the engine shutoff for which the information about
  the shutoff reaches the jet's head when it is at the edge of the
  star.}  If the engine stops working at a time {$t_e$}$<\tb$, the
head will ``feel" this cutoff while it is inside the star and will
stop propagating. In this case the jet will not break out and it will
not produce a regular GRB\footnote{A failed jet produces, most likely,
  a low-luminosity GRB when a shock wave generated by the dissipated
  energy breaks out from the seller envelope.}.  Since the information
travels outwards at very close to the speed of light, the breakout time is
related to the time at which the jet's head reaches the outer edge of the
star through
\begin{equation}\label{eq:t_b}
  \tb=\int_{_0}^{^{R_*}} \frac{dz}{\bh(z)c}-\frac{R_*}{c}\equiv
  \frac{R_*}{c}\frac{1-\bhbar}{\bhbar}\ ,
\end{equation}
where $\bh(z) c$ is the instantaneous velocity of the jet's head at a distance 
$z$ from the central source (along its symmetry axis, 
using cylindrical coordinates), and $\bhbar c$ is its average  velocity.

Simple analytic solutions to Eq.~(\ref{eq:t_b}) can be obtained in two
limits: (i) a non-relativistic limit, characterized by a proper speed (in units of $c$) 
$u_{\rm h}=\Gh\bh\ll1$ (where $\Gh$ is the head's Lorentz factor)
in which $\tb\simeq R_*/\bhbar c$, and (ii) the relativistic limit, characterized by $u_{\rm h}\gg1$,
in which $\tb\simeq R_*/2\Ghbar^2c$. The transition between these two limits
occurs when $\tb\simeq R_*/c$, which according to Eq.~(\ref{eq:t_b})
corresponds to $\bhbar\simeq1/2$. The jet's head is initially sub-relativistic, but it
accelerates in the steep density profile inside the star ($\xi = -{d\log\rho}/{d\log r}>2$). 
Therefore, if the jet becomes relativistic at some radius, $\rrel$, where
$u_{\rm h}\simeq1$, then it will remain so until it will break out.

In order to calculate the breakout time ones needs a proper model for
the propagation of the jet in the star. Such a model needs to consider
the evolution of the jet and the cocoon self-consistently, as they
affect one another. The propagation velocity of the head is determined
by its cross section, which is set by the collimation of
the jet.  The head's velocity, in turn, controls the energy
injection into the cocoon, which determines the collimating pressure.
The dynamics of this system can be described in a relatively simple
way in two extreme cases of a purely hydrodynamic jet and a purely
electromagnetic (Poynting flux dominated) jet.

\subsubsection{The Breakout Time of Collapsar Jets}

Close to the injection point the jet's internal pressure, $p_j$, is much
larger than the cocoon's pressure, $p_c$.  Therefore, initially the jet
material expands freely until the collimation point where the jet's
pressure equals the cocoon's pressure, $\pj=\pc$.  Above this point
the jet is collimated by the cocoon's pressure, and its behavior
depends on its magnetization.

In the hydrodynamic case, the collimation of the jet leads to the
formation of a collimation shock at the base of the jet
\citep[e.g.,][]{BL07}. Above this point the jet maintains a roughly
cylindrical shape due to a relatively uniform pressure in the cocoon
\citep[e.g.,][]{ZWM03,Matzner03,Bromberg11}. The jet material remains
relativistic with a roughly constant Lorentz factor $\Gj\sim
1/\theta_0$, where $\theta_0$ is the jet injection (or initial) half-opening angle. 
At the head of the jet the relativistic jet material
decelerates abruptly through a strong reverse shock. Since the jet is
roughly cylindrical upstream of the reverse shock, the width of the
head -- its cylindrical radius $r_j$ and its corresponding
cross-section $\Sigma_j = \pi r_j^2$, are set by the width and
cross-section of the jet at the collimation point, which are shown to
be
\begin{equation}
\Sigma_j = \pi r_j^2 \simeq \frac{L_j\theta_0^2}{4cp_c} \ ,
\end{equation}
where $L_j$ is the (one sided) jet luminosity.
The velocity of the head was shown \citep{Matzner03,Bromberg11} to follow
\begin{equation}
\beta_h = \frac{\beta_j}{1+\tilde{L}^{-1/2}}\ ,
\end{equation}
where the dimensionless parameter 
\begin{equation}
\tilde{L}=\frac{\rho_j h_j\Gamma_j^2}{\rho_a}\simeq \frac{L_j}{\Sigma_j\rho_a c^3}\ ,
\end{equation}
represents the ratio between the energy density of the jet
($L_j/\Sigma_j c$) and the rest-mass energy density of the surrounding
medium ($\rho_a c^2$) at the location of the head. Here $h_j = 1 + 4p_j/\rho_jc^2$ 
is the dimensionless specific enthalpy of the jet
material just upstream of the termination shock at the base of its head.

\begin{figure}
 \centerline{\includegraphics[width=0.9\textwidth,height=0.57\textwidth]{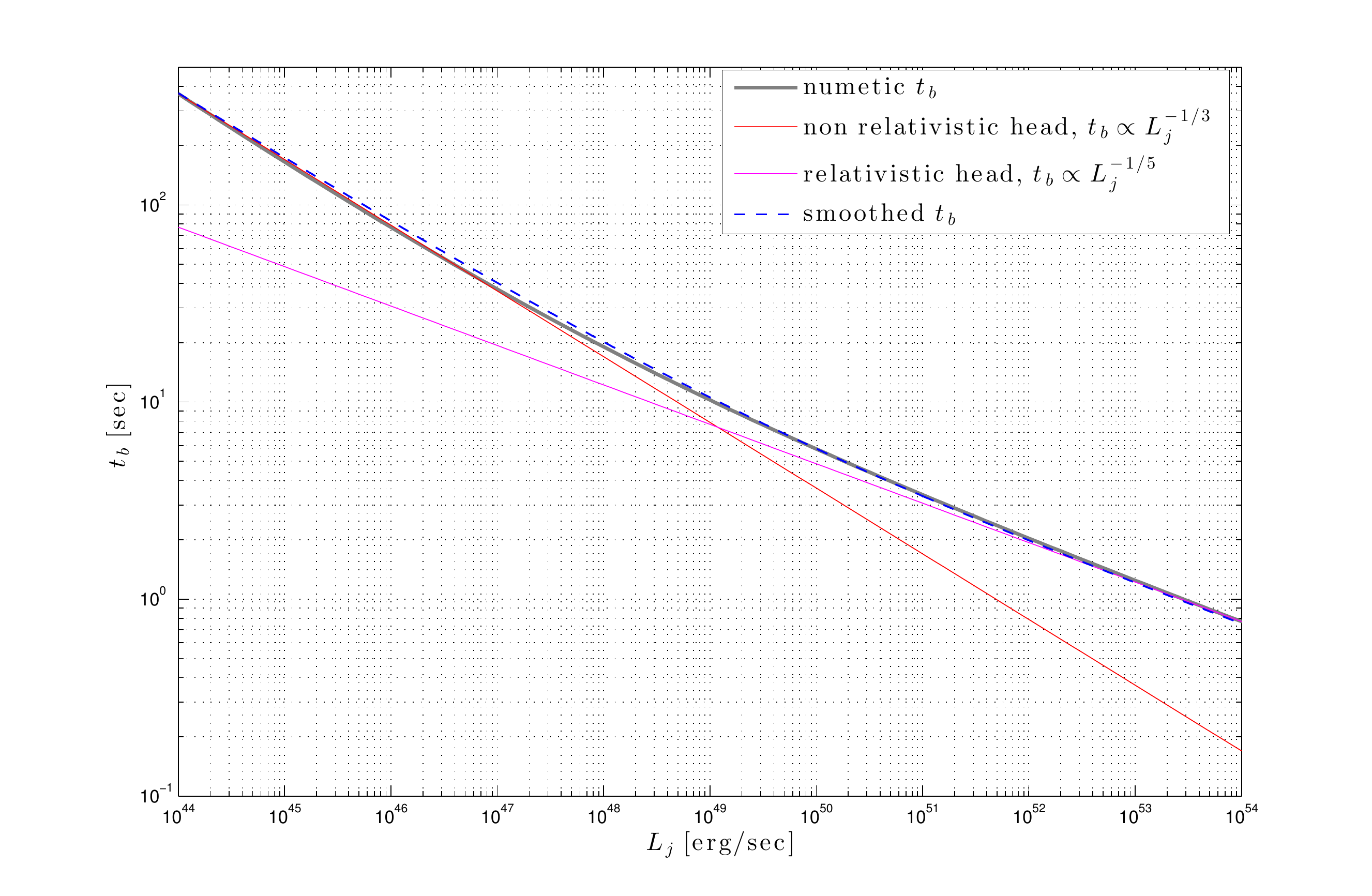}}  
 \caption{The breakout time, $\tb$, as a function of $\Lj$ calculated 
 for a jet with an opening angle $\theta_j=7^\circ$, and a star with a mass
 $M_*=15M_\odot$, radius $R_*=4R_\odot$ and a power-law density profile 
 $\rho\propto r^{-2.5}$. The gray solid curve tracks the exact integration of Eq.~(\ref{eq:t_b}),
 the red and magenta lines show the analytic approximation for the non relativistic
 and the relativistic cases respectively. 
 The dashed blue line follows the smoothed analytic solution for $\tb$
 from Eq.~(\ref{eq:t_b_smooth}). (This figure is taken from \citealt{BGP15}).}
\label{fig:tb_ann}
\end{figure}

\cite{Bromberg11} have obtained approximate analytic expressions for
the propagation velocity of a hydrodynamic jet, and demonstrated that
for typical stellar and jet properties, the jets head propagates at a
velocity that is at most mildly relativistic.  Therefore, in this case
the solution to Eq.~(\ref{eq:t_b}) is in the transition region between
the relativistic and the non-relativistic limits, and can be
approximated following \cite{BGP15}. In order to obtain a useful
analytic solution, they approximated the exact integration (shown by
the {\it dashed blue line} in Fig. \ref{fig:tb_ann}) by:
\begin{equation}\label{eq:t_b_smooth}
t_{\rm b, hyd}\simeq 6.5R_{*,4R_\odot}
  \left[\fracb{\Lj}{L_{\rm rel}}^{-2/3}+\fracb{\Lj}{L_{\rm rel}}^{-2/5}\right]^{1/2}\;{\rm s}\ ,
\end{equation}
where $L_{\rm rel}$ is the transition luminosity between a non-relativistic and a relativistic case:
\begin{equation}\label{eq:L_switch} 
 L_{\rm rel}\simeq 1.6\times10^{49}~R_{*,4R_\odot}^{-1}M_{*,15M_\odot}\theta_{0.84}^{4}
\fracb{3-\xi}{0.5}^{7/5}\fracb{5-\xi}{2.5}^{4/5}\fracb{7-\xi}{4.5}^{15/2}~{\rm erg\;s^{-1}}\ .  
\end{equation} 
As canonical parameters we have used here a stellar mass of $M_*=15
M_\odot$, a stellar radius $R_*=4R_\odot$ and we assume a power-law
density profile: $\rho_*\propto r^{-\xi}$ with $\xi=2.5$.  Hereafter
we measure masses and radii in units of solar mass and solar radius
respectively and use the subscript `$*$' to denote properties of the
progenitor star. For all other quantities we use the dimensionless
form $A_x\equiv A/10^x$ {measured in c.g.s units.}  For a typical
collapsar (one sided) jet luminosity of $L_j\sim2\times10^{49}\;{\rm
  erg\;s^{-1}}$, and injection angle of $\theta_j=7^\circ$ the
corresponding breakout time is $t_{\rm b}(L_{\rm rel})\simeq 9\;$s.

In a Poynting flux dominated jet the situation is different. The
cocoon's pressure is typically strong enough to collimate the jet
before it looses causal contact with the axis. In this case the
poloidal magnetic field is comparable to the toroidal field in the
comoving frame of the flow, and shocks are inhibited. This leads to a
smooth transition of the jet material from a free expansion state,
close to the engine, to a collimated state.  The jet material remains in
a strong causal contact also above the collimation point. Therefore as
it approaches the head, it does not shock. Moreover, it can be shown
\citep{BGLP14} that under these conditions the jet's proper velocity
$u_j$ is approximately equal to the ratio of its cylindrical radius
$\rj$ and the light cylinder radius $\rL$: $u_j\simeq \rj/\rL$, and
the same also holds at the jet's head, $u_{\rm h}\simeq
\rhead/\rL$.  Therefore, the jet material gradually decelerates and
becomes narrower as it approaches the head until at the head its
velocity matches that of the shocked stellar material just behind the
front tip of the bow shock. This deceleration and narrowing of the jet towards its
head is assisted by the fact that the cocoon's pressure becomes larger
closer to the head, as the bow shock is stronger there.  This
results in a jet head that is much narrower than in the hydrodynamic
case and therefore leeds to a much faster propagation speed, where the
head's proper speed $u_{\rm h}$ is given by \citep{BGLP14}:
\begin{equation} \label{Eq:u_a}
u_{\rm h} \sim\frac{\rhead}{\rL} \sim
\left\{\matrix{
a^{1/5} \quad (\uh\ll 1) \ ,\cr\cr
a^{1/6} \quad (\uh\gg 1) \ ,
}\right.
\end{equation}
where the dimensionless quantity
\begin{equation}\label{eq:a}
a\equiv\frac{\Lj}{\pi\rho_a c^3\rL^2}
=\frac{\pL}{\rho_a c^2}\approx 1.2\frac{L_{50}}{\rho_4 r_{_{\rm L7}}^2}\ ,
\end{equation}
is the ratio of the jet's magnetic pressure at the light cylinder and
the ambient medium's rest mass energy density near the
head.\footnote{This analysis does not account for 3D effects that can
  slow down the head's propagation speed (Bromberg \& Tchekhovskoy
  2015, in prep).}

Therefore, a Poynting flux dominated jet becomes relativistic at a
radius $\rrel$ deep inside the star, even with a modest power
\citep{BGLP14}:
\begin{equation}\label{r_1_mag} 
\frac{\rrel}{R_*}\simeq 1.4\times10^{-2} 
\left[L_{49.3}^{-1}M_{*,15M_\odot}R_{*,4R_\odot}^{-3}r_{_{L,7}}^2\fracb{3-\xi}{0.5} \right]^{1/\xi}\ .
\end{equation}
This implies that here only the relativistic asymptotic solution
($u_{\rm h}\approx a^{1/6}$) is relevant. The corresponding breakout
time is \citep{BGP15}:
\begin{equation}\label{eq:t_th_mag}
t_{\rm b,mag}\simeq 0.8~L_{49.3}^{-1/3}M_{*,15M_\odot}^{1/3}r_{_{L,7}}^{2/3}\fracb{0.5}{3-\xi}^{2/3}~ {\rm s}\ .
\end{equation}
This time is much shorter than the breakout time of a hydrodynamic jet with a similar luminosity.

\subsubsection{Observational Evidence for the Jet Breakout Time}

After the jet emerges from the stellar envelope it dissipates its
energy at a large distance and produces the GRB.  On average, the
overall behavior of the prompt emission does not vary significantly
during the burst (the second half of the prompt emission is rather
similar to the first one). This suggests that the prompt emission
arises at a more or less constant radius and not in a propagating
single shell. A single shell would have expanded by a factor of $\sim
10-100$ during the duration of a burst and it is unlikely to maintain
constant conditions as it emits the prompt $\gamma$-ray emission over
such a wide range of radii. This implies, in turn, that the GRB
activity follows the central engine's activity \citep{SP97}, and that
the GRB lasts as long as the central engine is active.  Therefore,
within the Collapsar model, the {observed} GRB duration (usually
denoted by $T_{90}$, which measures the time over which the central 
90\% of the prompt photon counts are detected) is the difference between 
the engine operation time, $t_e$, and the breakout time, $\tb$, namely
$T_{90} = t_e-\tb$ (not accounting for the cosmological time dilation here).

The breakout time essentially serves as a threshold time: a regular
GRB is formed only if $t_e>\tb$. \cite{Bromberg12} have shown that in
such a case one would expect a plateau in the duration distribution of
GRBs, $dN_{\rm GRB}/dT_{90}$, at durations that are shorter than
$\tb$. The logic behind this is as follow. At the time when the jet's
head breaches the edge of the star, it is already disconnected from the
engine and cannot transmit information backward to the engine. In
other words, the engine cannot ``tell'' when the jet breaks out of the
star and we do not expect that $t_e$ and $\tb$ will be strongly related
to each other.  In fact, for a given $\tb$ we expect to have a
distribution of engine activity times, where some are shorter
($t_e<\tb$) and some are longer ($t_e>\tb$) than $\tb$. In this case
the probability of observing a GRB with duration $T_{90}$ is equal to
the probability that the engine will work for a time $t_e=T_{90}+\tb$:
$P{_{\rm GRB}}(T_{90})\equiv P_e(t_e=T_{90}+\tb)$.  This probability
has a simple description in two limits:
\begin{equation}
  P_{\rm GRB}(T_{90})\approx \left\{\matrix{
P_e(\tb) &\quad (T_{90}\ll\tb) \ ,\cr\cr
P_e(T_{90}) &\quad (T_{90}\gg\tb)\ .
}\right.
\end{equation}
Now, if there is a dominant population of GRBs with a typical $\tb$, then at short durations 
$P{_{\rm GRB}}(T_{90})\rightarrow P_e(\tb)=\;$const,  
we expect to get a plateau at durations $T_{90}\ll\tb$.

\begin{figure}
\centerline{\includegraphics[width=0.95\textwidth,height=0.70\textwidth]{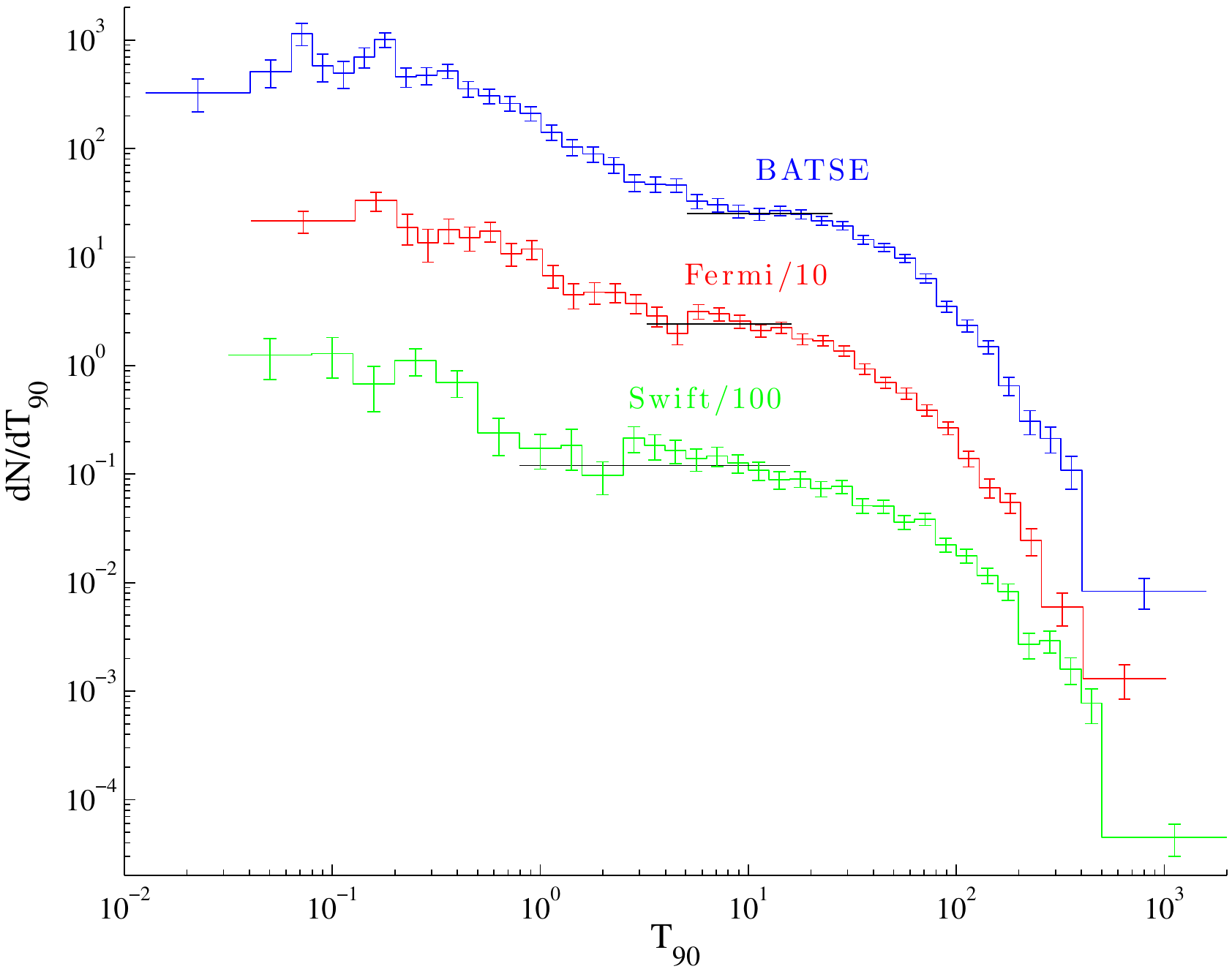}}
\caption{The duration distribution, $dN_{\rm GRB}/dT_{90}$ of BATSE
  (blue), {\it Fermi} (red) and {\it Swift} (Green) GRBs. The
  different curves are shifted in order to avoid overlap.  The {data
    bins} are evenly spaced in logarithmic scale with
  $\Delta\log(T_{90})=0.1$. Bins with less than 5 events are combined
  with their neighbors in order to achieve statistical significance. The black
  horizontal lines mark the bins that fit a plateau at a confidence
  interval up to $2 \sigma$. (This figure is taken from
  \citealt{BGP15}).}
  \label{fig:Plateau_all}
\end{figure}

Figure~\ref{fig:Plateau_all} depicts the duration distribution,
$dN_{\rm GRB}/dT_{90}$, of
BATSE\footnote{http://gammaray.msfc.nasa.gov/batse/grb/catalog/current/
  from April 21, 1991 until August 17, 2000.}  (2100 GRBs), {\it
  Fermi}-GBM\footnote{http://heasarc.gsfc.nasa.gov/W3Browse/fermi/fermigbrst.html,
  from July 17, 2008 until February 14, 2014.} (1310 GRBs) and {\it
  Swift}\footnote{http://{swift}.gsfc.nasa.gov/archive/grb\_table/,
  from December 17, 2004 until February 14, 2014.} (800 GRBs). To fit
a plateau in each data set \cite{BGP15} looked for the maximal number of bins
that are consistent with a plateau at a confidence level {$\leq95\%$}
($2\sigma$)\footnote{The confidence level is defined here as
  $\int_{_0}^{^{\chi^2}}P(x,\nu)dx$, where $P(\chi^2,\nu)$ is the
   probability density function of $\chi^2$ with $\nu$ degrees of freedom
  \citep{Press92}.}.  The extent of the best fitted plateaus is
5$\,$--$\,$25$\;$s in the BATSE data (7.19/4 $\chi^2/d.o.f$), 
2.5$\,$--$\,$17$\;$s in the {\it Fermi}-GBM data (10/5 $\chi^2/d.o.f$),
and 1$\,$--$\,$20$\;$s in the {\it Swift} data (15.85/9 $\chi^2/d.o.f$).  
\cite{BGP15} accounted for three free parameters in the fit:
the height of the plateau and the two opposite ends of the plateau
line.  The differences between the maximal durations of the plateaus
can be mostly attributed to the different sensitivity and triggering
algorithms of the different detectors.

\begin{figure}
\centerline{\includegraphics[width=0.95\textwidth,height=0.70\textwidth]{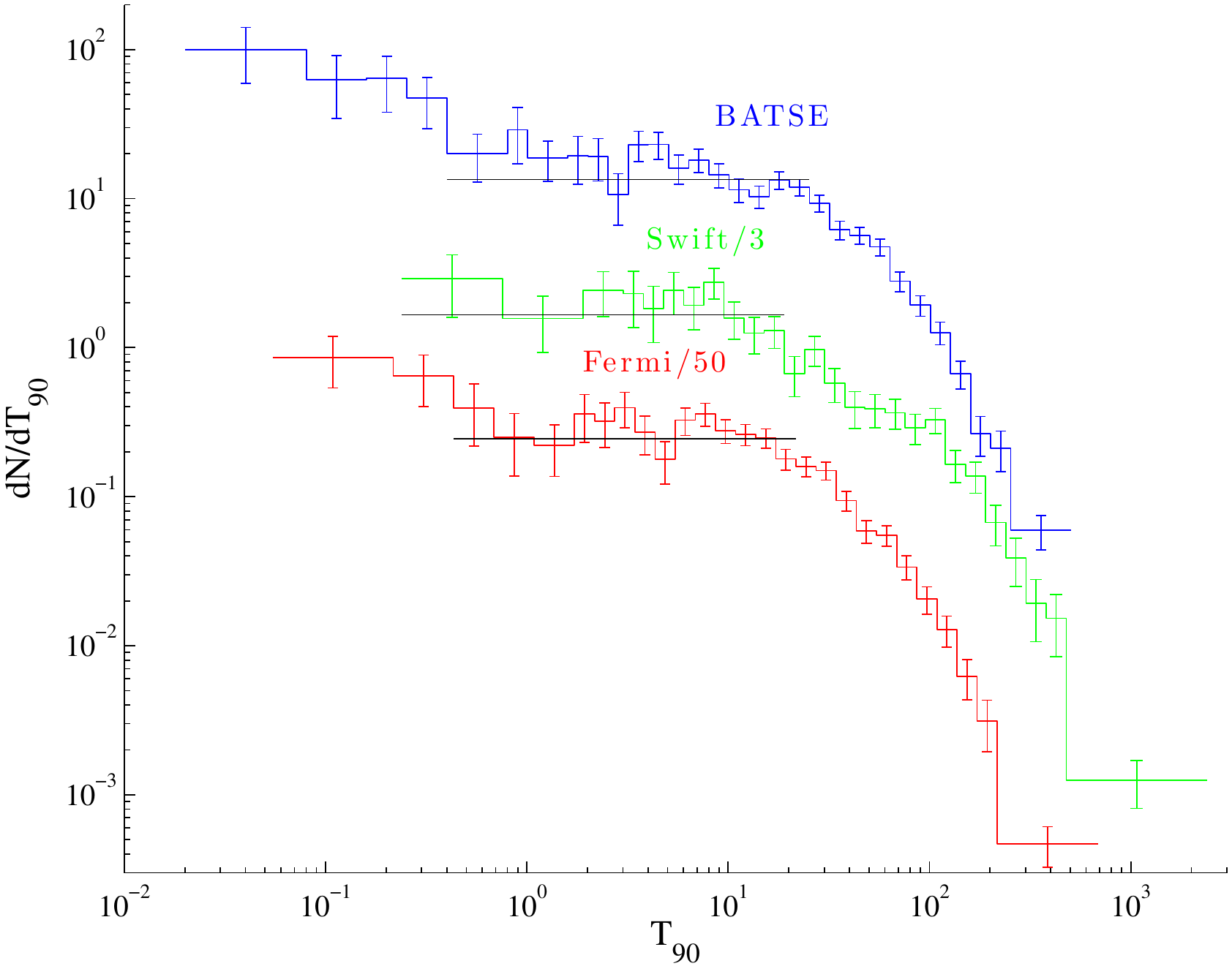}}  
\caption{The duration distribution, $dN_{\rm GRB}/dT_{90}$ of the soft
  GRBs. The analysis is the same as in Figure ~\ref{fig:Plateau_all},
  only the data from each satellite contains only events that are
  softer than the median hardness of the long GRBs with durations
  $T_{90}>20\;$s. For BATSE, this corresponds to GRBs having a
  hardness ratio $HR_{32}<2.6$, for {\it Fermi} the GRBs have a power
  law spectral index $<-1.5$, and for {\it Swift} the GRBs have a
  power law spectral index $<-1.7$. The analysis here updates the
  analysis in \citet{Bromberg13} using a more complete recent sample.
  (This figure is taken from \citealt{BGP15}.)}
  \label{fig:Plateau_soft}
\end{figure} 

At short durations, the plateau is concealed by the increasing number
of non-Collapsar (``short") GRBs having a typical duration of less
than a few seconds \citep{Bromberg13}.  As non-Collapsars have
on-average harder spectra than Collapsars
\citep[e.g.][]{Kouveliotou93}, the relative number of non-Collapsars
can be reduced by choosing a hardness threshold (for each sample) and
selecting only the events that are softer than this threshold.  This
should lead to a less prominent ``bump" at short duration. If the
plateau is indeed an intrinsic property of the (softer) Collapsars
duration distribution, it should extend to shorter durations in a
softer subsample.  To examine this effect they selected in each 
sample all the events that are softer than the median hardness of long GRBs
($T_{90}>20\;$s) in the sample \citep[see][for further
details]{Bromberg13}.  Figure \ref{fig:Plateau_soft} shows the duration
distribution of the soft GRB subsamples. The plateaus indeed extend to
much shorter durations than in the complete samples, {supporting} our
hypothesis.  The extent of the best fitted plateaus is 0.4$\,$--$\,$25$\;$s
in the BATSE data (20.75/12 $\chi^2/d.o.f$), 0.4$\,$--$\,$17$\;$s
in the {\it Fermi}-GBM data (8.7/10 $\chi^2/d.o.f$) and
0.2$\,$--$\,$20$\;$s in the {\it Swift} data (9.04/8 $\chi^2/d.o.f$).

Taking a median redshift of $z\simeq 2$ for {\it Swift} GRBs and
$z\simeq1$ for {\it Fermi}-GBM and BATSE bursts, \cite{BGP15} find that 
in the GRBs' rest frame, these plateaus extend up to intrinsic durations of
7$\,$--$\,$12$\;$s, consistent with the results obtained by
\cite{Bromberg12}.  Note that the actual $\tb$ may be somewhat longer
than the duration that marks the end of the plateau, but it cannot be
shorter. We use the duration interval of 7$\,$--$\,$12$\;$s as our
best estimate for the typical $\tb$.

\subsubsection{Implications: the Jet Composition at Early Times}

Eqs.~(\ref{eq:t_b_smooth}), (\ref{eq:L_switch}) and (\ref{eq:t_th_mag}) 
use parameter values inferred from typical GRB jets, after accounting for the 
jet opening angle \citep[e.g.,][]{BFK03}. From these equations it can 
be seen that a breakout time of 7$\,$--$\,$12$\;$s is consistent with that
expected for a hydrodynamic jet from a WR star with a radius of
$\sim(3\,-\,6)R_\odot$.  On the other hand, these breakout times are
too long for typical parameters expected for a Poynting dominated
jet. To account for these breakout times, the light cylinder of the
compact object at the base of the jet should be of the order of
$\rL\simeq(2.5-5)\times10^8\;$cm, corresponding to an angular
frequency of $\Omega_m\simeq60-120\;$rad/s at the base of the
jet. Such a frequency is too low to allow the engine to power a typical
GRB jet \citep{BGP15}.

The conclusion arising from this analysis is that during most of its
propagation within the star the jet has a low magnetization and it
propagates as a hydrodynamic jet (unless 3D effects significantly 
increase $t_{\rm b,mag}$).  This result leads to some
interesting implications for the properties of long GRB central
engines and the conditions at the base of the jets. One possibility is
that the jet is launched hydrodynamically at the source. The most
probable process for that is neutrino$\,$--$\,$anti-neutrino annihilation above the
rotational axis of the central engine \citep[e.g.][]{ELPS89,LE93}.
This scenario can work only if the accretion rate is $\gtrsim
0.1\,M_\odot~\textrm{s}^{-1}$, so that neutrino emission is large
  enough to power the observed jets \citep{KPK13,LG13}. The high
  accretion rate must be sustained throughout the entire duration of
  the GRB, which can last from tens to hundreds of seconds. Though a
  duration of $\lesssim 30\;$s seems to be consistent with such a
  model \citep[e.g.][]{Lindner10}, its seems unlikely to be capable of
  powering longer duration GRBs.

  A second possibility is that the jet is launched Poynting flux
  dominated {but} it dissipates {most of} its magnetic energy close to
  the source, {and} it then propagates as a hydrodynamic jet. An
  appealing process for such efficient dissipation is the kink
  instability \citep{Lyub92, Eichler93, Spruit97, Begelman98, Lyub99,
    GS06}.  \cite{BGLP14}, however, have shown via analytic
  considerations, that collapsar jets are less likely to be disrupted
  by the kink instability.  Thus a different process, possibly
  internal to the jet, may be needed to dissipate the jet energy.  A
  definite answer will be obtained only via 3D numerical simulations,
  which are underway.  In one such work, Bromberg \& Tchekhovskoy
  (2015, in prep.) show that indeed kink instability is unlikely to
  disrupt a typical collapsar jet. Nevertheless, kink modes can grow
  internally in the jet and lead to efficient dissipation of the
  magnetic energy via reconnection of the magnetic field lines without
  compromising the jets's integrity. The outcome of such dissipation
  is a jet with an equipartition between thermal and magnetic energy,
  which propagates more or less like a hydrodynamic jet.

  A third possibility is that the jet changes its character with
  time. Our conclusion concerning the jet composition applies only to
  the initial phase, while its head is still within the stellar
  envelope. This phase, which lasts $\sim 10$ s, must be predominantly
  hydrodynamic. Once the jet has breached the star it can be Poynting
  flux dominated. This would require a more complicated central engine
  that switches from one mode to another. While this seems contrived,
  remarkably, some magnetar models suggest such a possibility
  \citep{Metzger11}.  One can also imagine accretion disk models that
  initially cool via neutrinos and later on as the accretion rate
  decreases, become Poynting flux dominated \citep{KPK13}. However,
  all such models require some {degree of coincidence} as the central
  engine does not receive any feedback from the propagating jet and
  there is no {\it a priori} reason that the transition from one composition
  to the other would take place {just} at the right stage.

\subsection{Jet steady State Acceleration}

Magnetic acceleration and thermal acceleration are the two main
competing mechanisms for the acceleration of GRB jets or
outflows. Thus, the acceleration mechanism is tightly related to the
outflow composition and in particular its degree of magnetization,
which is both highly uncertain and of great interest.  In other
sources of relativistic jets or outflows, there are currently better
constraints on the composition. Pulsar winds are almost certainly
Poynting flux dominated near the central source. This most likely also
holds for active galactic nuclei (AGN) and tidal disruption events, as
in these sources the central accreting black hole is supermassive, and
therefore even close to it the Thompson optical depth, $\tau_T$, may
not be high enough for thermal acceleration by radiation pressure (the
main competition to magnetic acceleration) to work efficiently
\citep[e.g.,][]{Ghisellini12}.  In GRBs or micro-quasars, however,
thermal acceleration could also work (since $\tau_T\gg 1$ is possible,
or even likely close enough to the source), and the dominant
acceleration mechanism is less clear.

First, let us consider the thermal acceleration of a steady,
axisymmetric, and unmagnetized flow that is initially relativistically hot
with $p\gg\rho c^2$. Let the jet cross section be $\Sigma\propto r^2$
where $r$ is its cylindrical radius. The relativistic equation of
state implies $p\propto\rho^{4/3}$, while mass and energy conservation
read $\Gamma\rho c\Sigma = \;$const and $\Gamma^2(\rho c^2+4p)c\Sigma
=\;$const, respectively (where we have assumed a relativistic
velocity, $\beta = v/c\approx 1$). The ratio of the two last
expressions gives the Bernoulli equation -- the total energy per unit
rest energy (which is conserved without any significant energy losses
from the system), $(1+4p/\rho c^2)\Gamma = \;$const. As long as the
flow is relativistically hot ($p\gg\rho c^2$) it accelerates as
$\Gamma\propto \rho/p\propto\rho^{-1/3}\propto \Sigma^{1/2}\propto
r$. This reproduces the familiar result for a spherical or conical
flow for which $\Gamma\propto r\propto z$, i.e. the Lorentz factor
grows linearly with the distance $z$ from the central
source. Therefore, thermal acceleration is relatively fast, efficient
and robust.

Let us now do a similar simple analysis for a cold and initially
highly-magnetized flow, with $\sigma_0 = B_0^2/4\pi\rho_0c^2\gg 1$
(e.g., \citealt{Kom11}). Let the flow be steady, axisymmetric,
and ideal MHD (i.e. without magnetic dissipation). Let us consider the
flow between two magnetic flux surfaces defined by $r$ and $r+\delta
r$ (which are both functions of $z$). Flux freezing (ideal MHD)
implies that the poloidal and tangential magnetic field components
scale as $B_p\propto 1/r\delta r$ and $B_\phi\propto 1/\delta r$,
respectively, in the lab frame.  Therefore, the tangential field
component rapidly dominates far from the source, so that $B\approx
B_\phi \approx\Gamma B'$ where $B'$ in the magnetic field in the
comoving frame of the outflowing plasma (in which the electric field
vanishes).  Altogether this gives $B = \Gamma B'\propto 1/\delta
r$. Mass and energy conservation read $\Gamma\rho c\Sigma = \;$const
and $\Gamma^2(\rho c^2+B'^2/4\pi)c\Sigma =\;$const, respectively,
where $\Sigma\propto r\delta r$.  Their ratio implies a total energy
per unit rest energy of $(1+\sigma)\Gamma = (1+\sigma_0)\Gamma_0 =
\Gamma_{\rm max}$ where $\sigma = B'^2/4\pi\rho c^2\propto
r/\Gamma\delta r$ is the magnetization parameter. Therefore, this
results in the following Lorentz factor evolution:
\begin{equation}\label{eq:gamma}
\frac{\Gamma}{\Gamma_0} = 1+\sigma_0\left(1-  \frac{\delta r_0}{r_0}\frac{r}{\delta r}\right)\ ,
\quad\quad  \frac{\Gamma}{\Gamma_{\rm max}} = 
1 - \left(1-\frac{\Gamma_0}{\Gamma_{\rm max}}\right)\frac{\delta r_0}{r_0}\frac{r}{\delta r}\ . 
\end{equation}
This immediately implies that for a conical (or spherical) flow, in
which $\delta r\propto r$ and $\delta r/r = \delta r_0/r_0$, the
Lorentz factor essentially remains constant, $\Gamma\approx\;$const,
and the flow hardly accelerates.  This result can be understood by
simple energy considerations. As long as there is no expansion along
the direction of motion, the volume of a fluid element scales as
$\propto r^2$ while its magnetic energy density scales as $\propto
B^2\propto r^{-2}$, implying a constant magnetic energy and no
conversion into kinetic energy.

More generally, Eq.~(\ref{eq:gamma}) implies that in order for the
flow to accelerate, $r/\delta r$ must decrease, i.e. streamlines must
diverge faster than conical. For power-law streamlines, $z =
z_0(r/r_0)^\alpha = z_0[(r+\delta r)/(r_0+\delta r_0)]^\alpha$, one
has $r/\delta r = r_0/\delta r_0$ so there is still no
acceleration. If one allows the power law index to vary with 
$r_0 =r(z_0)$, i.e.  $\alpha = \alpha(r_0)$, then one finds 
$\delta{r}/r=(\delta r_0/r_0)[1-r_0\alpha'(r_0)\alpha^{-2}(r_0)\ln(z/z_0)]$,
and the condition for acceleration becomes
$\alpha'=d\alpha/dr_0<0$. Altogether one can see that such
steady-state, axisymmetric ideal MHD acceleration is quite
delicate and requires a very particular configuration of the magnetic
field lines.  Satisfying this requirement is not trivial, and in
particular it requires lateral causal contact across the jet.

A key open question regarding outflows that start out highly
magnetized near the central source is how they convert most of their
initial electromagnetic energy to other forms, namely bulk kinetic
energy or the energy in the random motions of the particles that also
produce the radiation we observe from these sources. It is suggested
by observations of relevant sources, such as AGN, GRBs or pulsar wind
nebulae that the outflow magnetization is rather low at large
distances from the source. This is the essence of the well-known
$\sigma$ problem -- how to transform from $\sigma\gg 1$ near the
source to $\sigma\ll 1$ very far from the source.

It has been shown early on that a highly magnetized steady spherical
flow accelerates only up to an asymptotic Lorentz factor
$\Gamma_\infty\sim\sigma_0^{1/3}$, and magnetization
$\sigma_\infty\sim\sigma_0^{2/3}$ \citep{GJ70} where $\sigma_0\gg 1$
is the initial value of the magnetization parameter $\sigma$, implying
that most of the energy remains in electromagnetic form (a Poynting
flux dominated flow). This is valid for any such {\bf unconfined}
flow, i.e. where the external pressure is effectively negligible.  A
sufficiently large external pressure can help collimate and accelerate
the flow. It has been found \citep{Lyub09,Lyub10a,Kom09} that for a
power law external pressure profile, $p_{\rm ext}\propto z^{-\kappa}$,
the collimation and acceleration can proceed in two distinct
regimes. 

For $\kappa > 2$, the {\bf weak confinement} regime, the
external pressure drops fast enough such that the flow loses lateral
causal contact while it is still highly magnetized, and from that
point on it becomes conical and essentially stops accelerating. This
collimation-induced acceleration increases $\Gamma_\infty$ and
decrease $\sigma_\infty$ by up to a factor of $\sim\theta_{\rm
  j}^{-2/3}$ compared to the unconfined (quasi-spherical) case, where
$\theta_{\rm j}$ is the asymptotic jet half-opening angle. This arises
because lateral causal contact in the jet is maintained as long as
$\theta_{\rm j}$ does not exceed the Mach angle, $\theta_{\rm
  j}\lesssim\theta_{\rm M}\sim\sigma^{1/2}/\Gamma$, where energy
conservation implies $\sigma\Gamma\sim\sigma_0$ (for $\sigma_0\gg 1$
and $\Gamma_0\sim 1$) as long as the flow remains highly magnetized
($\sigma\gg 1$). 

For $\kappa\leq 2$, the {\bf strong confinement}
regime, the external pressure drops slowly enough that the jet
maintains lateral causal contact throughout its collimation-induced
acceleration process. In this case about half of the initial magnetic
energy is converted into kinetic energy and the flow bacomes only
mildly magnetized, $\sigma_\infty\sim 1$, while the Lorentz factor
approaches its maximal possible value, $\Gamma_\infty\sim\sigma_0$.
In this regime the collimation and acceleration proceed as
$\Gamma\propto r\propto z^{\kappa/4}$ and the jet remains narrow,
$\Gamma_\infty\theta_{\rm j}\sim 1$.

The main problem with this picture, however, is that even under the
most favorable conditions the asymptotic magnetization is
$\sigma_\infty\geq 1$, which does not allow efficient energy
dissipation in internal shocks within the outflow
\citep{Lyub09,Lyub10a,Kom09}. It has been found \citep{Tchek10,Kom10}
that a sudden drop in the external pressure, as may occur when a
GRB jet exits its progenitor star, can result in a sudden additional
acceleration that can lead to $\Gamma_\infty\theta_{\rm j}\gg 1$ as
inferred in GRBs, but still with $\sigma_\infty\geq 1$.

These important limitations of the ``standard'' steady, axisymmetric
and non-dissipative (or ideal MHD) magnetic acceleration have, on the
one hand, led to the suggestion that the jets might remain Poynting
flux dominated at large distances from the source and the observed
emission is the result of magnetic reconnection events rather than
internal shocks \citep{Blandford02,LB03,Lyut06}.  On the other hand, other
models suggested increasing the acceleration efficiency by relaxing
one of the standard assumptions, such as axi-symmetry -- leading to
non-axisymmetric instabilities that randomize the magnetic field
orientation \citep{HB00}.  Since a highly tangled magnetic field
effectively behaves like a relativistic fluid (with an adiabatic index
of 4/3) this leads to efficient acceleration, similar to thermal
acceleration of relativistic outflows. What is more, both the kink
instability mentioned above \citep{DS02}, as well as other
instabilities \cite[such as the Kruskal-Schwarzschild instability in a
striped wind;][]{Lyub10b} can lead to magnetic reconnection,
i.e. gradual magnetic dissipation, which in turn enhances the
acceleration due to the conversion of magnetic to thermal energy,
where the thermal pressure efficiently accelerates the outflow.

\subsection{Impulsive Magnetic Acceleration}

Replacing the usual steady-state assumption by strong time-dependence
is a natural alternative. This impulsive regime was sparsely studied,
and mainly in the non-relativistic case \citep{Cont95}. Recently, a
new impulsive magnetic acceleration mechanism was found that operates
in the relativistic case \citep{GKS11}, which can be much more
efficient than magnetic acceleration in steady flows, and can lead to
low magnetizations, $\sigma\ll 1$, thus enabling efficient dissipation
in internal shocks. This qualitatively different behavior of impulsive
outflows can be very relevant for GRBs, as well as for other
relativistic jet sources such as tidal disruptions or flares in AGN or
micro-quasars, or even giant flares in soft gamma repeaters (SGRs,
thought to be magnetars), which also triggered renewed interest in this
topic \citep[e.g.,][]{Levinson10,Lyut11,Granot12a,Granot12b,Kom12}.

\begin{figure}
\includegraphics[height=4.23cm]{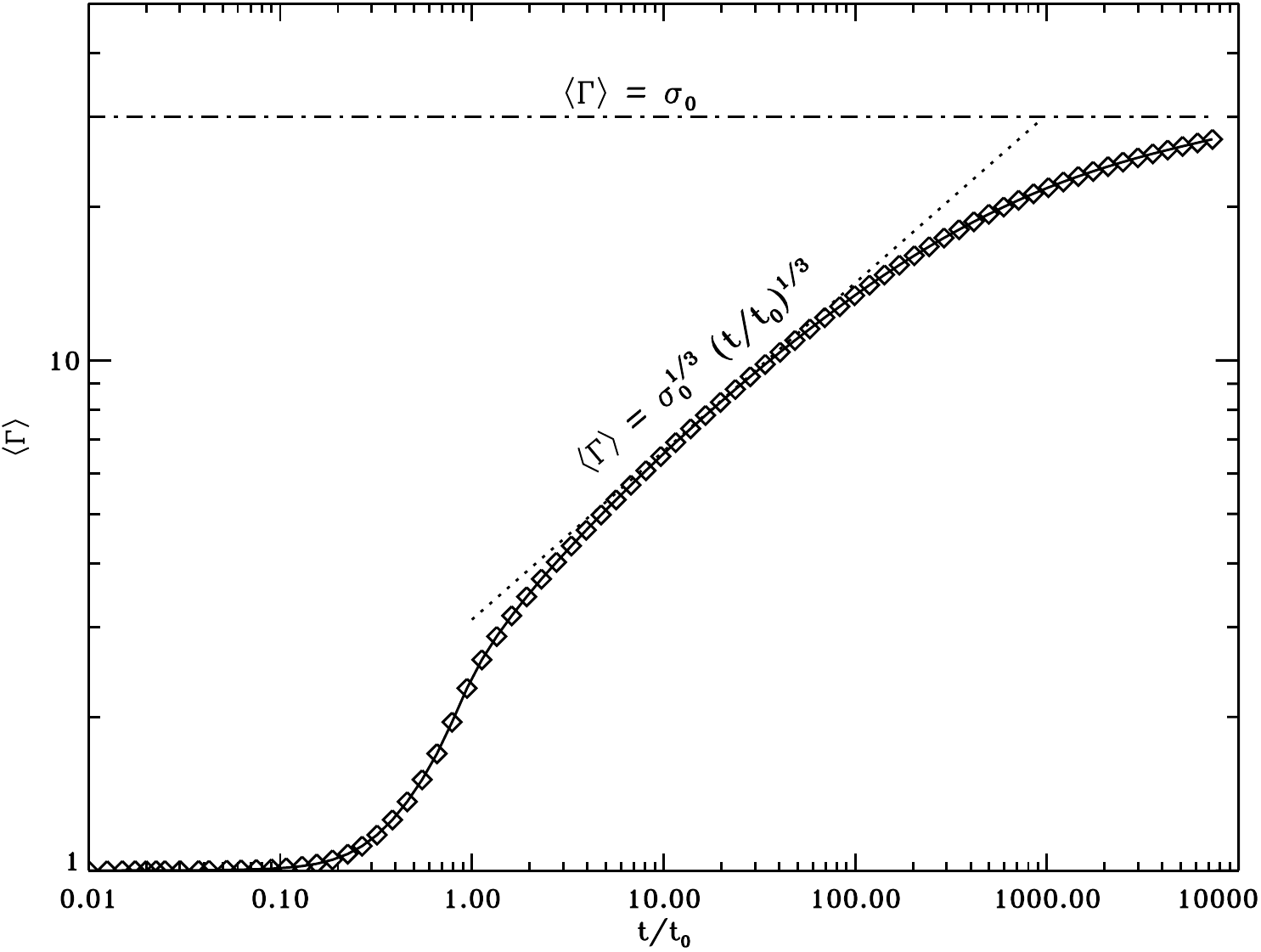}
\hspace{0.35cm}
\includegraphics[height=4.23cm]{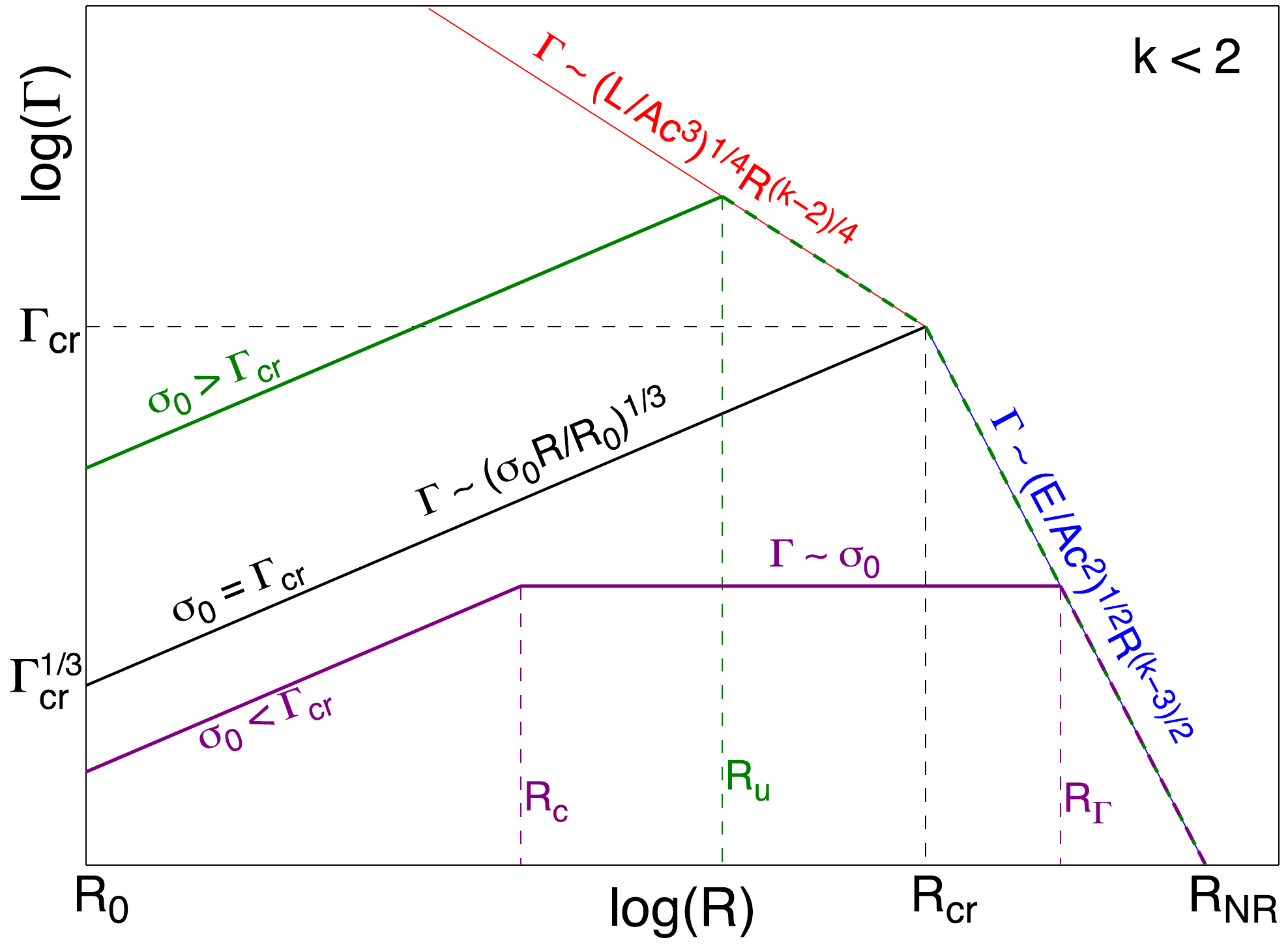}
\caption{\label{fig:acc} {\bf Left}: Test case for impulsive magnetic
  acceleration: the energy-weighted mean Lorentz factor
  $\langle\Gamma\rangle$ of a finite cold shell of plasma initially
  uniform (with width $l_0$, rest mass density $\rho_0$ and magnetic
  field $B_0$), highly magnetized ($\sigma_0 = B_0^2/4\pi\rho_0c^2 \gg
  1$; $\sigma_0 = 30$ was used here) and at rest, whose back leans
  against a conducting ``wall'' while its front faces vacuum, versus
  the time $t$ in units of the shell's initial fast magnetosonic
  crossing time $t_0\approx l_0/c$.  The analytic expectations ({\it
    dotted} and {\it dashed-dotted} lines) and the results of
  numerical simulations ({\it diamond symbols} joined by a {\it solid
    line}) are in very good agreement.  (This Figure is taken from
  \citealt{GKS11}).  {\bf Right}: Evolution of the typical (or
  energy-weighted average) Lorentz factor $\Gamma$ with the distance
  $R\approx ct$ from the central source, for a finite shell similar to
  that described in the {\it left panel}, but for a spherical shell
  propagating into an external medium with a power-law density
  profile, $\rho_{\rm ext} = AR^{-k}$. \citep[This figure is taken
  from][]{Granot12a}.}
\end{figure}

Figure~\ref{fig:acc} ({\it left panel}) shows results for an
impulsive magnetic acceleration test case: a cold, initially uniform 
plasma shell (of with width $l_0$, rest mass density $\rho_0$ and magnetic field $B_0$), 
highly magnetized ($\sigma_0=B_0^2/4\pi\rho_0c^2 \gg 1$) and at rest, 
with  a conducting ``wall'' at its back and vacuum in front of it.  
A strong, self-similar rarefaction wave forms at its front (vacuum
interface) and propagates towards its back, reaching the wall at $t
=t_0 \approx l_0/c$. By this time the shell's energy-weighted mean
Lorentz factor and magnetization are $\mean{\Gamma}\sim\sigma_0^{1/3}$
and $\mean{\sigma}\sim\sigma_0^{2/3}$. At $t >t_0$ the shell detaches
from the wall, keeps an almost constant width ($l\approx 2l_0$) and
accelerates as $\mean{\Gamma}\sim\sigma_0/\mean{\sigma} \sim (\sigma_0
t/t_0)^{1/3}$ up to the coasting time $t_c = \sigma_0^2 t_0$. At $t >
t_c$ the shell coasts at $\mean{\Gamma}\sim \sigma_0$, 
its width grows ($l/2l_0 \sim t/t_c$) and its magnetization rapidly
decreases ($\mean{\sigma}\sim t_c/t$), leading to complete
conversion of magnetic to kinetic energy that allows strong 
internal shocks to form that can lead to large radiative efficiencies.

\subsection{Interaction with the External Medium and the Reverse Shock}

Let us now consider the evolution of a similar shell in spherical
geometry that propagates into an external medium with a power-law
density profile, $\rho_{\rm ext} = AR^{-k}$, following
\cite{Granot12a}.  The main results are shown in the {\it right panel} of
Figure~\ref{fig:acc}. The initial shell magnetization $\sigma_0$ and
density $\rho_0\propto1/\sigma_0$ are allowed to vary while keeping
fixed the values of the initial time or length scale ($t_0 \approx
R_0/c$ or $R_0$), energy ($E \sim Lt_0\approx LR_0/c$ or power $L$),
and external density ($k<2$ in this figure, and $A$ or $\rho_{\rm
  ext}(R_0) = AR_0^{-k}$), which imply fixed $\Gamma_{\rm cr}\sim
(f_0\sigma_0)^{1/(8-2k)}$ where $f_0 = \rho_0/\rho_{\rm ext}(R_0)$ and
$R_{\rm cr}\sim R_0\Gamma_{\rm cr}^2$.  Shown are the two dynamical
regimes most relevant for GRBs.  The purple line shows regime I
($1<\sigma_0 < \Gamma_{\rm cr}$ or a sufficiently low external
density) where the shell initially expands as if into vacuum (as
described in the {\it left panel}) and only after becoming kinetically
dominated and expanding radially is it significantly decelerated by
the external medium through a strong relativistic reverse shock, which
can produce a bright emission that peaks on a timescale larger than
the duration of the prompt GRB emission \citep[the familiar
low-$\sigma$ ``thin shell'';][]{SP95}. Eventually, most of the energy
is transfered to the shocked external medium and the flow approaches
the \cite{BM76} self-similar solution.

\begin{figure}
\centerline{\includegraphics[height=13.0cm,width=0.80\textwidth]{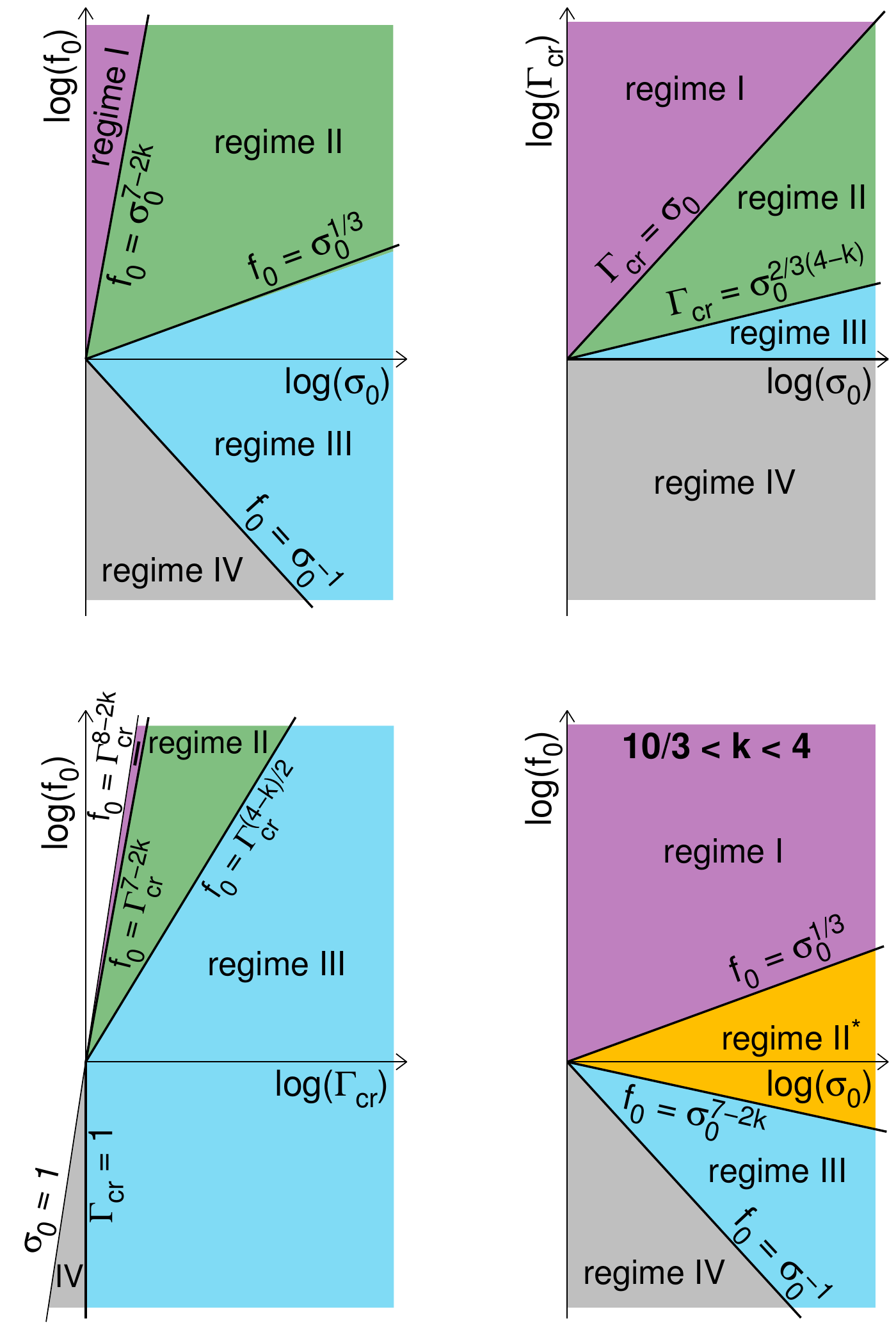}}
\caption{Phase space diagrams of the different dynamical regimes: in
  the $f_0\,$--$\,\sigma_0$ plane for $k<10/3$ ({\it top left panel}),
  $\Gamma_{\rm cr}\,$--$\,\sigma_0$ plane for $k<10/3$ ({\it top right
    panel}), $f_0\,$--$\,\Gamma_{\rm cr}$ plane for $k<10/3$ ({\it
    bottom left panel}), and in the $f_0\,$--$\,\sigma_0$ plane for
  $10/3<k<4$ ({\it bottom right panel}). Each regime is labeled and
  denoted by a different color, and the borders between the different
  regimes are indicated by labeled thick black lines. (This figure is
  taken from \citealt{Granot12a}.)}
\label{fig:regimes}
\end{figure}

\begin{figure}
\centerline{\includegraphics[width=\textwidth]{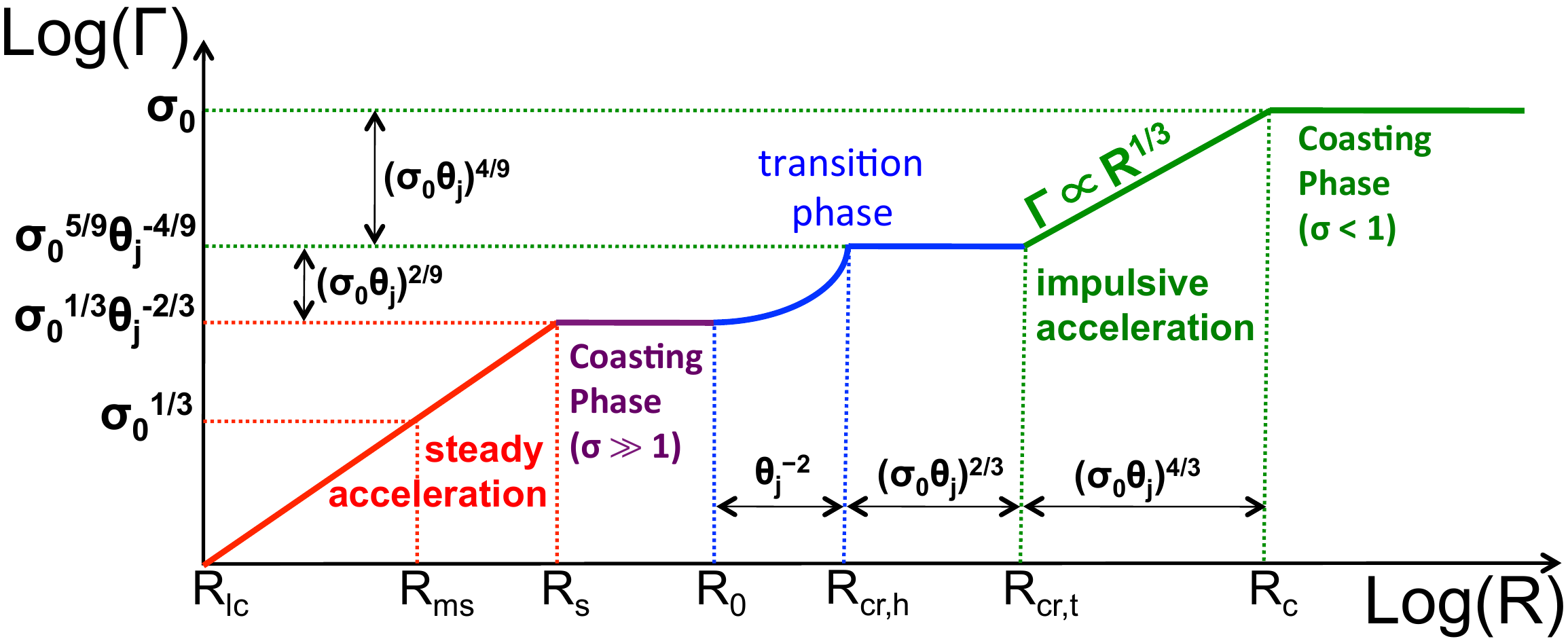}}
\caption{ An illustration of the expected transition from (quasi-)
  steady collimation-induced acceleration near the central source (in
  {\it red}) to impulsive acceleration further away from the source
  (in {\it green}).  The energy weighted mean Lorentz factor $\Gamma$
  is shown against the distance $R$ from the central source, and a few
  critical radii and Lorentz factors are indicated following the
  notations of \cite{GKS11} and \cite{Granot12a}.}
\label{fig:steady_to_impulsive}
\end{figure}

In regime II ($1 <\Gamma_{\rm cr} < \sigma_0 <\Gamma_{\rm
  cr}^{3(4-k)/2}$), depicted by the green line in the {\it right panel} of
Figure~\ref{fig:acc}, the shell is significantly affected by the
external medium while it is still Poynting dominated (at $R > R_u \sim
R_0(f_0\sigma_0^{-1/3})^{3/(10-3k)}$), thus suppressing the reverse
shock (which is either non-existent or very weak). The shell remains
highly magnetized and gradually transfers its energy to the shocked
external medium through $pdV$ work across the contact discontinuity up
to $R_{\rm cr}$, after which the flow approaches the Blandford-McKee
solution.  In this regime no significant reverse shock emission is
expected, and the onset of the afterglow (i.e. the peak of the
emission from the shocked external medium) is expected to be on a
timescale comparable to the prompt GRB duration (i.e. a high-$\sigma$
``thick shell'').

In addition, there are other regimes not shown in this figure, but all
of the regimes are mapped in the relevant parameter space in
Figure~\ref{fig:regimes}. In regime III ($1<\Gamma_{\rm cr}^{3(4-k)/2}
< \sigma_0$) the external density is high enough that there is no
impulsive acceleration stage where $\mean{\Gamma}\propto R^{1/3}$, and
instead $\mean{\Gamma}\sim\sigma_0/\mean{\sigma}\propto R^{(k-2)/4}$
at $R_0 <R < R_{\rm cr}\sim R_{\rm dec}$, and then approaches the
Blandford-McKee solution (its observational signatures are expected to
be similar to regime II).  In regime IV ($\Gamma_{\rm cr} < 1$) the
external density is so high that the flow remains Newtonian all along
(as might happen while the GRB jet is propagating inside a massive
star progenitor).  There is also an ``exotic" regime II* that exists
only in a highly stratified external medium ($10/3 < k < 4$).

Under realistic conditions, GRB variability times are in practice
typically large enough that the flow should first undergo quasi-steady
collimation-induced acceleration that saturates, and only later the
impulsive acceleration kicks in and operates until the flow becomes
kinetically dominated (see Figure
~\ref{fig:steady_to_impulsive}). Moreover, one typically expects the
outflow from the central source to consist of many sub-shells rather
than a single continuous shell. The effects of such multiple
sub-shells in the outflow can be important, and the collisions between
them may provide efficient energy dissipation that can power the GRB
emission \citep{Granot12b,Kom12}.  They may also allow a low-$\sigma$
``thick shell'', i.e. a strong relativistic reverse shock peaking on a
timescale comparable to the prompt GRB emission, which is not possible
for a single shell. For a long-lived source (e.g. AGN) with initial
sub-shell widths $l_0$ and separations $l_{\rm gap}$, each sub-shell
can expand by a factor of $1+l_{\rm gap}/l_0$.  Its magnetic energy
decreases by the same factor (where $\sigma_\infty\sim l_0/l_{\rm
  gap}$), and may be converted to kinetic or internal energy, or
radiation.  For a finite source activity time, the merged shell can still
expand further and convert more magnetic energy into other forms (even
without interaction with an external medium). Important related points
that warrant further study are the transition from quasi-steady 
collimation-induced acceleration to impulsive acceleration, both in a 
single shell and in multiple sub-shells, as well as the dissipation in 
the interaction between sub-shells and its effect on the outflow
acceleration and the resulting emission, such as a possible
photospheric spectral component.

\section{Dissipation and Prompt Emission\label{sec:prompt}}

As discussed above, GRBs must be associated with relativistic outflows
ejected by a stellar mass compact source, with a bulk Lorentz factor
$\Gamma\gtrsim 100$ in order to avoid the \textit{compactness problem}
\citep{baring:97,lithwick:01,granot:08,hascoet:12a}.  This also
naturally explains the afterglow through the deceleration of the
ejecta by the external medium, whereas the observed fast prompt
variability implies that the prompt emission must be produced by
internal dissipation within the ejecta \citep{SP97}.  Therefore, the
analysis of the GRB prompt emission may provide valuable information
on magnetic fields within an ultra-relativistic jet.  It can put
unique constraints on the state of the jet at the end of the
acceleration phase, and more specifically on the geometry of the
magnetic field and the magnetization at a large distance to the central
source, where the $\gamma$-ray emission is produced. This is, however, a
difficult task as it requires a full understanding of the nature of the
dissipative mechanisms and of the radiative processes at work.

There are several possible emission sites for the GRB prompt emission:
(i) a component can be emitted at the photosphere, where the ejecta
becomes transparent to its own photons; another component can be
produced above the photosphere in the optically thin regime,
associated with either (ii) internal shocks propagating within the
ejecta \citep{rees:94}; or (iii) magnetic reconnection
\citep{thompson:94,spruit:01}. These three dissipative mechanisms
extract energy, respectively, from the thermal, kinetic or magnetic
reservoirs. The expected prompt emission components are therefore 
strongly related to the composition of the ejecta.
 
\subsection{Photospheric Emission}
\subsubsection{Non-Dissipative Photospheres}
When internal dissipation below the photosphere is negligible, the
expected thermal emission at the photosphere is well understood with
precise predictions
\citep{paczynski:86,goodman:86,shemi:90,meszaros:93}. Only few
theoretical uncertainties remain, mainly related to the lateral
structure of the jet \citep[see e.g.][]{lundman:13,deng:14}. Assuming
that the photosphere is above the saturation radius, the photospheric
radius is given by
\citep{meszaros:00,meszaros:02,daigne:02,hascoet:13}
\begin{equation}
R_\mathrm{ph}\simeq 
\frac{\kappa \dot{E}}{8\pi c^3 \Gamma^3 (1+\sigma) }
\simeq
3\times 10^{13}\, \frac{\kappa_{0.2}\dot{E}_\mathrm{53}}{\Gamma_2^3(1+\sigma)}\, \mathrm{cm}\ ,
\label{eq:Rph}
\end{equation}
where $\kappa_{0.2}$ is the matter opacity in units of $0.2\,
\mathrm{cm^2\;g^{-1}}$, $\Gamma=100\Gamma_2$ is its bulk Lorentz factor, 
$\dot{E}=10^{53}\dot{E}_\mathrm{53}\, \mathrm{erg\;s^{-1}}$ is the isotropic 
equivalent jet power, and $\sigma$ is the magnetization parameter at the end 
of the acceleration, so that $\dot{E}/(1+\sigma)$ is the isotropic equivalent
kinetic power.  The observed photospheric luminosity and temperature are
\begin{equation}
L_\mathrm{ph} 
\simeq
\epsilon_\mathrm{th} \dot{E} \left(\frac{R_\mathrm{ph}}{\Gamma R_0}\right)^{-2/3}
\quad\mathrm{and}\quad
T_\mathrm{ph} 
\simeq
\frac{T_0}{1+z}\left(\frac{R_\mathrm{ph}}{\Gamma R_0}\right)^{-2/3}\, ,
\end{equation}
where the initial temperature equals $T_0 = \left( \epsilon_\mathrm{th} \dot{E} / 4 \pi \sigma R_0^2 \right)^{1/4}$. 
Here $\epsilon_\mathrm{th}$ is the thermal fraction of the jet power at the base of the flow, 
located at the initial radius $R_0$, and $1-\epsilon_\mathrm{th}$ is therefore the initial magnetic fraction. 
In the case of a passive magnetic field carried by the outflow without contributing to the acceleration, the initial thermal 
fraction $\epsilon_\mathrm{th}$ and the magnetization at the photosphere $\sigma$ are related by 
$\sigma_\mathrm{passive}=\sigma_0=(1-\epsilon_\mathrm{th})/\epsilon_\mathrm{th}$. 
An efficient magnetic acceleration leads to $\sigma<\sigma_\mathrm{passive}$ \citep{spruit:01}. 
The predicted spectrum is quasi-thermal, with an exponential cutoff at high-energy and a power law at low-energy 
with a photon-index of $\alpha\simeq 0.4$, which differs from the $\alpha=1$ slope of the Raleigh-Jeans spectrum 
due to the peculiar geometric shape of a relativistic photosphere \citep{goodman:86,beloborodov:11}. 

\subsubsection{Dissipative Photospheres}
If dissipation occurs below the photosphere, the emitted spectrum can
be significantly different than the previous case: a high-energy tail
can be produced by comptonization due to the presence of relativistic
electrons \citep{thompson:94,meszaros:00, rees:05,
  giannios:07,beloborodov:10}, and the low-energy slope can be
modified by synchrotron radiation \citep{peer:06,vurm:11}.  The
resulting observed spectrum may now appear non-thermal, with several
components. This scenario is more uncertain than the previous one. The
nature of the sub-photospheric dissipative mechanism must be
identified, with several candidates: early internal shocks, gradual
magnetic reconnection \citep{thompson:94,giannios:07}, neutron-proton
collisions \citep{beloborodov:10}, etc.  An important relevant issue
is related to the photon production efficiency and 
thermalization deep within the ejecta \citep{vurm:13}.
 
\subsection{Non-Thermal Emission in the Optically Thin Regime}
\subsubsection{Electron Acceleration and Synchrotron Radiation}
Non-thermal emission can be produced above the photosphere if some
internal dissipation processes can lead to efficient electron
acceleration. In this case, two natural candidates for the dominant
radiative process are the synchrotron radiation and the inverse
Compton scatterings of synchrotron photons by relativistic electrons
(SSC). However, the measurement of the prompt $\gamma$-ray spectrum
over a broad spectral range (keV-GeV) in a few bursts by
\textit{Fermi}/GBM+LAT can rule out the possibility of SSC being
dominant in the soft $\gamma$-ray range, as it would lead either to a
strong synchrotron peak at lower energy, or a strong second inverse
Compton peak at higher energy, which are not observed
\citep{bosnjak:09,piran:09}. Therefore, the discussion is focussed on
the synchrotron radiation of relativistic electrons, with several
possible dissipation mechanisms responsible for the acceleration of
electrons.

\subsubsection{Internal Shocks}
If the magnetization at a large distance to the central source is
sufficiently low, strong internal shocks are expected to
form and propagate within the ejecta due to the variability of the
ejected outflow. A large range of radii is expected,
\begin{equation}
R_\mathrm{is}\simeq 2\Gamma^2 c \left(t_{\rm var}\to t_\mathrm{GRB}\right) \simeq 
\Gamma_2^2 \left( 6\times 10^{12}\,\, t_{\rm var,-2}\to 6\times 10^{15}\,\, t_{\rm GRB,1}\right)\,{\rm cm}\, ,
\label{eq:Ris}
\end{equation}
where $t_{\rm var,-2}$ is the shortest timescale of variability, in units of $10^{-2}\;$s\, and $t_{\rm GRB,1}$ is the total 
duration of the relativistic ejection, in units of $10\, \mathrm{s}$. The shocks are expected to be mildly relativistic, except 
for a very large amplitude of variation of the initial Lorentz factor. The dynamics of the internal shocks phase has been studied 
in detail \citep{kobayashi:97,daigne:98,daigne:00}, from a simple ballistic approximation to a full hydrodynamical code, 
and is well understood. Up to $f_\mathrm{d}\simeq 40\%$ of the kinetic energy can be dissipated (for a low magnetization 
outflow, $\sigma\ll 1$), depending again on the initial distribution of the Lorentz factor.

On the other hand, large uncertainties on the emission remain, due to
the poor understanding of the microphysics of mildly relativistic
shocks \citep[for a recent review of relativistic collisionless shocks see][]{SKL15}. 
It is usually para\-metrized by assuming that a fraction
$\epsilon_\mathrm{B}$ of the internal energy is injected into an
amplified random magnetic field at the shock, whose structure is not
known, and a fraction $\epsilon_\mathrm{e}$ is injected into a
fraction $\zeta$ of electrons, which are therefore accelerated into a
non-thermal distribution with slope $p$
($dN_e/d\gamma_e\propto\gamma_e^{-p}$).  To reach the soft
$\gamma$-ray domain by synchrotron radiation, the fraction of
accelerated electrons must be low, $\zeta\lesssim 10^{-2}$
\citep{daigne:98,bosnjak:09,daigne:11,beniamini:13}. On the other
hand, values of $\epsilon_\mathrm{e}$ close to equipartition
($\epsilon_\mathrm{e}=1/3$) are required to explain the huge
luminosities of GRBs.  If the radiative efficiency is high, a fraction
$f_\mathrm{IS}\simeq f_\mathrm{d}\epsilon_\mathrm{e}\simeq 0.01-0.1$
of the initial kinetic power can be converted into radiation in
internal shocks.  With such assumptions, the non-thermal emission in
the comoving frame of the shocked regions can be computed with a
detailed radiative model including all the relevant processes, namely
synchrotron radiation and slef-absorption, inverse Compton scatterings
and photon-photon annihilation. The contributions of each internal
shock can then be added with an integration over equal-arrival time
surface of photons to the observer in the source frame to produce
synthetic light curves and spectra that can be directly compared to
observations. The predicted spectrum shows several components, a
strong synchrotron peak in the soft $\gamma$-ray range and a weaker
inverse Compton peak at higher energy \citep{bosnjak:09}.

\subsubsection{Magnetic Reconnection}
If the magnetization at large distances remains high ($\sigma\gtrsim
1$), then internal shocks are either significantly suppressed or in
some cases cannot form altogether \citep{mimica:10,narayan:11}. In
such cases, electrons may be accelerated predominantly in magnetic
reconnection sites. This scenario is even more uncertain (less
understood) than the two previous ones (photosphere, internal
shocks), but it is under extensive investigation \citep[for a recent review 
on relativistic magnetic reconnection see][]{Kagan15}. 
Some authors considered a gradual reconnection starting at a
small radius and extending up to $R_\mathrm{rec}\sim 10^{13}\,
\mathrm{cm}$ \citep{DS02,giannios:08}.  In this case, most of the
dissipation occurs below the photosphere, corresponding to the
dissipative photosphere scenario discussed above.  If, on the other
hand, reconnection remains inefficient below the photosphere, it could
occur at larger radii in the optically thin regime.  In the
simulations by \citet{mckinney:12}, a catastrophic dissipation of the
magnetic field occurs at $R_\mathrm{rec}\simeq 10^{13}-10^{14}\,
\mathrm{cm}$ when reconnection enters a rapid collisionless mode.
\citet{ZhangYan11} proposed another scenario where reconnection is
triggered by internal shocks, the so-called ICMART model.  The typical
radius may be as large as $R_\mathrm{rec}\simeq 10^{15}\,
\mathrm{cm}$. The microphysics in the reconnection sites is also
uncertain. One expects many electron acceleration sites, which may
move relativistically in the outflow's rest frame.  The non-thermal
electron distribution may be somewhat harder than in shock
acceleration \citep[see e.g.][]{sironi:14,Kagan15}, and the acceleration
process slower, which can lead to a different shape for the
synchrotron emission. Contrary to internal shocks, detailed
calculations of the light curves and spectra based on a detailed
radiative model coupled to a dynamical simulation are not yet
available.

\subsection{Magnetic Field in Emission Sites}

The models discussed above have very different implications for the
magnetic field:
\begin{itemize}
\item {\bf Case 1}: most of the prompt emission is due to a
  dissipative photosphere. Then, the magnetic field must be large
  enough at the photosphere to produce synchrotron radiation and
  affect the low-energy spectrum. If this magnetic field is generated by the 
  dissipation process (e.g. shocks; \citealt{SKL15}), it is most probably
  random. Otherwise, an ordered field must be present. The corresponding
  initial magnetization must either be low (otherwise the photospheric
  emission is weak), or high with very efficient reconnection below
  the photosphere, which then leads to a possible candidate for the
  sub-photospheric dissipation process.
\item {\bf Case 2}: the prompt emission is mostly non-thermal, from an optically thin region: 
\begin{itemize}
\item For internal shocks to be the dominant dissipation process, the
  magnetization at large distances from the source must be low.  A
  random field is generated locally at the shock front, where the
  electrons are accelerated. However, the magnetic field felt by the
  radiating electrons must be considered far behind the shock front
  (as the radiative cooling length is much larger than the plasma skin
  depth) where its strength and structure are not well known. 
\item If reconnection dominates the dissipation then $\sigma$ must be large far from the source. 
The ordered field is destroyed at the reconnection sites, but if electrons have enough time to migrate from 
their acceleration site before radiating, their emission may still be mostly in the large-scale ordered field. 
\end{itemize} 
\end{itemize} 
Observations of the GRB prompt emission, discussed in the next
subsection, can put strong constraints on these various emission
models.

\subsection{Constraints from the Observed Prompt Soft $\gamma$-ray Emission}

\subsubsection{Light Curves}
All the scenarios discussed above can reproduce the observed variable
light curves. There are, however, important differences:
\begin{itemize}
\item (Dissipative) {\bf photosphere}: the emission radius is low (see
  Eq.~\ref{eq:Rph}). Therefore the curvature effect, i.e. the
  spreading of photon arrival times from different angles with respect
  to the line of sight over an angular timescale $t_\theta=R /
  2\Gamma^2 c$ of a flash of photons emitted at the same time and
  radius, is negligible: the observed light curve directly traces the
  activity of the central engine.
\item {\bf Internal shocks}: the light curves trace the source
  activity \citep{kobayashi:97,daigne:98}, but two effects now affect
  the observed pulse shapes: the curvature effect (due to a larger
  radius) dominates the pulse decay \citep{genet:09,willingale:10},
  and the radial or hydrodynamic timescale due to shock propagation,
  $t_r=\Delta R/2\Gamma^2c$, dominates the pulse rise and overall
  shape \citep{daigne:98,daigne:03,bosnjak:14}.
\item {\bf Reconnection}: again, the light curve traces the source
  activity, with new effects due to relativistic bulk motion in the
  local jet's frame. Relativistic motions of emitting plasma in the
  jet's frame cause rapid variability (that should show up as a
  distinct component in the Fourier power spectrum), while a slower
  envelope may arise from their combined effect \citep{zhang:14} or
  from slower emitting plasma. This can be tested by characterizing
  the observed variability. Analysis of GRB light curves shows a
  continuum of timescales \citep[see
  e.g.][]{beloborodov:00,guidorzi:12}, which does not support the
  reconnection model of \cite{zhang:14} (see however
  \citealt{gao:12}). A possible concern appears if the emission is
  produced by many relativistically moving emitters: the predicted
  pulse shape may be too symmetric compared to observations
  \citep{lazar:09}.  However, both concerns (the power spectrum and
  pulse shapes) may be solved if the reconnection occurs in relatively
  ordered thin layers located between anti-parallel regions in the
  outflow (with a geometry of thin quasi-spherical shells) and the
  relativistic motions in the jet's frame are limited to these layers
  (Beniamini \& Granot 2015, in prep.). Such a model may also account
  for many of the correlations that are observed in the prompt
  emission.
\end{itemize}

\subsubsection{Polarization\label{subsec:pol}}
Measuring the polarization in the $\gamma$-ray domain remains
challenging. A very large degree of polarization was claimed by
\cite{CoburnBoggs03}, but it was later refuted by others
\citep{RutledgeFox04,Wigger04} as not being statistically significant.
Only a few later measurements (by INTEGRAL, GAP) are available,
however with a low or moderate significance \citep{mcglynn:07,
  gotz:09,yonetoku:11,yonetoku:12,gotz:14}. Such measurements (if
reliable) can put constraints on the magnetic field geometry in the
emission sites \citep[e.g.,][]{GK03,Granot03,LPB03,NPW03}. Current
observations seem to favor synchrotron radiation in an ordered field
with patches, which would favor emission in the optically thin regime
above the photosphere. It is however not trivial to justify a highly
ordered field in the internal shocks model (a turbulent field at the
shock is required for particle acceleration; The structure of the
field on intermediate scales between the plasma and the dynamical
scales is less known), or in the reconnection model (in principle the
structured field is destroyed by reconnection, but the remaining field
can still possess significant structure and electrons may also radiate
somewhat outside of these localized reconnection regions).  Either way,
more definitive polarization observations are needed before strong
conclusions can be drawn.

\subsubsection{Spectrum}

The prompt soft $\gamma$-ray spectrum is usually fitted by a
phenomenological model introduced by \citet{Band93}, which consists of
two power laws with low- and high-energy photon indices of $\alpha$ and
$\beta$, smoothly connected at the peak energy $E_\mathrm{p}$.  This
eliminates non-dissipative photospheres for the dominant contribution
to the emission, which puts a strong constraint on the initial
magnetization (low $\epsilon_\mathrm{th}$, \citealt{daigne:02}),
favoring magnetic acceleration of the outflow. The remaining scenarios
for the prompt GRB emission are either a dissipative photosphere, or a
combination of a weak photospheric emission and a non-thermal
component due to shocks or reconnection. The discussion is then
focussed on the general shape of the spectrum, and the low-energy
photon index $\alpha$, which is observed to be close to $\alpha\simeq
-1$ \citep{preece:00,kaneko:06,nava:11,gruber:14}.
\begin{itemize}
\item {\bf Dissipative photospheres}: the value of $\alpha$ can be
  reproduced by adjusting the magnetization, which controls the
  synchrotron emission at low energies. The theoretical instantaneous
  spectral peak is narrower than the observed time-integrated spectral
  peak, but the comparison should be made using a theoretical
  time-integra-ted spectrum, which should broaden it.
\item {\bf Internal shocks}: to reproduce the high luminosities and
  the short timescale variability of GRBs, the radiating electrons
  must be in the fast cooling regime
  \citep{cohen:97,SPN98,ghisellini:00}, i.e. their radiative timescale
  must be shorter than the dynamical timescale that governs the
  adiabatic cooling for the spherical expansion. This leads to a
  predicted photon index $\alpha \le -3/2$, in contradiction with
  observations (the so-called ``synchrotron line-of-death",
  \citealt{Preece98}).  Another potential problem is that the
  resulting spectrum is too broad around the peak. The two problems
  are naturally connected.  Several possibilities have been discussed
  to solve this issue: (i) inverse Compton scatterings in the
  Klein-Nishina regime affect the cooling of electrons, leading to
  photon indices $\alpha\lesssim -1$
  \citep{derishev:01,wang:09,nakar:09,daigne:11}.  This puts a
  constraint on the strength of the magnetic field, which should be
  small, with $\epsilon_\mathrm{B}\lesssim 10^{-3}$
  \citep{daigne:11,barniolduran:12}; (ii) in the marginally fast
  cooling regime \citep{daigne:11,beniamini:13}, where the radiative
  timescale is close to the dynamical timescale but still below, the
  electron radiative efficiency can remain large enough
  ($\gtrsim50\%$) to explain the observed luminosities, but the
  synchrotron spectrum is strongly affected: the intermediate region
  of the spectrum below the peak with a photon index $-3/2$ disappears
  and the slope $\alpha=-2/3$ usually associated with the inefficient
  slow cooling regime is measured. This regime is also favored by weak
  magnetic fields; (iii) in the fast cooling regime, electrons radiate
  on timescales which are long compared to the plasma scale at the
  shock front, but small compared to the dynamical timescale.  Then
  they experience a magnetic field that is not necessarily the same as
  the turbulent field just behind the shock \citep[see the structure
  of the magnetic field in the simulations, e.g.][]{keshet:09}. If the
  field is decaying on this intermediate scale, it will affect the
  synchrotron spectrum and can lead to a hard spectrum, with
  $-1\lesssim\alpha\lesssim-2/3$
  \citep{derishev:07,Lemoine13,uhm:14,zhao:14}.
\item {\bf Reconnection}: the slow electron heating in the turbulent
  field can lead to hard synchrotron spectra with $\alpha\simeq -1$
  \citep{uhm:14}. It is unclear if the expected hard power-law index
  $p \lesssim 1.5$ of the non-thermal electron distribution
  \citep[e.g.,][]{sironi:14} can be identified in the observed
  spectrum. A potential issue is that the presence of many emitting
  regions that move relativistically in random directions in the jet's fame
  might lead to much broader spectra than observed.  This may be
  alleviated, however, in models where these regions move
  predominantly along the thin reconnection layer that is located
  between regions of oppositely-directed magnetic field in the flow, normal to
  the jet's bulk motion (Beniamini \& Granot 2015, in prep.).
\end{itemize}

\begin{figure}
\begin{center}
\rotatebox{90}{\includegraphics[height=0.46\linewidth]{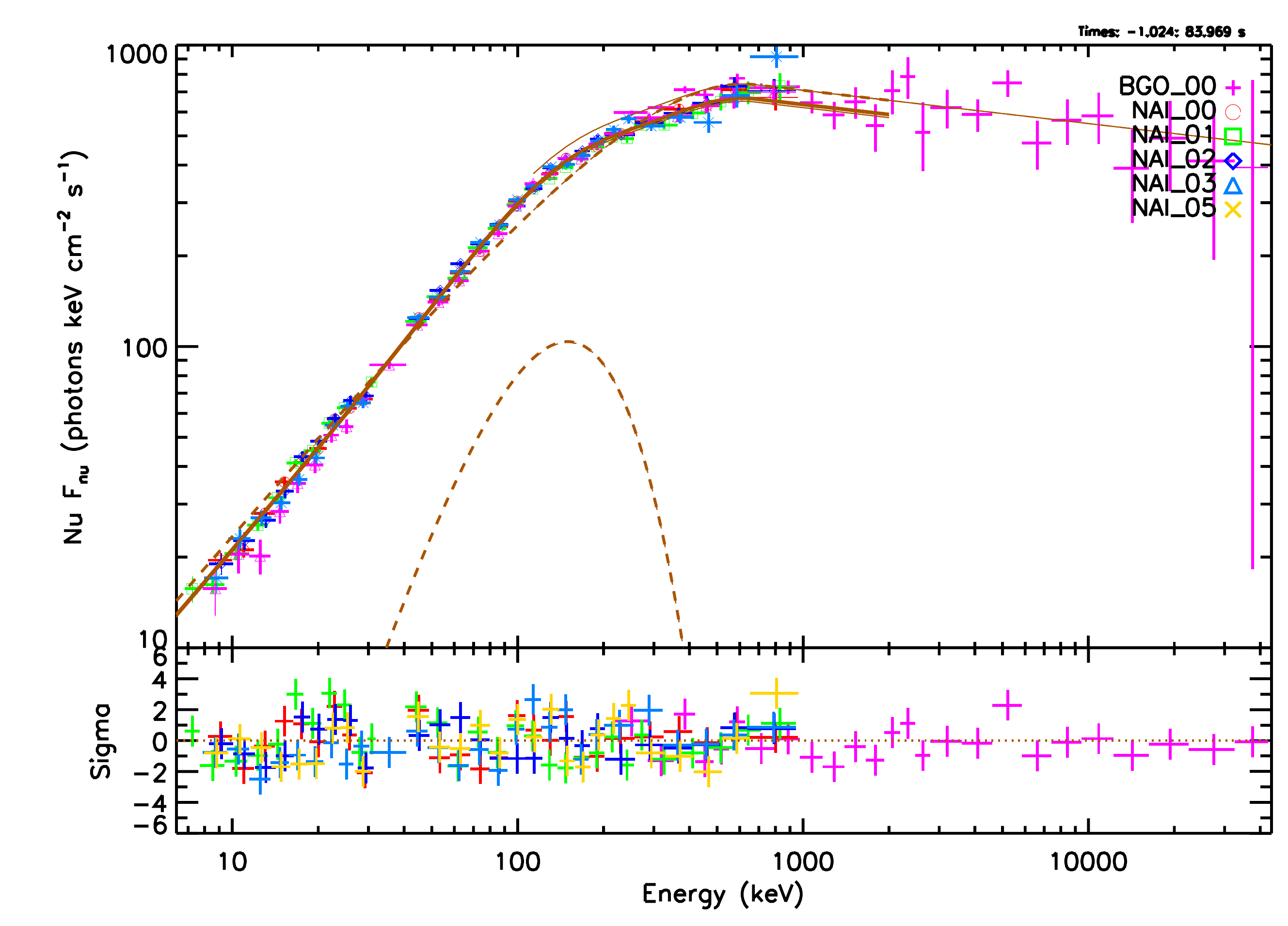}}
\hspace{0.6cm}
\includegraphics[width=0.46\linewidth]{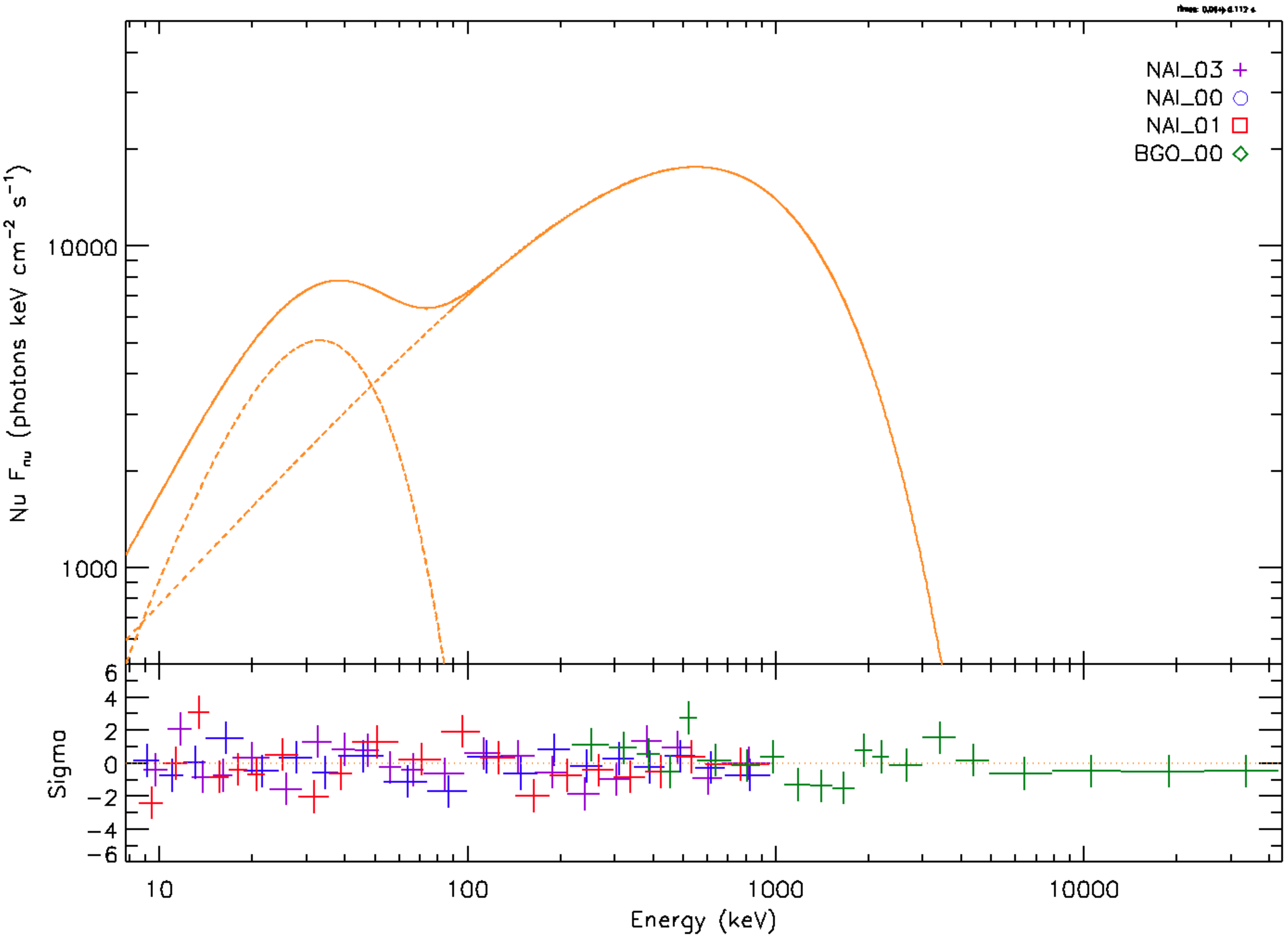}
\end{center}
\caption{Two examples of quasi-thermal components detected in GRB prompt spectra. 
{\bf Left}: a weak quasi-thermal component at $T\simeq 38$ keV in the long GRB 100724B ( from \citealt{Guiriec11}). 
{\bf Right}: a quasi-thermal component at $T\simeq 12$ keV in the short GRB 120323A (from \citealt{Guiriec13}).}
\label{fig:nt_ph_obs}
\end{figure}
\begin{figure}
\begin{center}
\vspace{1.0cm}
\begin{tabular}{cc}
\includegraphics[width=0.48\linewidth,height=4.2cm]{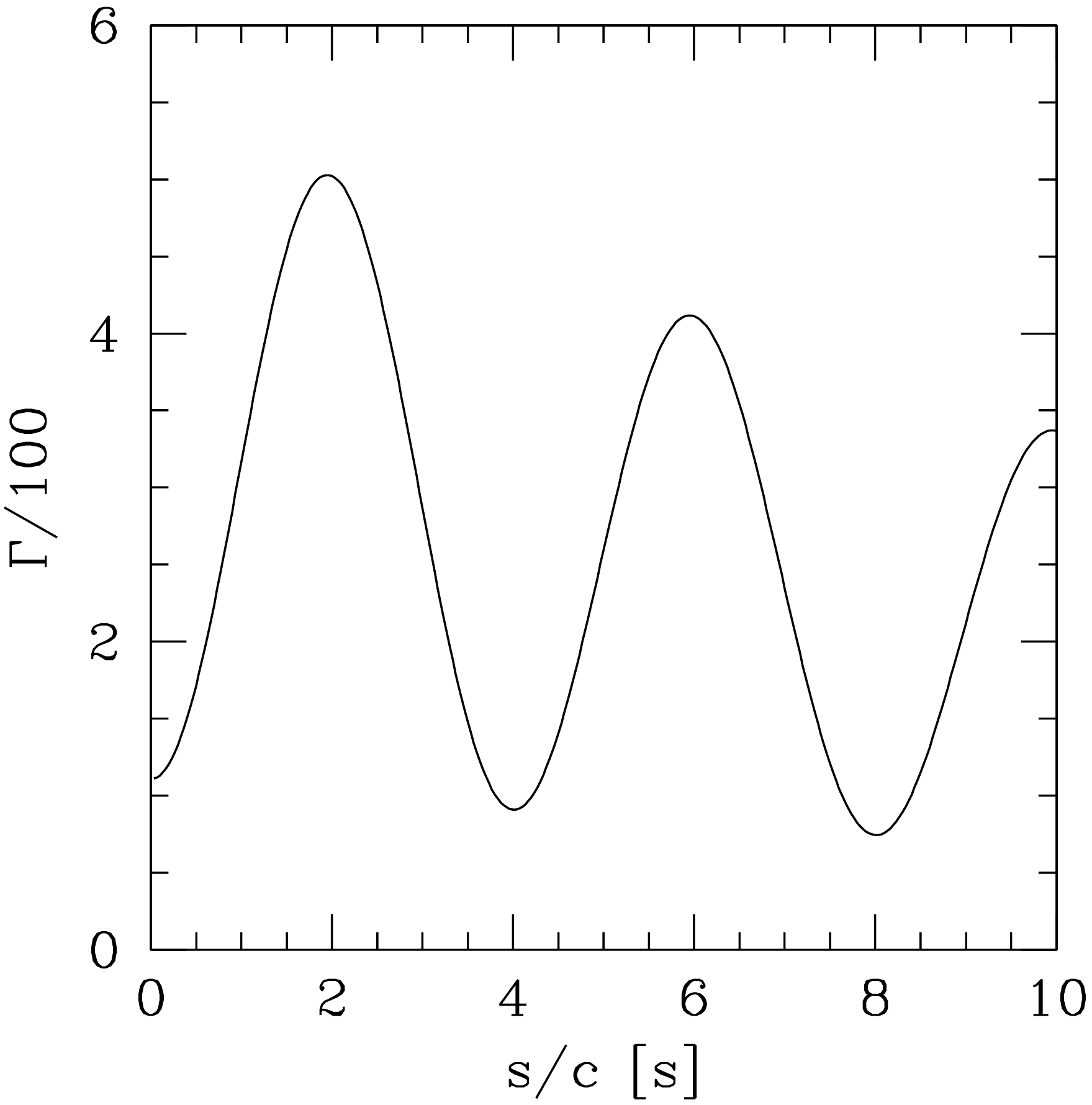} &
\includegraphics[width=0.48\linewidth,height=4.2cm]{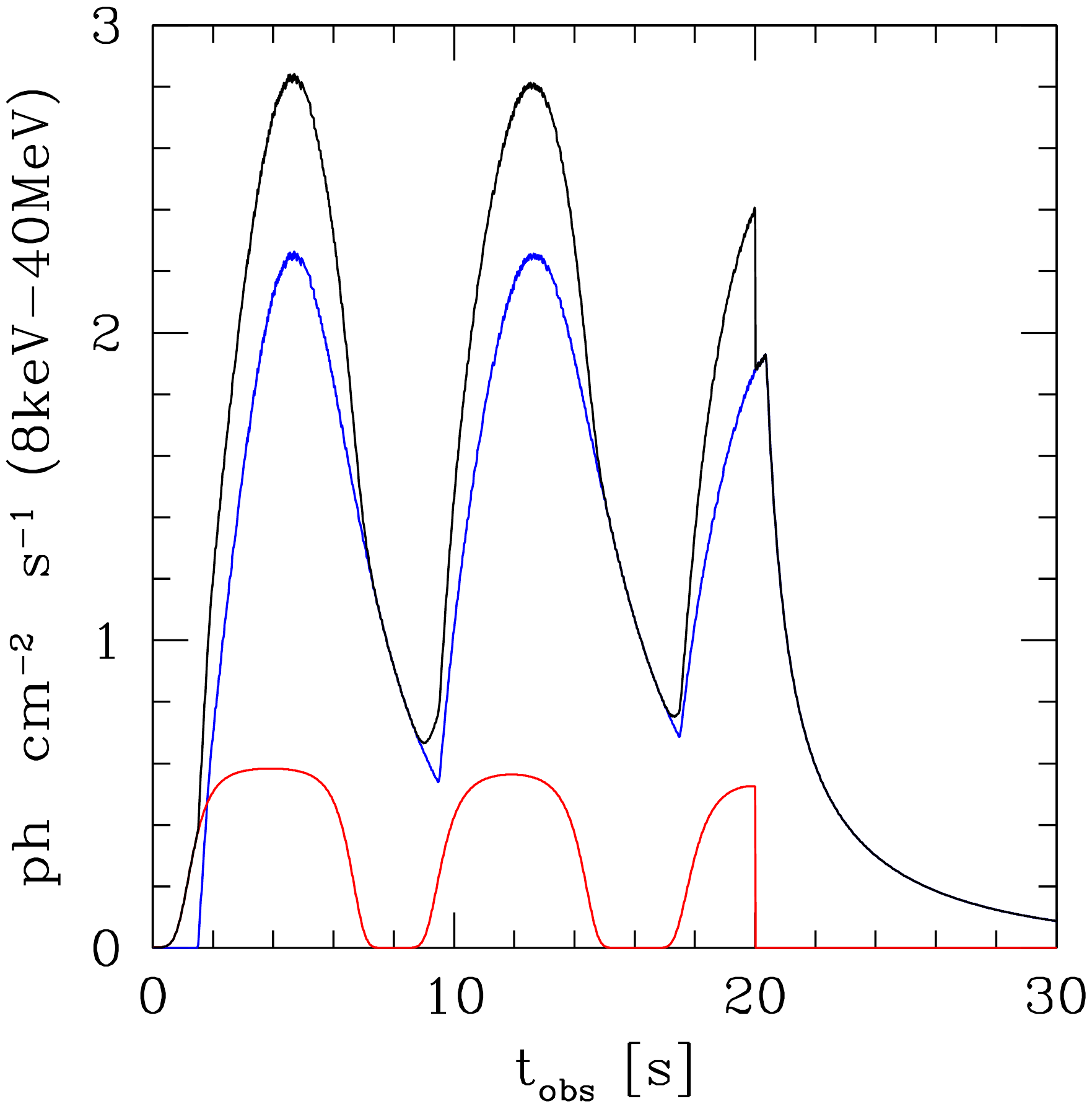} \\ \\
\includegraphics[width=0.48\linewidth,height=4.2cm]{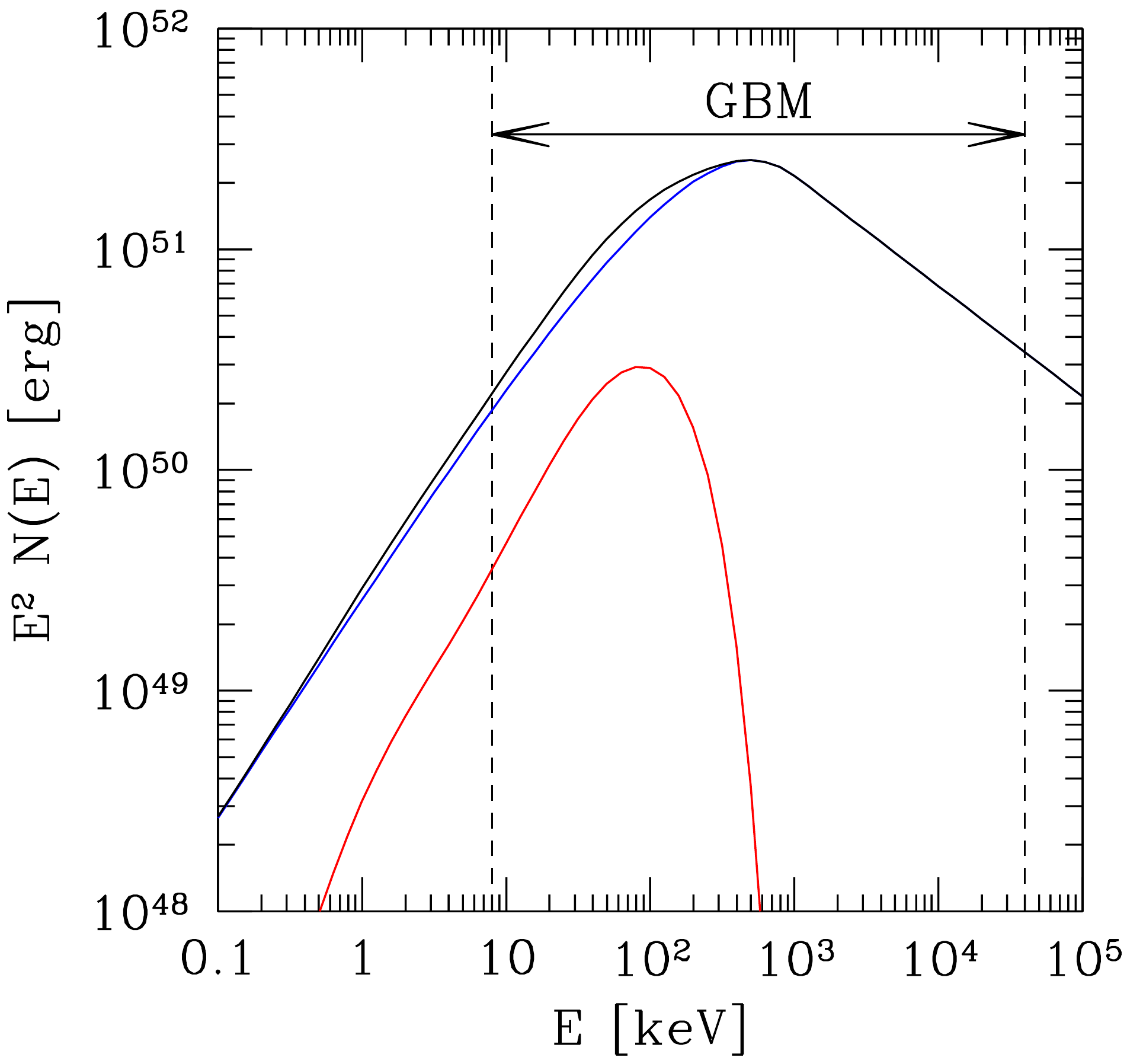} &
\includegraphics[width=0.48\linewidth,height=4.2cm]{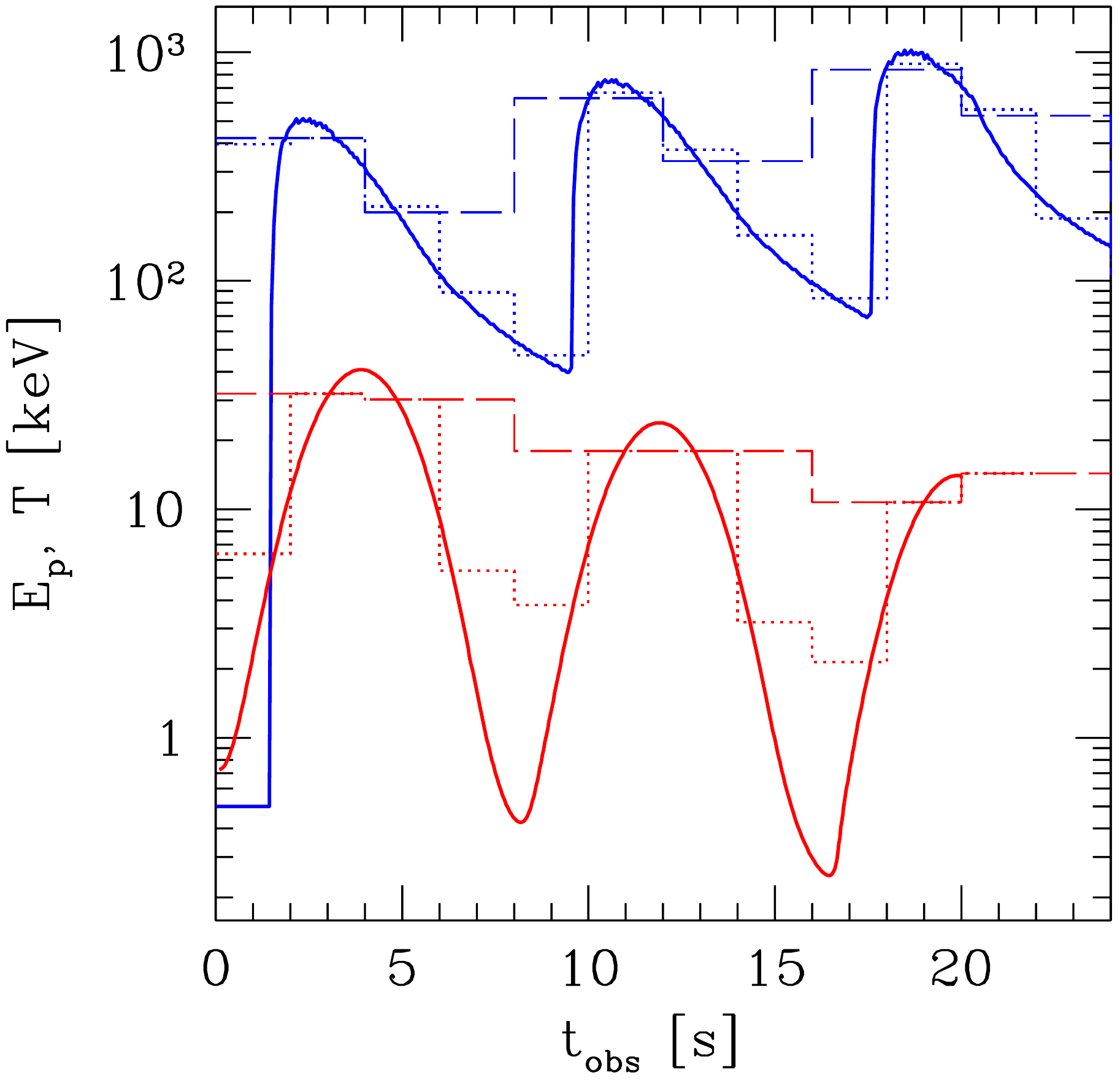} \\
\end{tabular}
\end{center}
\caption{An example of a synthetic GRB with the contribution from the photosphere and  internal shocks computed self-consistently.
The photospheric emission is plotted in red, the non-thermal emission from internal shocks in blue, the total in black. {\bf Top-left}: 
initial distribution of the outflow Lorentz factor at the end of the acceleration phase. {\bf Top-right:} light-curves in the GBM 
energy range. {\bf Bottom-left}: spectrum. {\bf Bottom-right}: spectral evolution (observed peak energy of the non-thermal 
component and temperature of the photosphere). The dashed and dotted lines show the expected result when integrating the 
spectrum over different timescales. The parameters are $\dot{E}=10^{53}\, \mathrm{erg/s}$, $\epsilon_\mathrm{th}=0.03$ 
(high initial magnetization, $\sigma_0=32.3$), $\sigma=0.1$ (low magnetization at large distance), $R_0=3\times 10^7\, \mathrm{cm}$. 
(All panels are taken from \citealt{hascoet:13}).}
\label{fig:nt_ph_theo}
\end{figure}

Recently the description of observed GRB spectra in the soft
$\gamma$-ray range has been greatly improved by \textit{Fermi}/GBM
observations. An important result is the identification of significant
deviations from the Band spectrum, which seem to be related to the
presence of a weak thermal component\footnote{Or possibly even a
  dominant photospheric component in the case of GRB~090902B.}  below
the dominant non-thermal one \citep[see
e.g.][]{Ryde10,Ryde11,Guiriec11,Burgess11,Axelsson12,Guiriec13,Burgess14,Guiriec15},
as illustrated in Figure ~\ref{fig:nt_ph_obs}. A natural explanation
is to associate the quasi-thermal weak component to a (non-dissipative) 
photosphere and the Band component to synchrotron
radiation from electrons accelerated either in shocks or in
reconnection: Figure \ref{fig:nt_ph_theo} shows an example of a
synthetic burst with these two contributions in the case of internal
shocks.  The weakness of the photospheric emission puts interesting constraints on
the initial magnetization of the outflow \citep{daigne:02,hascoet:13},
favoring an efficient magnetic acceleration, with a large range of initial
magnetization in the GRB population, $\epsilon_{\rm th}\lesssim 0.01$ ($\sigma_0\gtrsim100$)
in most cases where no detection is made and 
$\epsilon_{\rm th}\simeq0.01-0.1$ ($\sigma_0\simeq10-100$)
in less frequent cases like GRB 100724B \citep{hascoet:13}. 
GRB~090902B with $\epsilon_{\rm th}\simeq 0.3-1$ ($\sigma_0\lesssim 2.3$) remains an exception
within long GRBs, and the short GRB~120323A appears as an intermediate
case between GRBs 100724B and 090902B with $\epsilon_{\rm th}\simeq
0.1-0.5$ ($\sigma_0\simeq 1-9$) \citep{Guiriec13}. The fact that the photospheric emission seems brighter in 
the only case of detection in a short GRB (GRB 120323A, \citealt{Guiriec13}) may 
indicate a different acceleration mechanism. If this interpretation is correct, these recent 
detections rule out purely thermal acceleration (standard fireball) at least in long GRBs.

\subsection{Constraints on a Poynting Flux Dominated Outflow\label{subsec:paz}} 

Several authors \citep{Lyut06,GS06, ZhangYan11} have proposed that GRB
jets are Poynting flux dominated all the way up to the emission
region. The prompt $\gamma$-rays arise, in this case, from a process
that converts this magnetic energy to radiation.  Obviously, this
cannot take place directly and one has to invoke some sort of magnetic
dissipation (e.g. reconnection) that converts the magnetic energy to
accelerated electrons (or electron-positron pairs) that emit the
observed $\gamma$-rays.  Particular support for this idea came with
the claim of of strong polarization in the prompt emission by
\cite{CoburnBoggs03}, which was later refuted
\citep{RutledgeFox04,Wigger04}. Such polarization could arise if the
magnetic field is ordered and this will arise naturally if the
magnetic field is dominant \citep{GK03,Granot03,LPB03}.

However, the efficiency of the synchrotron emission process poses
serious constraints on models in which the emission region is Poynting
flux dominated \citep{Beniamini14}. Consider a Poynting flux dominated
outflow and an observed (isotropic equivalent) $\gamma$-ray luminosity
$L_\gamma$.  This luminosity immediately sets a lower limit on the
strength of the magnetic field $B$ in the rest frame\footnote{The
  magnetic field in the jet's frame is $B'=B/\Gamma$, where $\Gamma$
  is its bulk Lorentz factor.} of the central source, $L_\gamma <
R_{\rm em}^2B^2c$, where $R_{\rm em}$ is the emission
radius. Accelerated electrons effectively emit synchrotron
radiation. The critical issue here is that synchrotron emission is too
efficient. The accelerated electrons cool so rapidly in a strong
magnetic field that their lower bands (X-rays and optical) synchrotron
emission would produce a signal that is much stronger than the
observed emission in these bands.

\begin{figure}
\vspace{0.5cm}
\hspace{0.7cm}\includegraphics[width=0.98\textwidth]{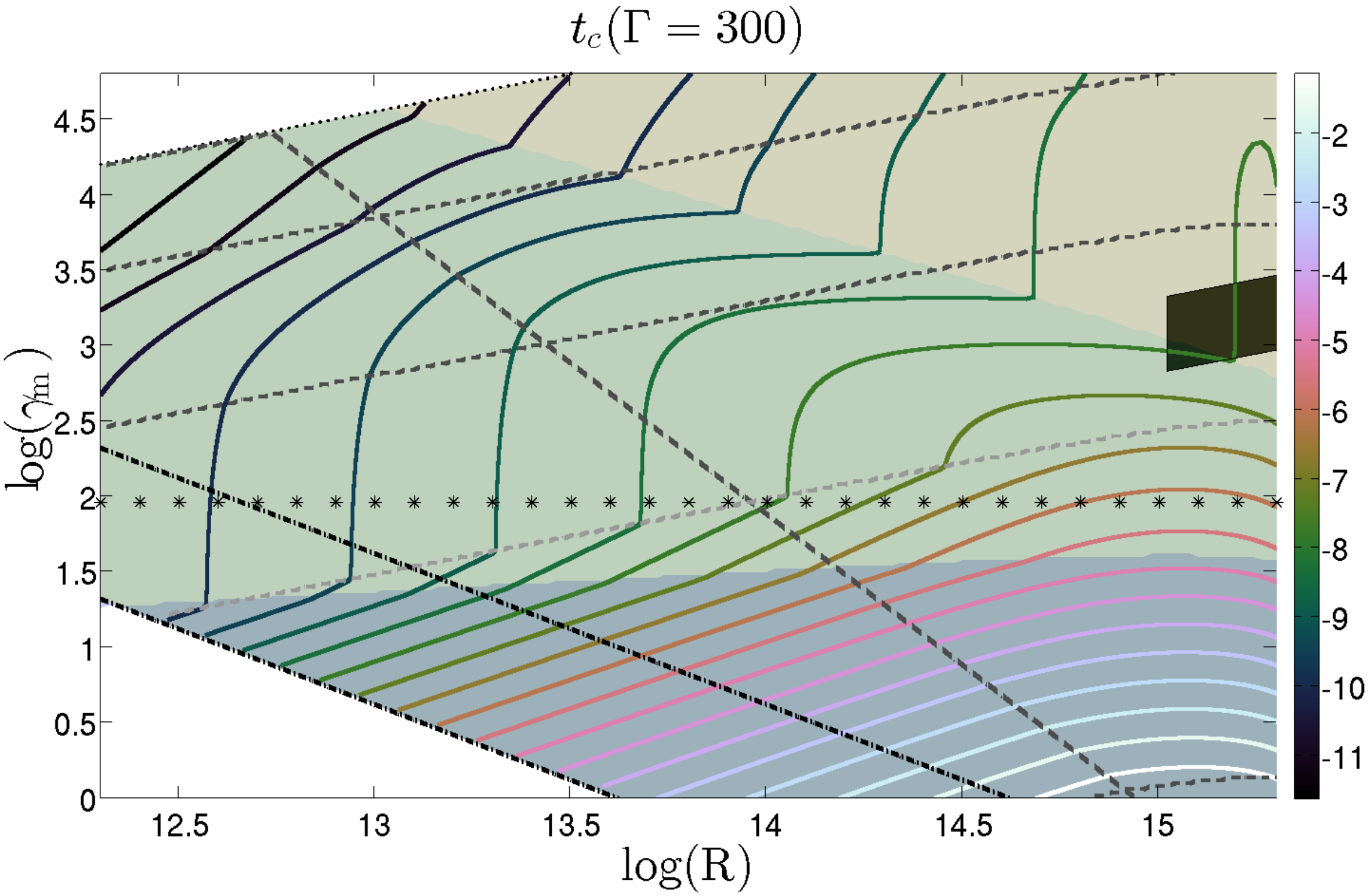}
\vspace{-0.5cm}
\caption{ If the dominant $\gamma$-ray emission mechanism is not
  synchrotron, then in order for it to be able to tap a significant
  fraction of the electrons' energy its cooling time, $t_{\rm c}$,
  must be shorter than that due to synchrotron radiation, $t_{\rm
    c,syn}$, which is depicted here by the contour lines. The
  observational constraints $F_{\nu,{\rm syn,opt}}<1\;$mJy,
  $F_{\nu,{\rm syn,X-ray}}<1\;$mJy and $\nu_{\rm syn,LAT}F_{\nu,{\rm
      syn,LAT}}< 10^{-7}\;{\rm erg\;s^{-1}\;cm^{-2}}$ further
  constrain $t_{\rm c}$.  Beyond the corresponding lines, $t_{\rm c}$
  should be significantly shorter than $t_{\rm c,syn}$ in order for
  the synchrotron not to overproduce the upper limits on the optical,
  X-ray or GeV fluxes. Within the black region the synchrotron
  emission produces the observed prompt $\gamma$-rays. The conditions
  $\tau_T<1,\,10$ (dot-dashed lines; a Thompson optical depth that is
  not too large) define general limits on the parameter space.  (see
  \citealt{Beniamini14} for more details).}
\label{fig:synch}
\end{figure}

The observed prompt upper limits in the optical or the X-rays set
strong constraints on the conditions within the emitting region.
First, if the observed $\gamma$-rays are due to some other
(non-synchrotron) emission process then this process must be extremely
efficient and its cooling time should be significantly shorter than
the relevant synchrotron cooling time (see e.g. Figure
~\ref{fig:synch}).

Alternatively, if the observed prompt $\gamma$-ray emission is
synchrotron then there must be a rapid reaccelerating process that
keeps the electrons with the right Lorentz factor so that they would
not cool too much and emit strongly in lower energy bands, in
particular in soft X-rays
\citep{GhiselliniCelotti99,KumarMcMahon08,Fan10}. This requires strong
fine tuning as the Lorentz factor range in which the electrons must be
kept is rather narrow (a factor of $\sim 3\,$--$\,$10) .  Multi-zone
configurations in which the electrons escape the emitting region
before cooling and over-producing X-ray or optical emission are also a
possibility \citep[e.g. as in the ICMRAT model][]{ZhangYan11}.
\cite{Beniamini14} considered several such two-zone toy models (in
which electrons are accelerated in one region and emit in the other)
but proper conditions could not be found in any of them.
 
 These considerations pose severe constraints on prompt emission
 models that involve Poynting flux dominated outflows.  Any emission
 model in such a regime should satisfy these constraints. Lacking a
 model that satisfies all these constraints, it is likely that if the
 outflow is initially Poynting flux dominated then the magnetic energy
 is dissipated before the emitting region, where it must be
 subdominant.

\subsection{Spectral Diversity -- Spectral Evolution}

The peak energy $E_\mathrm{p}$ varies a lot from one GRB to another, from a few to
tens of keV (X-ray Flashes, X-ray Rich GRBs; \citealt{sakamoto:05}) to
over 10$\;$MeV \citep{Axelsson12}. An important property is that 
short GRBs are harder with larger peak energies
\citep{Kouveliotou93,guiriec:10}. Spectral evolution is also found:
when time-resolved spectroscopy is possible, the GRB spectrum 
is always found to strongly evolve during the prompt phase (see
e.g. \citealt{lu:12,Burgess14,Guiriec15} for recent analyses of
\textit{Fermi} GRBs, or \citealt{preece:14} for a very bright case
where the spectral evolution can be studied in great detail); 
$E_\mathrm{p}$ typically varies over more than a factor of 30 within 
an individual pulse. Spectral and temporal properties appear correlated
within GRB pulses: hardness following the intensity, pulses being
narrower and peaking earlier at higher energies, etc.

Both the spectral diversity between different GRBs and the spectral 
evolution within individual GRBs are hard to reproduce by models. 
Reconnection models are barely developed enough to allow 
discussion of these observations.  In dissipative photospheres, 
variations in $E_\mathrm{p}$ are related to changes in the properties 
of the outflow ejection leading to a change in the location of the 
photosphere \citep{peer:08,beloborodov:13,deng:14}. A potential issue 
is to explain how the dissipative process adjusts to always remain located
just below the photosphere (unless it always occurs over a wide range
of radii, in which case it should also occur above the photosphere, so
this would not be a pure photospheric model).  In internal shocks, the
spectral evolution is reproduced qualitatively
\citep{daigne:98,daigne:03,asano:11,bosnjak:14}, and can even be
reproduced quantitatively with some constraints on microphysics
parameters \citep{bosnjak:14}, which may indicate non-universal values
in mildly relativistic shocks, as suggested for instance by
\citet[][see also \citealt{bykov:12}]{bykov:96}. The spectral
diversity is also naturally explained by variations in the lifetime
and variability of the central engine \citep{barraud:05}. The
hardness-duration relation is well reproduced
\citep{daigne:98,bosnjak:14}.

\subsection{Constraints at Other Wavelengths}
The discussion above was centered on observations in the soft $\gamma$-ray range, where the prompt emission is observed in most 
GRBs. We discuss here briefly some additional constraints coming from observations of the prompt emission at other wavelengths.

\subsubsection{The End of the Prompt Emission: the X-ray Early Steep Decay}

\textit{Swift}/XRT discovered in most GRB X-ray afterglows an early
steep decay at the end of the prompt phase, before
recovering a plateau and/or a standard afterglow decay
\citep{nousek06,zhang06,obrien06}. A natural explanation is provided by the
high-latitude tail of the prompt emission, once the on-axis emission
has stopped \citep{kumar:00,genet:09}.  It can reproduce the observed
temporal decay and spectral evolution
\citep{liang:06,willingale:10}. This puts a strong constraint on the
emission radius at the end of the prompt phase, which may be fulfilled
by internal shocks, and also possibly by reconnection models if the
radius is large enough \citep{hascoet:12b}. On the other hand, it is
incompatible with photospheric models, which must instead explain the
rapid decay phase by a universal behavior of the central engine when
it is switching off.

\subsubsection{Prompt GeV Emission}

\textit{Fermi}/LAT detects GeV emission in some GRBs \citep{ackermann:13}. 
As detection requires enough photons in its energy range (tens of MeV to 
$\gtrsim 300\;$GeV), it detects mainly very bright GRBs, in terms of both their 
GeV fluence and their total fluence (and thus also in terms of $E_{\gamma,{\rm iso}}$). 
For the same reason, LAT detects a smaller fraction of short GRBs compared to
soft $\gamma$-ray instruments, since their fluence is typically much
smaller than that of long GRBs. Bright enough LAT GRBs show a
distinct high-energy spectral component, usually fitted by a power law
\citep[e.g.,][]{GRB090902B,GRB090510,ackermann:13}.  The observed
variability in the prompt LAT light curve indicates an internal origin. 
It is followed by a long-lasting emission (with a power law in
time and energy) that likely originates from the deceleration phase or 
early afterglow.  In dissipative photospheric models, it is hard to produce 
GeV photons due to strong $\gamma\gamma$ annihilation.  
However, additional processes such as later scatterings
of prompt photons by the external medium can explain this GeV emission
\citep[see e.g.][]{beloborodov:14}. In reconnection models, spectral
models cannot make such predictions yet.  In internal shocks, such
multi-component spectra are expected
\citep{GG03,bosnjak:09,asano:12,bosnjak:14}; the fact that the GeV
component is usually weaker than the soft $\gamma$-ray component 
constrains the strength of the magnetic field, implying that it must
be weak, $\epsilon_\mathrm{B}\lesssim 10^{-2}$
\citep{daigne:11,bosnjak:14}.

\subsection{Prompt Emission Summary}

The dissipation mechanism and radiative processes responsible for the
prompt GRB emission are still not well understood due to the complex
physics involved, both on large and micro-scales. The lack of strong thermal 
components in GRB spectra suggests a high initial magnetization in GRB outflows, 
while prompt GRB observational constraints imply
a low magnetization in the emission region. Put together, this
strongly suggests either very efficient conversion of magnetic to
kinetic energy, which leaves a low magnetization in the emission
region, and allows for efficient internal shocks
\citep{GKS11,Granot12b}, or strong magnetic reconnection that converts
magnetic energy to thermal energy and accelerates particles, and yet
somehow leaves a low enough magnetization where these particle radiate
most of their energy.

A weak thermal emission can be produced at the photosphere,
followed by a dominant non-thermal emission at larger radii. 
Depending on the efficiency of the acceleration and the resulting
magnetization at large distances, the dissipation leading to the
emission can occur either in shocks or in magnetic reconnection.  In
both cases, the dominant radiative process should be synchrotron
emission. Only in the first case (internal shocks), detailed
simulations coupling a dynamical calculation with a detailed radiative
model are available.  To have a good agreement between the observed
spectrum and the predicted one, detailed modeling is needed, where
the strength and structure of the magnetic field play a crucial
role: moderately efficient inverse Compton scatterings in the
Klein-Nishina regime are needed, which requires a weak field, and a
decay of the magnetic field far from the shock front is also probably
required.

An alternative is to explain the whole soft $\gamma$-ray emission by a
dissipative photosphere.  Its nature, however, must then be elucidated. 
The magnetic field could again play a vital role, 
via the magnetic reconnection below the photosphere, which is a
natural candidate.  Further progress can come from more observations
over a broad spectral range with time-resolved spectroscopy,
additional and firmer polarization measurements, and improvements
in the modeling of the expected spectrum and spectral evolution
in each model. Unfortunately, it remains limited by the current 
knowledge of the microphysics (structure of the magnetic field, particle 
acceleration) in mildly relativistic shocks and magnetic reconnection
\citep{SKL15,Kagan15}.

\section{Magnetic Fields in the Afterglow\label{afterglow}}

Eventually, the GRB outflow is decelerated by the external medium. It
drives a strong relativistic blast wave -- the afterglow (or external
forward) shock -- into the surrounding medium. It transfers most of
its energy to the shocked external medium (via $pdV$ work across the
contact discontinuity that separates them) at a distance $R_{\rm dec}$
from the central source -- the deceleration radius. Radiation from
$R_{\rm dec}$ reaches the observer at the deceleration time, $T_{\rm
  dec}$. At $R>R_{\rm dec}$ the original outflow composition no longer
affects the dynamics (or emission) of the afterglow shock. However,
the outflow magnetization can greatly affect the reverse shock (or
external reverse shock, as it is formed due to the interaction with
the external medium), whose strength and emission can be greatly
suppressed if the outflow is highly magnetized, $\sigma(R_{\rm
  dec})\gtrsim 1$.

\subsection{The Afterglow Emission}
\label{sec:AG_emission}

\begin{figure}
\centerline{\includegraphics[width=0.95\textwidth]{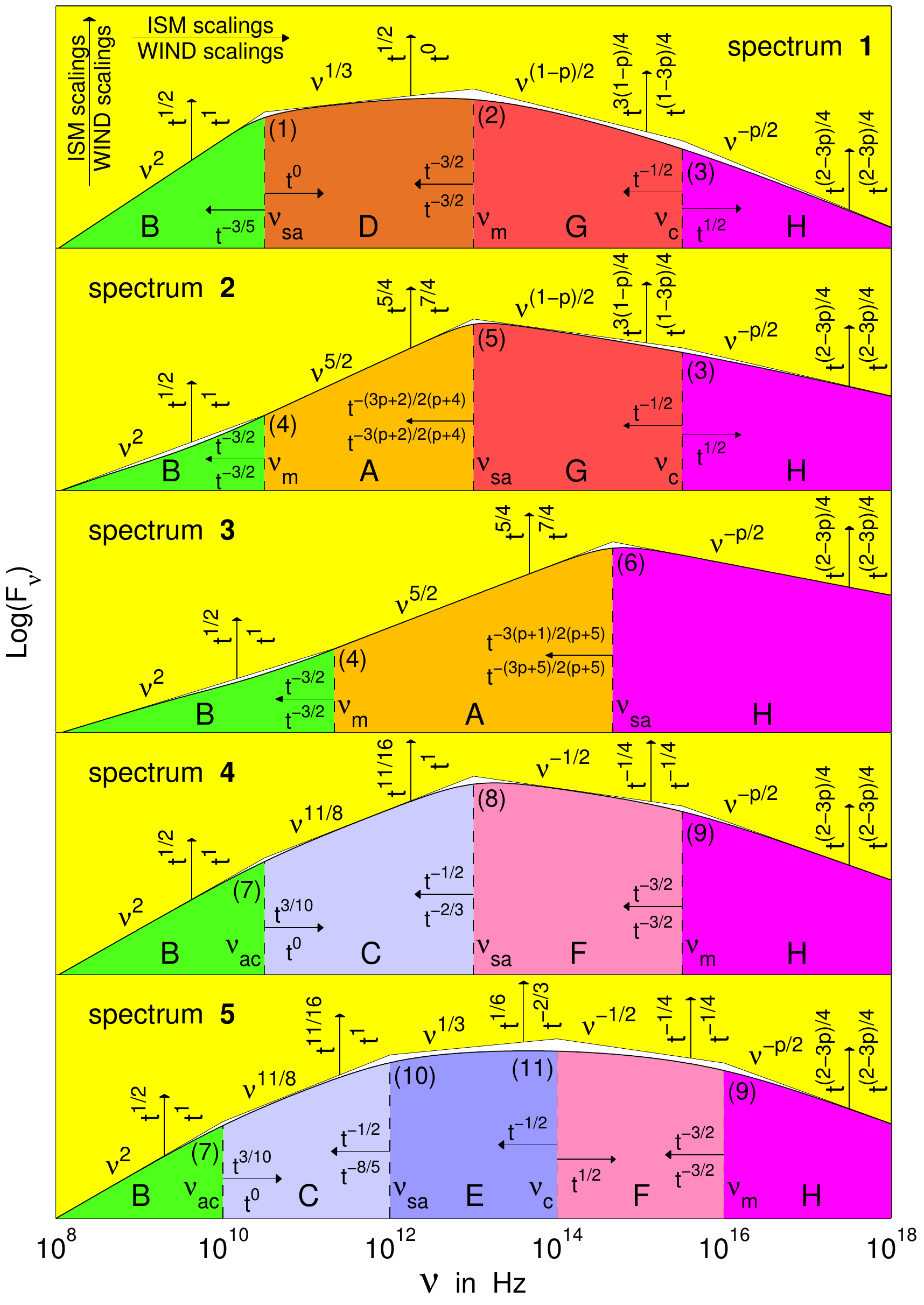}} 
\caption{The afterglow synchrotron spectrum, calculated for the \cite{BM76} spherical self-similar solution, under
standard assumptions, using the accurate form of the synchrotron spectral emissivity and integration over the emission 
from the whole volume of shocked material behind the forward (afterglow) shock 
\citep[for details see][from which this figure is taken]{GS02}. 
The different panels show the five possible broad band spectra of the afterglow synchrotron emission, each corresponding 
to a different ordering of the spectral break frequencies. Each spectrum consists of several power-law segments (PLSs; 
each shown with a different color and labeled by a different letter A--H) that smoothly join at the break frequencies 
(numbered 1--11). The broken power-law spectrum, which consists of the asymptotic PLSs that abruptly join at the break 
frequencies (and is widely used in the literature), is shown for comparison.  Most PLSs appear in more than one of the five 
different broad band spectra. Indicated next to the arrows are the temporal scaling of the break frequencies and the flux 
density at the different PLSs, for a uniform (ISM; $k=0$) and stellar wind (WIND; $k=2$) external density profile.}
\label{fig:spectrum}
\end{figure}

The dominant emission mechanism in the afterglow is thought to be
synchrotron radiation, which is produced by relativistic electrons
accelerated at the afterglow shock that gyrate in the magnetic fields
within the shocked external medium.\footnote{In an alternative scenario, 
the afterglow emission is dominated at early times by the contribution of 
a long-lived reverse shock \citep{uhm:07,genet:07}, which allows to reproduce 
more easily the observed diversity and variability, such as X-ray plateaus
\citep{uhm:12,hascoet:14} or X-ray flares \citep{hascoet:15}, though in this 
scenario a transition to forward shock dominance is expected at late times but 
not observed.}  Such a synchrotron origin of the
afterglow emission is strongly supported by the detection of linear
polarization at the level of $\sim\,$1$\,$--$\,$3\% in several optical
or NIR afterglows (see \S\ref{sec:pol}), and by the shape of the
broadband spectrum, which consists of several power law segments that
smoothly join at some typical break frequencies.
Figure~\ref{fig:spectrum} shows the possible resulting afterglow
spectra.  Broadband (radio to $\gamma$-ray) afterglows fit such
synchrotron spectra far better than the prompt emission.  The broad
and mostly featureless smoothly broken power-law shapes of afterglow
spectra evolve and fade more slowly over time, and have characteristic
frequencies that vary as a power law with time, roughly according to
the theoretically expected power-law indices \citep{SPN98,GS02}.
Synchrotron self-Compton (SSC) -- the inverse-Compton scattering of
the synchrotron photons to (much) higher energies by the same
population of relativistic electrons that emits the synchrotron
photons -- can sometimes dominate the afterglow flux in the X-rays
\citep{SE01,Harrison01}, and may affect the synchrotron emission by
increasing the electron cooling.

Relativistic collisionless shock physics (e.g., how they amplify the 
magnetic field and accelerate a non-thermal population of relativistic 
particles) are still not well understood from first principles 
\citep[e.g.][]{SKL15}. Thus, simple assumptions are usually made 
that conveniently parameterize our ignorance. The electrons are 
assumed to be (instantly) shock-accelerated into a power-law
distribution of energies, $dN/d\gamma_e \propto \gamma_e^{-p}$ for
$\gamma_e > \gamma_m$, and then cool both adiabatically and due to
radiative losses.\footnote{It is usually also further assumed that
  practically all of the electrons take part in this acceleration
  process and form such a non-thermal (power-law) distribution,
  leaving no thermal component \citep[which is not at all clear or
  justified; e.g.][]{EW05}.} The relativistic electrons are assumed to
hold a fraction $\epsilon_e$ of the internal energy immediately behind
the shock, while the magnetic field is assumed to hold a fraction
$\epsilon_B$ of the internal energy everywhere in the shocked
region. Both the temporal and spectral indices depend on the power law
index $p$ of the electron energy distribution. The temporal index
(i.e. the rate of flux decay) also depends on the circumburst density
profile, which is parameterized in Figure~\ref{fig:spectrum} as a
power law of index $k$ with the distance $R$ from the central source,
$\rho_{\rm ext}\propto R^{-k}$, with $k=0$ and $k=2$, respectively,
corresponding to an ISM and a stellar wind -- WIND.  The temporal
index can also be affected by other factors, such as energy losses or
injection into the afterglow shock, the afterglow jet angular
structure and the viewing angle relative to the jet symmetry axis, or
time evolution of the shock microphysics parameters $\epsilon_e$
and/or $\epsilon_B$.

\subsection{Polarization: Afterglow and Reverse Shock}
\label{sec:pol}

\begin{figure}
\begin{center}
\includegraphics[width=0.45\textwidth]{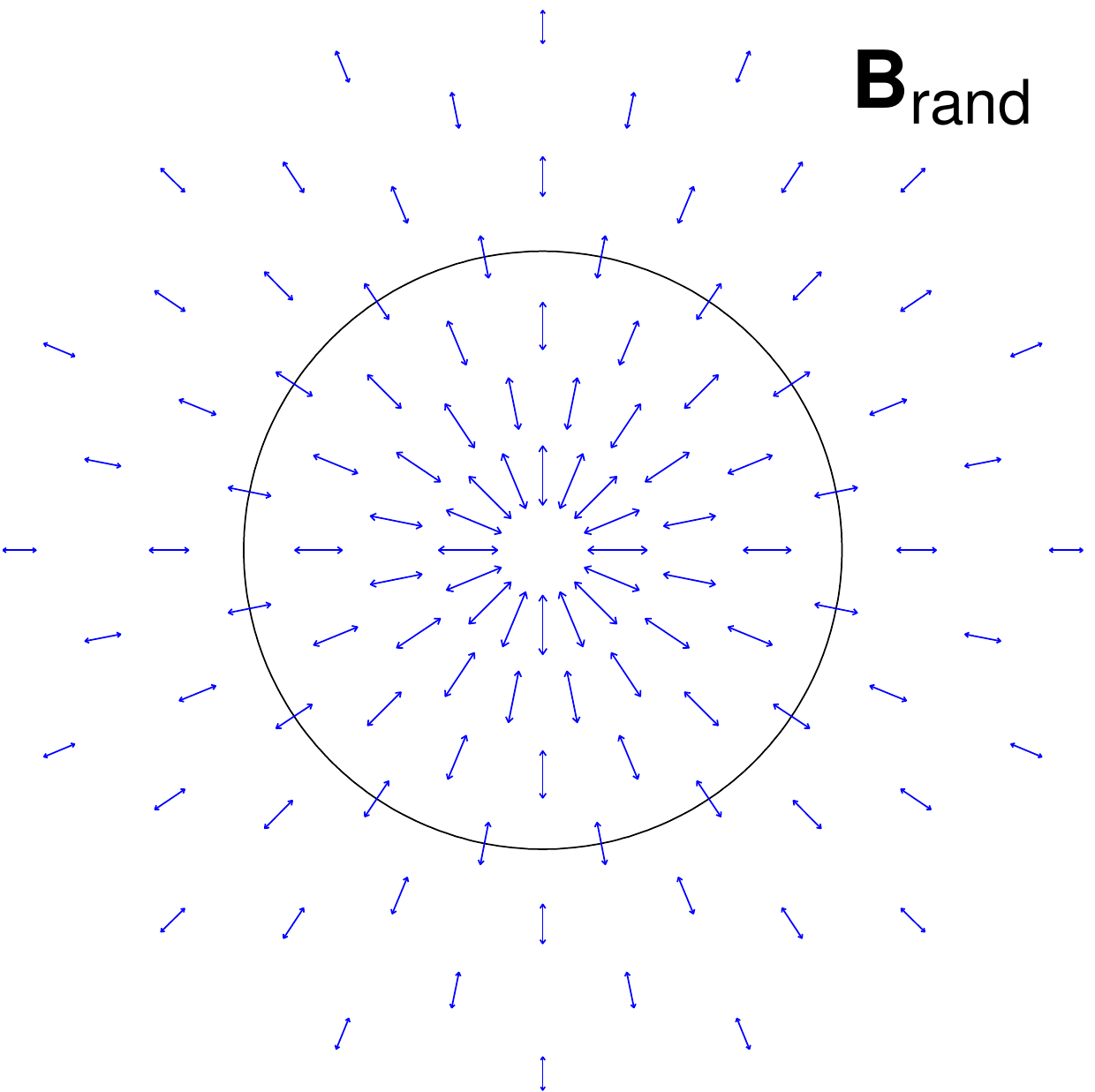} 
\hspace{0.9cm}
\includegraphics[width=0.45\textwidth]{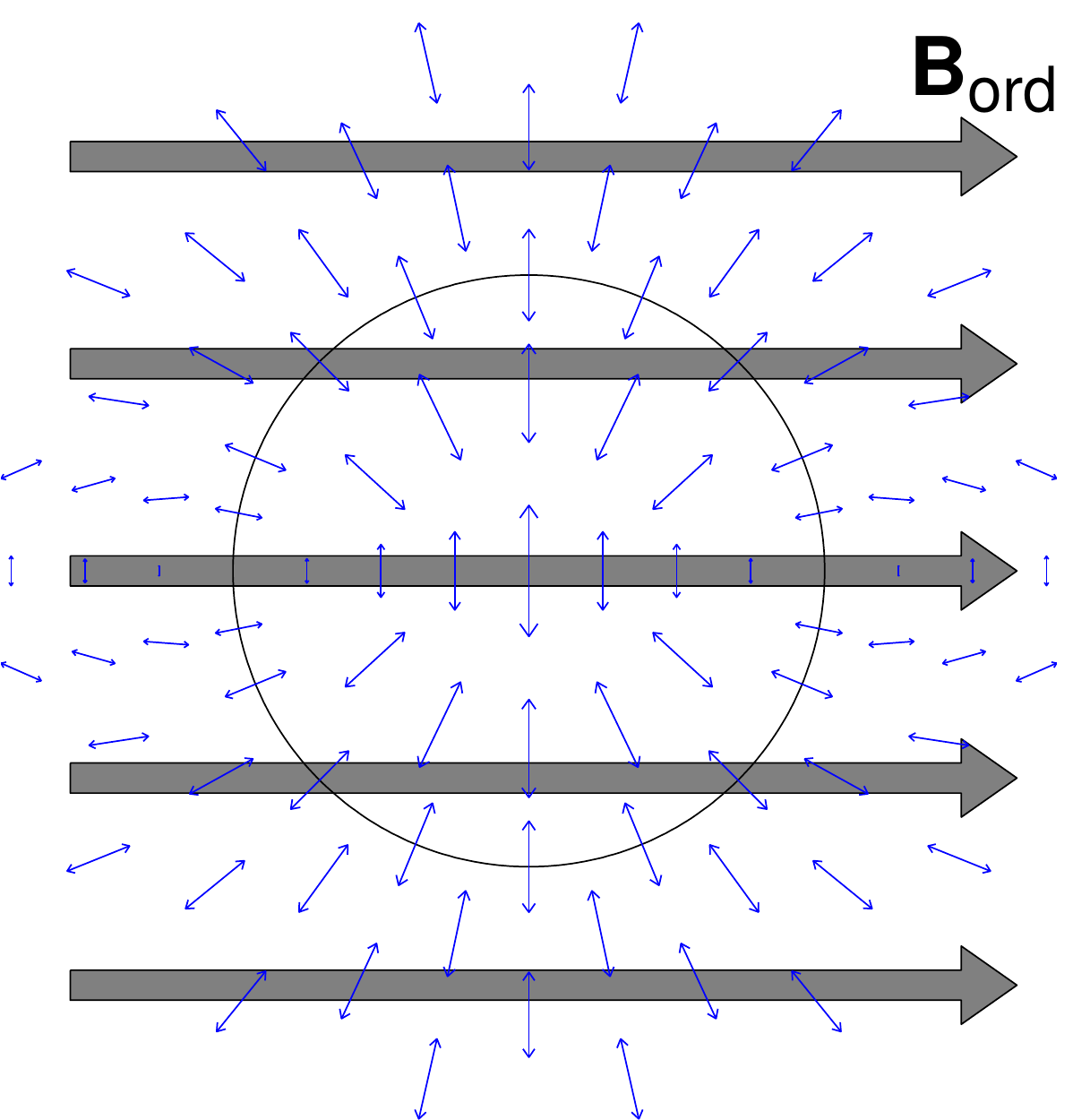}
\end{center} 
\caption{The predicted polarization map for synchrotron emission from a thin spherical ultra-relativistic shell expanding with 
a Lorentz factor $\Gamma \gg 1$. The double-sided arrows show the direction of the linear polarization (the wave electric vector), 
while their length depends monotonically on the polarized intensity (in a non-trivial way, for display purposes). 
The circle indicates an angle of $1/\Gamma$ around the line of sight to the central source, and contains the region responsible 
for most of the observed flux. {\bf Left}: for a magnetic field that is random within the plane of the shell 
(normal to the radial direction), for which the polarization direction always points at the center of the image, 
where the polarization vanishes (due to symmetry consideration). {\bf Right}: for an ordered magnetic field within the 
plane of the shell that is coherent over angular scales $\gg 1/\Gamma$ \citep{GK03}. In this case the direction of the ordered 
magnetic field clearly breaks the symmetry around the center of the image, resulting in a large net polarization. For simplicity, 
the map is for a constant emission radius, rather than for a constant photon arrival time.}
\label{fig:plo_ord-rand}
\end{figure}

The detection of linear polarization of a few percent in the optical
and NIR {\bf afterglow} of several GRBs \citep[see][and references
therein]{Covino04} was considered as a confirmation that synchrotron
radiation is the dominant afterglow emission mechanism. The
synchrotron emission from a fluid element with a locally uniform
magnetic field is linearly polarized in the direction perpendicular to
the projection of the magnetic field onto the plane normal to the wave
vector. Since the source moves relativistically, one must account for
aberration of light when calculating the observed local direction of
polarization. Figure~\ref{fig:plo_ord-rand} shows the predicted local
polarization map from emission by an ultra-relativistic expanding
shell, for two different magnetic field structures: a magnetic field
that is random within the plane normal to the radial direction ({\it
  left panel}) as could be expected from a shock-produced field 
  \citep[e.g.,][]{ML99}, and an ordered magnetic field normal to
the radial direction ({\it right panel}; as could be expected in the prompt or
reverse-shock emission for a magnetic field coherent on angular scales
$\gg 1/\Gamma$ that is advected from the central source).

The afterglow image is almost always unresolved, so we can only
measure the (weighted) average polarization over the whole image.  Therefore, a
shock produced magnetic field that is symmetric about the shock normal
will procure no net polarization for a spherical flow (as in this case
the polarization pattern across the image is symmetric around its
center, and the polarization averages out to zero when summed over the
the whole image).  For a shock-produced magnetic field, one thus needs
to break this symmetry of the emission to produce net polarization.  A
simple and natural way of doing this is considering a jet, or narrowly
collimated outflow \citep[e.g.,][]{Sari99,GL99}. In this picture a jet
geometry together with a line of sight that is not along the jet
symmetry axis (but still within the jet aperture, in order to see the
prompt GRB) is needed to break the symmetry of the afterglow image
around our line of sight. Other models for afterglow polarization
include a magnetic field that is coherent over patches of a size
comparable to that of causally connected regions \citep{GW99},
polarization that is induced by microlensing \citep{LP98} or by
scintillations in the radio \citep{ML99}, a small ordered magnetic
field component originating from the circumburst medium \citep{GK03},
clumps in the external medium \citep{GK03}, or a very inhomogeneous
jet angular structure -- a ÔÔpatchy shellÕÕ with
ÔÔhot spotsÕÕ \citep{GK03,NO04}.
The many possible causes of polarization, and the degeneracy with other factors makes it difficult to robustly determine 
the magnetic field structure in the emitting region from afterglow polarization measurements. Nonetheless, a high degree
of linear polarization with a stable position angle is hard to produce without a magnetic field that is ordered on large scales.

The {\bf reverse shock} has two main observational signatures: a
sharply-peak ``optical flash" \citep[e.g.,][]{Akerlof99} on a
timescale comparable to the prompt GRB $T_{90}$, and a ``radio flare"
\citep[e.g.,][]{Kulkarni99,Frail00,Berger03} that peaks on a timescale
of a day or so after the GRB.  In such cases, if the relatively bright
observed emission is indeed from the reverse shock, this implies that
the outflow was not strongly magnetized near the deceleration epoch,
$\sigma(R_{\rm dec})\ll 1$. Moreover, the polarization properties of
the synchrotron emission from the reverse shock provide a powerful and
unique probe for the magnetic field structure in the original
outflow. Early optical polarization measurements from $ T\gtrsim
T_{\rm dec}$ have finally been obtained in the last eight years or so
\citep{Mundell07,Mundell13,Steele09}.

On the one hand, there is a strict upper limit on the degree of linear
polarization from GRB~060418 of $P<8\%$ (2-$\sigma$) at $T=203\;$s
after the GRB trigger, while the deceleration time suggested by the
early optical lightcurve of this GRB is $T_{\rm dec} = 153\pm10\;$s
\citep{Molinari07} and its prompt emission lasted only $T_{90} =
52\pm 1\;$s.  The fact that $T_{\rm dec}/T_{90}\approx 3$ suggest a
``thin shell" in this case, which is consistent with a moerate
magnetization ($\sigma(R_{\rm dec})\ll 1$) that allows a strong
reverse shock with bright emission. However, the polarization is
fairly low near $T_{\rm dec} = 153\pm10\;$s (at $T=203\;$s), which
suggests that either the reverse shock emission even near its peak is
for some reason greatly sub-dominant compared to the (very weakly polarized) 
forward shock emission, or more likely that in this GRB there is hardly any 
ordered magnetic field in the ejecta on angular scales $\gtrsim 1/\Gamma$ 
that cover most of the visible region.

On the other hand, GRB~090102 had a prompt duration of $T_{90} =
27\;$s and an optical linear polarization of $P=10.2\pm 1.3\%$ in a
$60\;$s exposure starting at $T=161\;$s after the trigger time
\citep{Steele09}.  Its optical lightcurve shows a power-law decay
$F_\nu\propto t^{-\alpha}$ with $\alpha = 1.50\pm0.06$ from $T\sim
40\;$s to $T\sim1000\;$s and then flattens to $\alpha = 0.97\pm0.03$
\citep{Gendre10}.  This suggests a deceleration time $T_{\rm
  dec}\lesssim 40\;$s, well before the polarization measurement.
However, the optical emission may be dominated by the reverse shock up
to the break time of $\sim 1000\;$s, and in particular during the
polarization measurement. In the latter case this might possibly
explain the measured polarization as arising from an ordered magnetic
field component in the ejecta, though a purely ordered field on the
scale of the whole emitting region (angular scale $\gtrsim 1/\Gamma$)
would produce a significantly larger polarization of several tens of
percent \citep{GK03,Granot03,LPB03,NPW03}, which would suggest either
a smaller magnetic field coherence length or a dominant contribution
from the much less polarized external forward shock emission.

Finally, GRB~120308A that lasted $T_{90}\sim 100\;$s (between
$T\approx-30\;$s and $T\approx70\;$s post-trigger) showed an optical
linear polarization of $P=28\pm4\%$ in an exposure between $T=240\;$s
and $323\;$s \citep{Mundell13}, which gradually decreased 
to $P=16^{+5}_{-4}\%$ over the next ten minutes, while keeping an
approximately constant position angle (to within an accuracy of about
$15^\circ$). The optical lightcurve peaked at around $T\sim 300\;$s,
during the time bin in which the largest polarization was measured,
and subsequently decayed, with a possible transition from reverse to
forward shock domination of the optical emission around $\sim
1000\;$s. This strongly suggests the presence of a large-scale ordered
magnetic field in the original GRB ejecta.

Observations of radio flares at roughly a day after the GRB have so
far produced no detection of polarization. However, these observations
have enabled to set strict upper limits on a possible linear or
circular polarization \citep{GT05}.  The strictest limits are for
GRB~991216, for which the 3-$\sigma$ upper limits on the linear and
circular polarization are $P_{\rm lin}<7\%$ and $P_{\rm circ}<9\%$,
respectively.  These limits provide interesting constraints on
existing GRB models \citep{GT05}, and in particular are hard to
reconcile with a predominantly ordered toroidal magnetic field in the
GRB outflow together with a ``structured" jet, where the energy per
solid angle drops as the inverse square of the angle from the jet
axis, as is expected in some models in which the outflow is Poynting
flux dominated.

Recently, the detection of circular polarization was reported in the
optical afterglow of GRB 121024A by \cite{Wiersema14}.  In particular,
they measured a circular polarization of $P_{\rm circ} = 0.61\pm0.13\%$
at $T=0.15\;$days after the GRB.  The linear polarization during that
time was $P_{\rm lin} \sim 4\%$ implying a circular to linear
polarization ratio of $P_{\rm circ}/P_{\rm lin}\sim 0.15$. A very
recent detailed study that examined different assumptions for the
magnetic field configuration, jet geometry and electron pitch-angle
distribution \citep{NNP15} concluded that such a relatively high
$P_{\rm circ}/P_{\rm lin}$ ratio cannot be produced by synchrotron
emission from the afterglow (i.e. forward external) shock, which
suggests an alternative origin.

\subsection{Maximum Synchrotron Photon Energy}
\label{sec:E_syn_max}

Since the afterglow emission from the shocked external medium is
independent of the outflow composition, it can much more ``cleanly"
probe the physics of relativistic collisionless shocks, and serve as a
testbed for how the very weak upstream magnetic fields in the pristine
surrounding medium are amplified in the shock, and how the particles
are accelerated in this shock and radiate in the downstream magnetic field.

\begin{figure}
\begin{center}
\includegraphics[width=0.75\columnwidth]{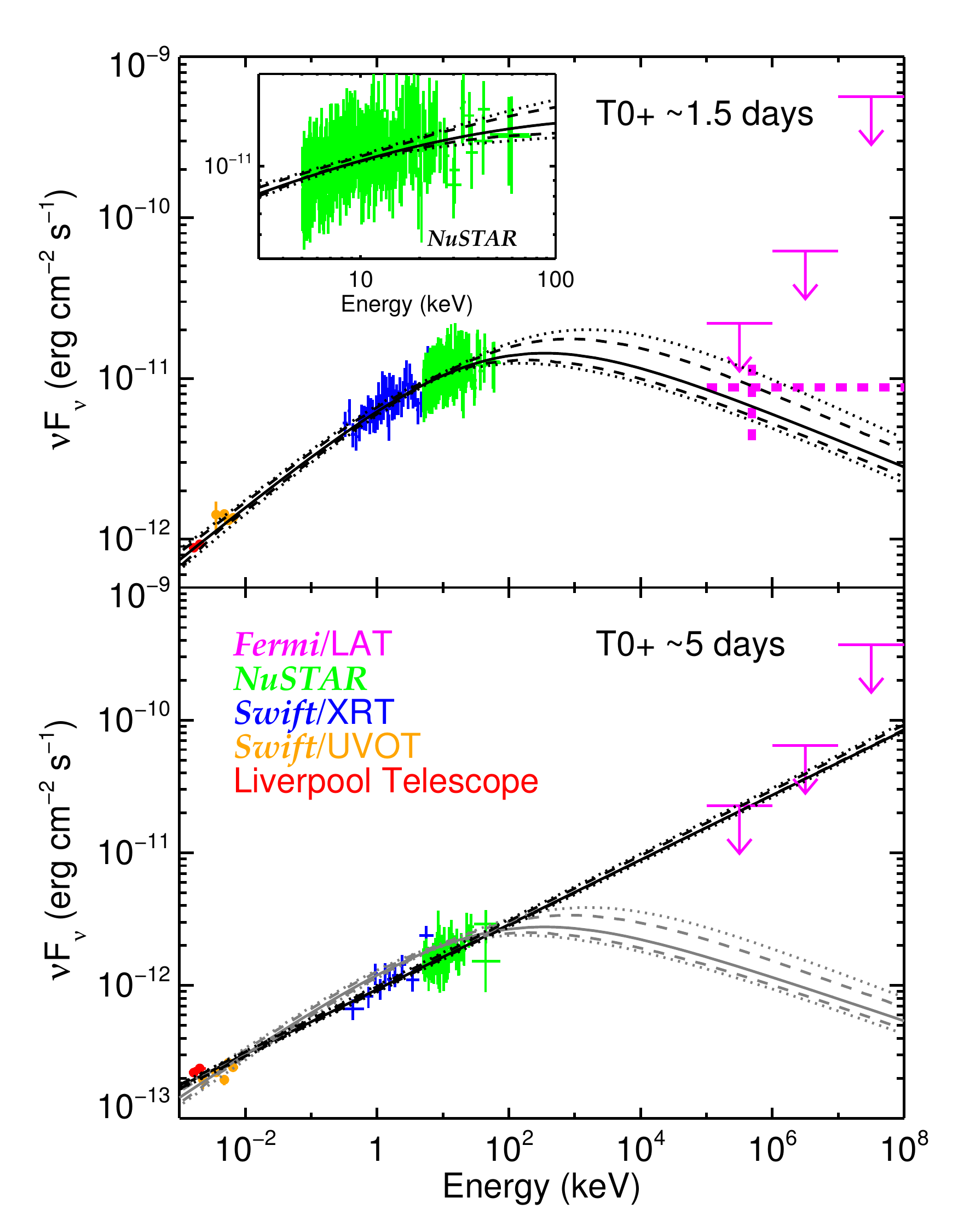}
\caption{The optical to GeV spectrum of GRB$\;$130427A fit with the
  afterglow synchrotron model of \cite{GS02}.  Broadband SEDs are
  shown during the first ({\it top-panel}) and the second ({\it
    bottom-panel}) NuSTAR epochs.  The {\it Fermi}/LAT upper-limits are
  shown as arrows and the extrapolation of the LAT flux light curve is
  shown as a dashed magenta cross (only during the first epoch). The
  second epoch ({\it bottom-panel}) is fit with a power law ({\it black
    lines}); the fit to the first epoch is scaled down and superposed
  on the second epoch data for comparison ({\it in gray}).  (This
  figure is taken from \citealt{Kouveliotou13}.)}
\label{fig:sed}
\end{center}
\end{figure}

A recent challenge to the standard synchrotron afterglow scenario was
raised by the exceptional GRB~130427A
\citep{Ackermann14,Perley14,Kouveliotou13,Maselli14}. This was a very
energetic GRB (with an isotropic equivalent $\gamma$-ray energy of
$E_{\gamma,{\rm iso}}=1.4\times 10^{54}\;$erg) and it occured relatively nearby
(at redshift $z=0.34$). Therefore, it was extremely bright and in
particular, it was detected by {\it Fermi}-LAT
(100$\;$MeV$\;$--$\;100\;$GeV) for nearly a day, including a
95$\;$GeV photon several minutes after the burst and a 32$\;$GeV
photon after 9~hours \citep{Ackermann14}. Altogether, this GRB has a
large number of high-energy photons that clearly violate
\citep{Ackermann14} the maximum synchrotron photon energy
limit,\footnote{The exact numerical coefficient depends on the exact
  assumptions, and in particular on whether the acceleration time is
  assumed to be a fraction of or a complete Larmor gyration time,
  which is in any case a very fast acceleration, and arguably even
  unrealistically so.  Here $\alpha\approx 1/137$ is the fine
  structure constant.}
\begin{equation} 
E_{\rm syn,max}\sim \frac{\Gamma}{(1+z)}\frac{m_e c^2}{\alpha} 
= 3.5\fracb{2}{1+z}\frac{\Gamma}{100}\;{\rm GeV}\ ,
\end{equation}
which is obtained by equating the electron acceleration time to its
synchrotron cooling time, assuming that it is accelerated and radiates
its synchrotron emission in same same magnetic field strength. This
has been argued in order to rule out an afterglow synchrotron origin
of the late-time high-energy LAT photons, and in particular motivated
suggestions for an origin in a distinct SSC spectral component
\citep[e.g.][]{Ackermann14,Tam13,Liu13,Fan13}.

However, \cite{Kouveliotou13} have shown that the optical to GeV
spectrum is consistent with a single spectral component that very
nicely matches the expectation for synchrotron afterglow emission \citep[][see 
Fig.~\ref{fig:spectrum}]{GS02}. Figure~\ref{fig:sed} shows their spectral
fit, and demonstrates that there is hardly any room for a distinct
(SSC) spectral component to dominate the observed flux at the highest
LAT energies (above several GeV or so, as is needed to avoid
violating $E_{\rm syn,max}$).  This conclusion is strengthened by
strict upper limits on the $>100\;$GeV flux measured by VERITAS at three
different epochs near the first NuSTAR observation \citep{Aliu14}.

Therefore, this comprises very compelling evidence for a genuine
violation of the $E_{\rm syn,max}$ in this case, which is much harder
to circumvent compared with previous {\it Fermi}/LAT GRBs
\citep{GRB090902B,PN10,pass8}. Thus, one should start to
seriously consider how this limit can indeed be violated. A possible
solution may lie in relaxing the assumption of a uniform magnetic
field and instead allowing for a lower magnetic field acceleration
region and a higher magnetic field synchrotron radiation region
\citep[e.g.,][]{Lyut10,Kumar12}. Such a situation might arise for
diffusive shock acceleration (Fermi Type I) if the tangled
shock-amplified magnetic field decays on a short length-scale behind
the shock front (where most of the high-energy radiation is emitted),
while the highest energy electrons are accelerated in the lower
magnetic field further downstream \citep{Kumar12}. In fact, such a
scenario has recently been suggested \citep{Lemoine13,LLW13}, and also
shown to significantly alleviate the previously very high $\gamma$-ray
radiative efficiencies inferred for most {\it Swift} GRBs
\citep{Beniamini15}.

\section{Conclusions\label{sec:conc}}

In this review we have demonstrated that magnetic fields clearly play
a vital role in GRBs, practically in every important aspect of
this phenomenon. Here we briefly summarize some of our main
conclusions, and stress both what was learned so far, as well as
what still needs to be carefully studied. Following the main text, the
discussion proceeds more or less in order of increasing distance from
the central source.

Magnetic fields most likely play a crutial role in the launching of GRB
jets. Moreover, hydromagnetic jet launching implies dynamically
strong magnetic fields near the central source, which can naturally
help avoid an excessive baryonic loading into the jet. Such a low
baryon loading is essential in order for the jet to be able to reach
sufficiently large Lorentz factors ($\Gamma\gtrsim 100-500$) that are
inferred from prompt GRB observations. The jet launching definitely
requires many further detailed studies, also (semi-) analytic, but
mainly numerical studies. The latter are, however, involved as they require
general-relativistic magneto-hydrodynamics (GRMHD) codes, coupled with
neutrino, plasma and radiation physics. Therefore, is it likely
to take many years before such studies will provide definitive
answers.

Millisecond-magnetar models for the GRB central engine, discussed in
\S\ref{sec:progenitors}, face serious challenges that still must be overcome. 
Models where the GRB arises from the delayed collapse of a supramassive 
millisecond magnetar \citep[such as the ``time reversal model";][]{Rezzolla15,Ciolfi15} 
face serious difficulty as this requires the formation of a disk during this
collapse, which was recently argued to not be possible \citep{Margalit15}. 
Producing both the prompt GRB emission and the X-ray plateau observed by 
{\it Swift} in long GRBs is challenging for millisecond-magnetar models. 
Models invoking millisecond magnetars in short GRBs also face many problems, 
such as how to produce the short GRB, or hide the huge amount of of 
rotational energy (a few $\times 10^{52}\;$erg) that is injected into the afterglow 
shock, while short GRB afterglows are very dim. More detailed studies of
the relevant physics, both analytic and numerical, are definitely
needed in order to produce more robust and realistic predictions that
could be tested in more detail against the relevant observations. A
relatively simple example is that a highly-relativistic pulsar-like
magnetar MHD wind is invoked to explain the X-ray plateaus observed by
{\it Swift}. However, most studies consider only the forward shock
emission and ignore the emission from the MHD wind itself 
\citep[see, however,][]{Dai04}, which may involve magnetic
reconnection and the resulting particle acceleration at the
termination shock of the pulsar wind, similar to pulsar wind nebulae
\citep[see, e.g.,][]{LK01,Lyub03,sironi:14}. The birth process of a
millisecond magnetar, either in the core-collapse of a massive star or
in a binary neutron star merger, and how it evolves in these messy
environments, is obviously very involved but there is definitely a lot
of room for improvement and new studies on such systems.

The GRB outflow composition, and in particular its degree of
magnetization is an important open question.  On the one hand, in
\S\ref{subsec:collapsar} we have shown very compelling evidence based
on the theoretical jet propagation time in the progenitor star of a
long GRB and the observed GRB duration distribution, that the jet
appears to have modest magnetization ($\sigma\lesssim 1$) throughout
most of the time it takes for it to bore its way out of the
star. Moreover, in \S\ref{subsec:paz} we argued based on the prompt
GRB observations (both $\gamma$-ray detections and upper limits in optical 
and soft X-ray) that the emission region has a low magnetization. On the 
other hand, hydromagnetic jet launching is much more promising than a 
pure thermal one, and can naturally help avoid excessive baryon loading,
suggesting a high magnetization near the source ($\sigma_0\gg 1$).  
A similar conclusion is also strongly suggested from the fact that
quasi-thermal components in the prompt GRB spectrum are typically
highly sub-dominant (see \S\ref{sec:prompt}). Taking these two lines
of evidence together, it appears that the magnetization significantly
decreases as the jet propagates from the source out to large
distances.  This can occur through two main channels (or some combination of 
the two): (i) very efficient conversion of magnetic to kinetic energy that leaves 
a low magnetization in the emission region and allows for efficient internal
shocks, through first quasi-steady and then impulsive magnetic acceleration
\citep[see Fig.~\ref{fig:steady_to_impulsive};][]{GKS11,Granot12b}, 
or (ii) strong magnetic reconnection that converts magnetic energy into 
thermal energy and accelerates particles, and yet somehow leaves a sufficiently 
low magnetization where these particles radiate most of their energy.

Magnetic fields can also play a very important role in the energy
dissipation that leads to the GRB emission.  On the one hand,
sufficiently low magnetization ($\sigma<1$ or even $\sigma\lesssim
0.1$) is needed for efficient energy dissipation in internal
shocks. On the other hand, while a large magnetization ($\sigma>1$)
effectively suppresses internal shocks, or the reverse shock,
it can lead to significant dissipation through magnetic reconnection
under appropriate conditions (e.g. if the source ejects outflow with a
magnetic field that occasionally changes its polarity, or through
certain instabilities).

Another vital role of magnetic fields in GRBs is in particle
acceleration. Within the outflow the magnetization can be high,
allowing efficient magnetic reconnection, which can directly convert a
good fraction of the dissipated magnetic energy into the random
motions of the particles that it accelerates in this process. 
The exact magnetic field strength and geometry throughout the emission 
region also greatly affect the radiation of the accelerated particles, and 
therefore their observable signatures.  In particular, magnetic fields are 
vital to the synchrotron emission, which dominates in the afterglow and 
also likely plays a key role in the prompt GRB emission.

Both optically thin internal shocks as well as the external forward
and reverse shocks are collisionless, and mediated by electromagnetic
fields (through collective plasma interactions). Moreover, while the
outflow itself typically has a large enough magnetic field for
efficient synchrotron radiation, the external medium has a very low
magnetization that must be significantly amplified at the afterglow 
shock front in order to produce the observed afterglow emission. 
Therefore, magnetic field amplification in relativistic collisionless shocks 
and its possible decay behind the shock are vital for understanding the
afterglow physics and interpreting afterglow observations. Moreover
they can strongly affect the particle acceleration in the afterglow
shock, and the resulting afterglow emission. Such physics might
hold the key to unravel a puzzle (see \S\ref{sec:E_syn_max}) arising
from observations of the very bright and relatively nearby long
GRB~130427A, which show an apparent violation of the maximum
synchrotron photon energy limit, $E_{\rm syn,max}$.

The different roles of magnetic fields in GRBs are numerous and
diverse. Many of them are only starting to be understood, while new
roles are still being occasionally discovered. Since magnetic fields appear in
almost all aspects of GRB physics, future studies of their properties
and effects are likely to greatly improve our understanding of GRBs, and
lead to fundamental progress.

\begin{acknowledgements}
The authors acknowledge support from the ISF grant 719/14 (JG),
as well as from the I-CORE Program -- ISF grant 1829/12, 
the ISA grant 3-10417, and an ISF-CNSF grant (TP).
\end{acknowledgements}

\vspace{1.5cm}
\bibliographystyle{aps-nameyear}      
\bibliography{grbbib}                

\begin{thebibliography}{250}
\ifx \bisbn   \undefined \def \bisbn  #1{ISBN #1}\fi
\ifx \binits  \undefined \def \binits#1{#1} \fi
\ifx \bauthor  \undefined \def \bauthor#1{#1} \fi
\ifx \bjtitle  \undefined \def \bjtitle#1{\textrm{#1}}\fi
\ifx \batitle  \undefined \def \batitle#1{#1} \fi
\ifx \bctitle  \undefined \def \bctitle#1{#1} \fi
\ifx \bvolume  \undefined \def \bvolume#1{\textbf{#1}}\fi
\ifx \byear  \undefined \def \byear#1{#1} \fi
\ifx \bissue  \undefined \def \bissue#1{#1} \fi
\ifx \bfpage  \undefined \def \bfpage#1{#1} \fi
\ifx \blpage  \undefined \def \blpage #1{#1} \fi
\ifx \burl  \undefined \def \burl#1{#1} \fi
\ifx \doiurl  \undefined \def \doiurl#1{#1} \fi
\ifx \betal  \undefined \def \betal{et al.} \fi
\ifx \binstitute  \undefined \def \binstitute#1{#1} \fi
\ifx \beditor  \undefined \def \beditor#1{#1} \fi
\ifx \bpublisher  \undefined \def \bpublisher#1{#1} \fi
\ifx \bbtitle  \undefined \def \bbtitle#1{\textit{#1}} \fi
\ifx \bedition  \undefined \def \bedition#1{#1} \fi
\ifx \bseriesno  \undefined \def \bseriesno#1{#1} \fi
\ifx \blocation  \undefined \def \blocation#1{#1} \fi
\ifx \bsertitle  \undefined \def \bsertitle#1{#1} \fi
\ifx \bsnm \undefined \def \bsnm#1{#1} \fi
\ifx \bsuffix \undefined \def \bsuffix#1{#1} \fi
\ifx \bparticle \undefined \def \bparticle#1{#1} \fi
\ifx \barticle \undefined \def \barticle#1{#1} \fi
\ifx \botherref \undefined \def \botherref #1{#1} \fi
\ifx \url \undefined \def \url#1{#1} \fi
\ifx \bchapter \undefined \def \bchapter#1{#1} \fi
\ifx \bbook \undefined \def \bbook#1{#1} \fi
\ifx \bcomment \undefined \def \bcomment#1{#1} \fi
\ifx \oauthor \undefined \def \oauthor#1{#1} \fi
\ifx \citeauthoryear \undefined \def \citeauthoryear#1{#1} \fi
\ifx \texttildelow  \undefined \def \texttildelow{\symbol{126}} \fi
\def \endbibitem {}
\ifx \bconflocation  \undefined \def \bconflocation#1{#1} \fi

\bibitem[\protect\citeauthoryear{{Abdo} et~al.}{2009}]{GRB090902B}
\begin{barticle}
\bauthor{\binits{A.A.} \bsnm{{Abdo}}}, \betal,
\batitle{{Fermi Observations of GRB 090902B: A Distinct Spectral Component in
  the Prompt and Delayed Emission}}.
\bjtitle{\apjl}
\bvolume{706},
\bfpage{138}--\blpage{144}
(\byear{2009}).
doi:\doiurl{10.1088/0004-637X/706/1/L138}
\end{barticle}
\endbibitem

\bibitem[\protect\citeauthoryear{{Ackermann} et~al.}{2010}]{GRB090510}
\begin{barticle}
\bauthor{\binits{M.} \bsnm{{Ackermann}}}, \betal,
\batitle{{Fermi Observations of GRB 090510: A Short-Hard Gamma-ray Burst with
  an Additional, Hard Power-law Component from 10 keV TO GeV Energies}}.
\bjtitle{\apj}
\bvolume{716},
\bfpage{1178}--\blpage{1190}
(\byear{2010}).
doi:\doiurl{10.1088/0004-637X/716/2/1178}
\end{barticle}
\endbibitem

\bibitem[\protect\citeauthoryear{{Ackermann} et~al.}{2013}]{ackermann:13}
\begin{barticle}
\bauthor{\binits{M.} \bsnm{{Ackermann}}}, \betal,
\batitle{{The First Fermi-LAT Gamma-Ray Burst Catalog}}.
\bjtitle{\apjs}
\bvolume{209},
\bfpage{11}
(\byear{2013}).
doi:\doiurl{10.1088/0067-0049/209/1/11}
\end{barticle}
\endbibitem

\bibitem[\protect\citeauthoryear{{Ackermann} et~al.}{2014}]{Ackermann14}
\begin{barticle}
\bauthor{\binits{M.} \bsnm{{Ackermann}}}, \betal,
\batitle{{Fermi-LAT Observations of the Gamma-Ray Burst GRB 130427A}}.
\bjtitle{Science}
\bvolume{343},
\bfpage{42}--\blpage{47}
(\byear{2014}).
doi:\doiurl{10.1126/science.1242353}
\end{barticle}
\endbibitem

\bibitem[\protect\citeauthoryear{{Akerlof} et~al.}{1999}]{Akerlof99}
\begin{barticle}
\bauthor{\binits{C.} \bsnm{{Akerlof}}}, \betal,
\batitle{{Observation of contemporaneous optical radiation from a
  {$\gamma$}-ray burst}}.
\bjtitle{\nat}
\bvolume{398},
\bfpage{400}--\blpage{402}
(\byear{1999}).
doi:\doiurl{10.1038/18837}
\end{barticle}
\endbibitem

\bibitem[\protect\citeauthoryear{{Aliu} et~al.}{2014}]{Aliu14}
\begin{barticle}
\bauthor{\binits{E.} \bsnm{{Aliu}}}, \betal,
\batitle{{Constraints on Very High Energy Emission from GRB 130427A}}.
\bjtitle{\apjl}
\bvolume{795},
\bfpage{3}
(\byear{2014}).
doi:\doiurl{10.1088/2041-8205/795/1/L3}
\end{barticle}
\endbibitem

\bibitem[\protect\citeauthoryear{{Asano} and
  {M{\'e}sz{\'a}ros}}{2011}]{asano:11}
\begin{barticle}
\bauthor{\binits{K.} \bsnm{{Asano}}},
\bauthor{\binits{P.} \bsnm{{M{\'e}sz{\'a}ros}}},
\batitle{{Spectral-Temporal Simulations of Internal Dissipation Models of
  Gamma-Ray Bursts}}.
\bjtitle{\apj}
\bvolume{739},
\bfpage{103}
(\byear{2011}).
doi:\doiurl{10.1088/0004-637X/739/2/103}
\end{barticle}
\endbibitem

\bibitem[\protect\citeauthoryear{{Asano} and
  {M{\'e}sz{\'a}ros}}{2012}]{asano:12}
\begin{barticle}
\bauthor{\binits{K.} \bsnm{{Asano}}},
\bauthor{\binits{P.} \bsnm{{M{\'e}sz{\'a}ros}}},
\batitle{{Delayed Onset of High-energy Emissions in Leptonic and Hadronic
  Models of Gamma-Ray Bursts}}.
\bjtitle{\apj}
\bvolume{757},
\bfpage{115}
(\byear{2012}).
doi:\doiurl{10.1088/0004-637X/757/2/115}
\end{barticle}
\endbibitem

\bibitem[\protect\citeauthoryear{{Atwood} et~al.}{2013}]{pass8}
\begin{barticle}
\bauthor{\binits{W.B.} \bsnm{{Atwood}}}, \betal,
\batitle{{New Fermi-LAT Event Reconstruction Reveals More High-energy Gamma
  Rays from Gamma-Ray Bursts}}.
\bjtitle{\apj}
\bvolume{774},
\bfpage{76}
(\byear{2013}).
doi:\doiurl{10.1088/0004-637X/774/1/76}
\end{barticle}
\endbibitem

\bibitem[\protect\citeauthoryear{{Axelsson} et~al.}{2012}]{Axelsson12}
\begin{barticle}
\bauthor{\binits{M.} \bsnm{{Axelsson}}}, \betal,
\batitle{{GRB110721A: An Extreme Peak Energy and Signatures of the
  Photosphere}}.
\bjtitle{\apjl}
\bvolume{757},
\bfpage{31}
(\byear{2012}).
doi:\doiurl{10.1088/2041-8205/757/2/L31}
\end{barticle}
\endbibitem

\bibitem[\protect\citeauthoryear{{Band} et~al.}{1993}]{Band93}
\begin{barticle}
\bauthor{\binits{D.} \bsnm{{Band}}}, \betal,
\batitle{{BATSE observations of gamma-ray burst spectra. I - Spectral
  diversity}}.
\bjtitle{\apj}
\bvolume{413},
\bfpage{281}--\blpage{292}
(\byear{1993}).
doi:\doiurl{10.1086/172995}
\end{barticle}
\endbibitem

\bibitem[\protect\citeauthoryear{{Baring} and {Harding}}{1997}]{baring:97}
\begin{barticle}
\bauthor{\binits{M.G.} \bsnm{{Baring}}},
\bauthor{\binits{A.K.} \bsnm{{Harding}}},
\batitle{{The Escape of High-Energy Photons from Gamma-Ray Bursts}}.
\bjtitle{\apj}
\bvolume{491},
\bfpage{663}--\blpage{686}
(\byear{1997})
\end{barticle}
\endbibitem

\bibitem[\protect\citeauthoryear{{Barniol Duran}
  et~al.}{2012}]{barniolduran:12}
\begin{barticle}
\bauthor{\binits{R.} \bsnm{{Barniol Duran}}},
\bauthor{\binits{{\v Z}.} \bsnm{{Bo{\v s}njak}}},
\bauthor{\binits{P.} \bsnm{{Kumar}}},
\batitle{{Inverse-Compton cooling in Klein-Nishina regime and gamma-ray burst
  prompt spectrum}}.
\bjtitle{\mnras}
\bvolume{424},
\bfpage{3192}--\blpage{3200}
(\byear{2012}).
doi:\doiurl{10.1111/j.1365-2966.2012.21533.x}
\end{barticle}
\endbibitem

\bibitem[\protect\citeauthoryear{{Barraud} et~al.}{2005}]{barraud:05}
\begin{barticle}
\bauthor{\binits{C.} \bsnm{{Barraud}}},
\bauthor{\binits{F.} \bsnm{{Daigne}}},
\bauthor{\binits{R.} \bsnm{{Mochkovitch}}},
\bauthor{\binits{J.L.} \bsnm{{Atteia}}},
\batitle{{On the nature of X-ray flashes}}.
\bjtitle{\aap}
\bvolume{440},
\bfpage{809}--\blpage{817}
(\byear{2005}).
doi:\doiurl{10.1051/0004-6361:20041572}
\end{barticle}
\endbibitem

\bibitem[\protect\citeauthoryear{{Begelman}}{1998}]{Begelman98}
\begin{barticle}
\bauthor{\binits{M.C.} \bsnm{{Begelman}}},
\batitle{{Instability of Toroidal Magnetic Field in Jets and Plerions}}.
\bjtitle{\apj}
\bvolume{493},
\bfpage{291}
(\byear{1998}).
doi:\doiurl{10.1086/305119}
\end{barticle}
\endbibitem

\bibitem[\protect\citeauthoryear{{Beloborodov}}{2010}]{beloborodov:10}
\begin{barticle}
\bauthor{\binits{A.M.} \bsnm{{Beloborodov}}},
\batitle{{Collisional mechanism for gamma-ray burst emission}}.
\bjtitle{\mnras}
\bvolume{407},
\bfpage{1033}--\blpage{1047}
(\byear{2010}).
doi:\doiurl{10.1111/j.1365-2966.2010.16770.x}
\end{barticle}
\endbibitem

\bibitem[\protect\citeauthoryear{{Beloborodov}}{2011}]{beloborodov:11}
\begin{barticle}
\bauthor{\binits{A.M.} \bsnm{{Beloborodov}}},
\batitle{{Radiative Transfer in Ultrarelativistic Outflows}}.
\bjtitle{\apj}
\bvolume{737},
\bfpage{68}
(\byear{2011}).
doi:\doiurl{10.1088/0004-637X/737/2/68}
\end{barticle}
\endbibitem

\bibitem[\protect\citeauthoryear{{Beloborodov}}{2013}]{beloborodov:13}
\begin{barticle}
\bauthor{\binits{A.M.} \bsnm{{Beloborodov}}},
\batitle{{Regulation of the Spectral Peak in Gamma-Ray Bursts}}.
\bjtitle{\apj}
\bvolume{764},
\bfpage{157}
(\byear{2013}).
doi:\doiurl{10.1088/0004-637X/764/2/157}
\end{barticle}
\endbibitem

\bibitem[\protect\citeauthoryear{{Beloborodov} et~al.}{2014}]{beloborodov:14}
\begin{barticle}
\bauthor{\binits{A.M.} \bsnm{{Beloborodov}}},
\bauthor{\binits{R.} \bsnm{{Hasco{\"e}t}}},
\bauthor{\binits{I.} \bsnm{{Vurm}}},
\batitle{{On the Origin of GeV Emission in Gamma-Ray Bursts}}.
\bjtitle{\apj}
\bvolume{788},
\bfpage{36}
(\byear{2014}).
doi:\doiurl{10.1088/0004-637X/788/1/36}
\end{barticle}
\endbibitem

\bibitem[\protect\citeauthoryear{{Beloborodov} et~al.}{2000}]{beloborodov:00}
\begin{barticle}
\bauthor{\binits{A.M.} \bsnm{{Beloborodov}}},
\bauthor{\binits{B.E.} \bsnm{{Stern}}},
\bauthor{\binits{R.} \bsnm{{Svensson}}},
\batitle{{Power Density Spectra of Gamma-Ray Bursts}}.
\bjtitle{\apj}
\bvolume{535},
\bfpage{158}--\blpage{166}
(\byear{2000}).
doi:\doiurl{10.1086/308836}
\end{barticle}
\endbibitem

\bibitem[\protect\citeauthoryear{{Beniamini} and {Piran}}{2013}]{beniamini:13}
\begin{barticle}
\bauthor{\binits{P.} \bsnm{{Beniamini}}},
\bauthor{\binits{T.} \bsnm{{Piran}}},
\batitle{{Constraints on the Synchrotron Emission Mechanism in Gamma-Ray
  Bursts}}.
\bjtitle{\apj}
\bvolume{769},
\bfpage{69}
(\byear{2013}).
doi:\doiurl{10.1088/0004-637X/769/1/69}
\end{barticle}
\endbibitem

\bibitem[\protect\citeauthoryear{{Beniamini} and {Piran}}{2014}]{Beniamini14}
\begin{barticle}
\bauthor{\binits{P.} \bsnm{{Beniamini}}},
\bauthor{\binits{T.} \bsnm{{Piran}}},
\batitle{{The emission mechanism in magnetically dominated gamma-ray burst
  outflows}}.
\bjtitle{\mnras}
\bvolume{445},
\bfpage{3892}--\blpage{3907}
(\byear{2014}).
doi:\doiurl{10.1093/mnras/stu2032}
\end{barticle}
\endbibitem

\bibitem[\protect\citeauthoryear{{Beniamini} et~al.}{2015}]{Beniamini15}
\begin{botherref}
\oauthor{\binits{P.} \bsnm{{Beniamini}}},
\oauthor{\binits{L.} \bsnm{{Nava}}},
\oauthor{\binits{R.} \bsnm{{Barniol Duran}}},
\oauthor{\binits{T.} \bsnm{{Piran}}},
{Energies of GRB blast waves and prompt efficiencies as implied by
  self-consistent modeling of X-ray and LAT afterglows}.
ArXiv e-prints
(2015)
\end{botherref}
\endbibitem

\bibitem[\protect\citeauthoryear{{Berger} et~al.}{2003}]{Berger03}
\begin{barticle}
\bauthor{\binits{E.} \bsnm{{Berger}}},
\bauthor{\binits{A.M.} \bsnm{{Soderberg}}},
\bauthor{\binits{D.A.} \bsnm{{Frail}}},
\bauthor{\binits{S.R.} \bsnm{{Kulkarni}}},
\batitle{{A Radio Flare from GRB 020405: Evidence for a Uniform Medium around a
  Massive Stellar Progenitor}}.
\bjtitle{\apjl}
\bvolume{587},
\bfpage{5}--\blpage{8}
(\byear{2003}).
doi:\doiurl{10.1086/375158}
\end{barticle}
\endbibitem

\bibitem[\protect\citeauthoryear{{Blandford}}{2002}]{Blandford02}
\begin{bchapter}
\bauthor{\binits{R.D.} \bsnm{{Blandford}}},
\bctitle{{To the Lighthouse}},
in \bbtitle{Lighthouses of the Universe: The Most Luminous Celestial Objects
  and Their Use for Cosmology},
ed. by \beditor{\binits{M.} \bsnm{{Gilfanov}}},
\beditor{\binits{R.} \bsnm{{Sunyeav}}},
\beditor{\binits{E.} \bsnm{{Churazov}}},
\byear{2002},
p. \bfpage{381}.
doi:\doiurl{10.1007/10856495\_59}
\end{bchapter}
\endbibitem

\bibitem[\protect\citeauthoryear{{Blandford} and {McKee}}{1976}]{BM76}
\begin{barticle}
\bauthor{\binits{R.D.} \bsnm{{Blandford}}},
\bauthor{\binits{C.F.} \bsnm{{McKee}}},
\batitle{{Fluid dynamics of relativistic blast waves}}.
\bjtitle{Physics of Fluids}
\bvolume{19},
\bfpage{1130}--\blpage{1138}
(\byear{1976}).
doi:\doiurl{10.1063/1.861619}
\end{barticle}
\endbibitem

\bibitem[\protect\citeauthoryear{{Bloom} et~al.}{2003}]{BFK03}
\begin{barticle}
\bauthor{\binits{J.S.} \bsnm{{Bloom}}},
\bauthor{\binits{D.A.} \bsnm{{Frail}}},
\bauthor{\binits{S.R.} \bsnm{{Kulkarni}}},
\batitle{{Gamma-Ray Burst Energetics and the Gamma-Ray Burst Hubble Diagram:
  Promises and Limitations}}.
\bjtitle{\apj}
\bvolume{594},
\bfpage{674}--\blpage{683}
(\byear{2003}).
doi:\doiurl{10.1086/377125}
\end{barticle}
\endbibitem

\bibitem[\protect\citeauthoryear{{Bo{\v s}njak} and
  {Daigne}}{2014}]{bosnjak:14}
\begin{barticle}
\bauthor{\binits{{\v Z}.} \bsnm{{Bo{\v s}njak}}},
\bauthor{\binits{F.} \bsnm{{Daigne}}},
\batitle{{Spectral evolution in gamma-ray bursts: Predictions of the internal
  shock model and comparison to observations}}.
\bjtitle{\aap}
\bvolume{568},
\bfpage{45}
(\byear{2014}).
doi:\doiurl{10.1051/0004-6361/201322341}
\end{barticle}
\endbibitem

\bibitem[\protect\citeauthoryear{{Bo{\v s}njak} et~al.}{2009}]{bosnjak:09}
\begin{barticle}
\bauthor{\binits{{\v Z}.} \bsnm{{Bo{\v s}njak}}},
\bauthor{\binits{F.} \bsnm{{Daigne}}},
\bauthor{\binits{G.} \bsnm{{Dubus}}},
\batitle{{Prompt high-energy emission from gamma-ray bursts in the internal
  shock model}}.
\bjtitle{\aap}
\bvolume{498},
\bfpage{677}--\blpage{703}
(\byear{2009}).
doi:\doiurl{10.1051/0004-6361/200811375}
\end{barticle}
\endbibitem

\bibitem[\protect\citeauthoryear{{Bromberg} and {Levinson}}{2007}]{BL07}
\begin{barticle}
\bauthor{\binits{O.} \bsnm{{Bromberg}}},
\bauthor{\binits{A.} \bsnm{{Levinson}}},
\batitle{{Hydrodynamic Collimation of Relativistic Outflows: Semianalytic
  Solutions and Application to Gamma-Ray Bursts}}.
\bjtitle{\apj}
\bvolume{671},
\bfpage{678}--\blpage{688}
(\byear{2007}).
doi:\doiurl{10.1086/522668}
\end{barticle}
\endbibitem

\bibitem[\protect\citeauthoryear{{Bromberg} et~al.}{2015}]{BGP15}
\begin{barticle}
\bauthor{\binits{O.} \bsnm{{Bromberg}}},
\bauthor{\binits{J.} \bsnm{{Granot}}},
\bauthor{\binits{T.} \bsnm{{Piran}}},
\batitle{{On the composition of GRBs' Collapsar jets}}.
\bjtitle{\mnras}
\bvolume{450},
\bfpage{1077}--\blpage{1084}
(\byear{2015}).
doi:\doiurl{10.1093/mnras/stv226}
\end{barticle}
\endbibitem

\bibitem[\protect\citeauthoryear{{Bromberg} et~al.}{2011}]{Bromberg11}
\begin{barticle}
\bauthor{\binits{O.} \bsnm{{Bromberg}}},
\bauthor{\binits{E.} \bsnm{{Nakar}}},
\bauthor{\binits{T.} \bsnm{{Piran}}},
\bauthor{\binits{R.} \bsnm{{Sari}}},
\batitle{{The Propagation of Relativistic Jets in External Media}}.
\bjtitle{\apj}
\bvolume{740},
\bfpage{100}
(\byear{2011}).
doi:\doiurl{10.1088/0004-637X/740/2/100}
\end{barticle}
\endbibitem

\bibitem[\protect\citeauthoryear{{Bromberg} et~al.}{2012}]{Bromberg12}
\begin{barticle}
\bauthor{\binits{O.} \bsnm{{Bromberg}}},
\bauthor{\binits{E.} \bsnm{{Nakar}}},
\bauthor{\binits{T.} \bsnm{{Piran}}},
\bauthor{\binits{R.} \bsnm{{Sari}}},
\batitle{{An Observational Imprint of the Collapsar Model of Long Gamma-Ray
  Bursts}}.
\bjtitle{\apj}
\bvolume{749},
\bfpage{110}
(\byear{2012}).
doi:\doiurl{10.1088/0004-637X/749/2/110}
\end{barticle}
\endbibitem

\bibitem[\protect\citeauthoryear{{Bromberg} et~al.}{2013}]{Bromberg13}
\begin{barticle}
\bauthor{\binits{O.} \bsnm{{Bromberg}}},
\bauthor{\binits{E.} \bsnm{{Nakar}}},
\bauthor{\binits{T.} \bsnm{{Piran}}},
\bauthor{\binits{R.} \bsnm{{Sari}}},
\batitle{{Short versus Long and Collapsars versus Non-collapsars: A
  Quantitative Classification of Gamma-Ray Bursts}}.
\bjtitle{\apj}
\bvolume{764},
\bfpage{179}
(\byear{2013}).
doi:\doiurl{10.1088/0004-637X/764/2/179}
\end{barticle}
\endbibitem

\bibitem[\protect\citeauthoryear{{Bromberg} et~al.}{2014}]{BGLP14}
\begin{barticle}
\bauthor{\binits{O.} \bsnm{{Bromberg}}},
\bauthor{\binits{J.} \bsnm{{Granot}}},
\bauthor{\binits{Y.} \bsnm{{Lyubarsky}}},
\bauthor{\binits{T.} \bsnm{{Piran}}},
\batitle{{The dynamics of a highly magnetized jet propagating inside a star}}.
\bjtitle{\mnras}
\bvolume{443},
\bfpage{1532}--\blpage{1548}
(\byear{2014}).
doi:\doiurl{10.1093/mnras/stu995}
\end{barticle}
\endbibitem

\bibitem[\protect\citeauthoryear{{Bucciantini} et~al.}{2007}]{Bucciantini07}
\begin{barticle}
\bauthor{\binits{N.} \bsnm{{Bucciantini}}},
\bauthor{\binits{E.} \bsnm{{Quataert}}},
\bauthor{\binits{J.} \bsnm{{Arons}}},
\bauthor{\binits{B.D.} \bsnm{{Metzger}}},
\bauthor{\binits{T.A.} \bsnm{{Thompson}}},
\batitle{{Magnetar-driven bubbles and the origin of collimated outflows in
  gamma-ray bursts}}.
\bjtitle{\mnras}
\bvolume{380},
\bfpage{1541}--\blpage{1553}
(\byear{2007}).
doi:\doiurl{10.1111/j.1365-2966.2007.12164.x}
\end{barticle}
\endbibitem

\bibitem[\protect\citeauthoryear{{Bucciantini} et~al.}{2008}]{Bucciantini08}
\begin{barticle}
\bauthor{\binits{N.} \bsnm{{Bucciantini}}},
\bauthor{\binits{E.} \bsnm{{Quataert}}},
\bauthor{\binits{J.} \bsnm{{Arons}}},
\bauthor{\binits{B.D.} \bsnm{{Metzger}}},
\bauthor{\binits{T.A.} \bsnm{{Thompson}}},
\batitle{{Relativistic jets and long-duration gamma-ray bursts from the birth
  of magnetars}}.
\bjtitle{\mnras}
\bvolume{383},
\bfpage{25}--\blpage{29}
(\byear{2008}).
doi:\doiurl{10.1111/j.1745-3933.2007.00403.x}
\end{barticle}
\endbibitem

\bibitem[\protect\citeauthoryear{{Bucciantini} et~al.}{2009}]{Bucciantini09}
\begin{barticle}
\bauthor{\binits{N.} \bsnm{{Bucciantini}}},
\bauthor{\binits{E.} \bsnm{{Quataert}}},
\bauthor{\binits{B.D.} \bsnm{{Metzger}}},
\bauthor{\binits{T.A.} \bsnm{{Thompson}}},
\bauthor{\binits{J.} \bsnm{{Arons}}},
\bauthor{\binits{L.} \bsnm{{Del Zanna}}},
\batitle{{Magnetized relativistic jets and long-duration GRBs from magnetar
  spin-down during core-collapse supernovae}}.
\bjtitle{\mnras}
\bvolume{396},
\bfpage{2038}--\blpage{2050}
(\byear{2009}).
doi:\doiurl{10.1111/j.1365-2966.2009.14940.x}
\end{barticle}
\endbibitem

\bibitem[\protect\citeauthoryear{{Bucciantini} et~al.}{2012}]{Bucciantini12}
\begin{barticle}
\bauthor{\binits{N.} \bsnm{{Bucciantini}}},
\bauthor{\binits{B.D.} \bsnm{{Metzger}}},
\bauthor{\binits{T.A.} \bsnm{{Thompson}}},
\bauthor{\binits{E.} \bsnm{{Quataert}}},
\batitle{{Short gamma-ray bursts with extended emission from magnetar birth:
  jet formation and collimation}}.
\bjtitle{\mnras}
\bvolume{419},
\bfpage{1537}--\blpage{1545}
(\byear{2012}).
doi:\doiurl{10.1111/j.1365-2966.2011.19810.x}
\end{barticle}
\endbibitem

\bibitem[\protect\citeauthoryear{{Burgess} et~al.}{2011}]{Burgess11}
\begin{barticle}
\bauthor{\binits{J.M.} \bsnm{{Burgess}}}, \betal,
\batitle{{Constraints on the Synchrotron Shock Model for the Fermi GRB 090820A
  Observed by Gamma-Ray Burst Monitor}}.
\bjtitle{\apj}
\bvolume{741},
\bfpage{24}
(\byear{2011}).
doi:\doiurl{10.1088/0004-637X/741/1/24}
\end{barticle}
\endbibitem

\bibitem[\protect\citeauthoryear{{Burgess} et~al.}{2014}]{Burgess14}
\begin{barticle}
\bauthor{\binits{J.M.} \bsnm{{Burgess}}}, \betal,
\batitle{{Time-resolved Analysis of Fermi Gamma-Ray Bursts with Fast- and
  Slow-cooled Synchrotron Photon Models}}.
\bjtitle{\apj}
\bvolume{784},
\bfpage{17}
(\byear{2014}).
doi:\doiurl{10.1088/0004-637X/784/1/17}
\end{barticle}
\endbibitem

\bibitem[\protect\citeauthoryear{{Bykov} and {Meszaros}}{1996}]{bykov:96}
\begin{barticle}
\bauthor{\binits{A.M.} \bsnm{{Bykov}}},
\bauthor{\binits{P.} \bsnm{{Meszaros}}},
\batitle{{Electron Acceleration and Efficiency in Nonthermal Gamma-Ray
  Sources}}.
\bjtitle{\apjl}
\bvolume{461},
\bfpage{37}
(\byear{1996}).
doi:\doiurl{10.1086/309999}
\end{barticle}
\endbibitem

\bibitem[\protect\citeauthoryear{{Bykov} et~al.}{2012}]{bykov:12}
\begin{barticle}
\bauthor{\binits{A.} \bsnm{{Bykov}}},
\bauthor{\binits{N.} \bsnm{{Gehrels}}},
\bauthor{\binits{H.} \bsnm{{Krawczynski}}},
\bauthor{\binits{M.} \bsnm{{Lemoine}}},
\bauthor{\binits{G.} \bsnm{{Pelletier}}},
\bauthor{\binits{M.} \bsnm{{Pohl}}},
\batitle{{Particle Acceleration in Relativistic Outflows}}.
\bjtitle{\ssr}
\bvolume{173},
\bfpage{309}--\blpage{339}
(\byear{2012}).
doi:\doiurl{10.1007/s11214-012-9896-y}
\end{barticle}
\endbibitem

\bibitem[\protect\citeauthoryear{{Campana} et~al.}{2006}]{Campana06}
\begin{barticle}
\bauthor{\binits{S.} \bsnm{{Campana}}}, \betal,
\batitle{{The association of GRB 060218 with a supernova and the evolution of
  the shock wave}}.
\bjtitle{\nat}
\bvolume{442},
\bfpage{1008}--\blpage{1010}
(\byear{2006}).
doi:\doiurl{10.1038/nature04892}
\end{barticle}
\endbibitem

\bibitem[\protect\citeauthoryear{{Cenko} et~al.}{2010}]{Cenko10}
\begin{barticle}
\bauthor{\binits{S.B.} \bsnm{{Cenko}}}, \betal,
\batitle{{The Collimation and Energetics of the Brightest Swift Gamma-ray
  Bursts}}.
\bjtitle{\apj}
\bvolume{711},
\bfpage{641}--\blpage{654}
(\byear{2010}).
doi:\doiurl{10.1088/0004-637X/711/2/641}
\end{barticle}
\endbibitem

\bibitem[\protect\citeauthoryear{{Ciolfi} and {Siegel}}{2015}]{Ciolfi15}
\begin{barticle}
\bauthor{\binits{R.} \bsnm{{Ciolfi}}},
\bauthor{\binits{D.M.} \bsnm{{Siegel}}},
\batitle{{Short Gamma-Ray Bursts in the ''Time-reversal'' Scenario}}.
\bjtitle{\apjl}
\bvolume{798},
\bfpage{36}
(\byear{2015}).
doi:\doiurl{10.1088/2041-8205/798/2/L36}
\end{barticle}
\endbibitem

\bibitem[\protect\citeauthoryear{{Coburn} and {Boggs}}{2003}]{CoburnBoggs03}
\begin{barticle}
\bauthor{\binits{W.} \bsnm{{Coburn}}},
\bauthor{\binits{S.E.} \bsnm{{Boggs}}},
\batitle{{Polarization of the prompt {$\gamma$}-ray emission from the
  {$\gamma$}-ray burst of 6 December 2002}}.
\bjtitle{\nat}
\bvolume{423},
\bfpage{415}--\blpage{417}
(\byear{2003}).
doi:\doiurl{10.1038/nature01612}
\end{barticle}
\endbibitem

\bibitem[\protect\citeauthoryear{{Cohen} et~al.}{1997}]{cohen:97}
\begin{barticle}
\bauthor{\binits{E.} \bsnm{{Cohen}}},
\bauthor{\binits{J.I.} \bsnm{{Katz}}},
\bauthor{\binits{T.} \bsnm{{Piran}}},
\bauthor{\binits{R.} \bsnm{{Sari}}},
\bauthor{\binits{R.D.} \bsnm{{Preece}}},
\bauthor{\binits{D.L.} \bsnm{{Band}}},
\batitle{{Possible Evidence for Relativistic Shocks in Gamma-Ray Bursts}}.
\bjtitle{\apj}
\bvolume{488},
\bfpage{330}--\blpage{337}
(\byear{1997})
\end{barticle}
\endbibitem

\bibitem[\protect\citeauthoryear{{Contopoulos}}{1995}]{Cont95}
\begin{barticle}
\bauthor{\binits{J.} \bsnm{{Contopoulos}}},
\batitle{{A Simple Type of Magnetically Driven Jets: an Astrophysical Plasma
  Gun}}.
\bjtitle{\apj}
\bvolume{450},
\bfpage{616}
(\byear{1995}).
doi:\doiurl{10.1086/176170}
\end{barticle}
\endbibitem

\bibitem[\protect\citeauthoryear{{Covino} et~al.}{2004}]{Covino04}
\begin{bchapter}
\bauthor{\binits{S.} \bsnm{{Covino}}},
\bauthor{\binits{G.} \bsnm{{Ghisellini}}},
\bauthor{\binits{D.} \bsnm{{Lazzati}}},
\bauthor{\binits{D.} \bsnm{{Malesani}}},
\bctitle{{Polarization of Gamma-Ray Burst Optical and Near-Infrared
  Afterglows}},
in \bbtitle{Gamma-Ray Bursts in the Afterglow Era},
ed. by \beditor{\binits{M.} \bsnm{{Feroci}}},
\beditor{\binits{F.} \bsnm{{Frontera}}},
\beditor{\binits{N.} \bsnm{{Masetti}}},
\beditor{\binits{L.} \bsnm{{Piro}}}
\bsertitle{Astronomical Society of the Pacific Conference Series},
vol. \bseriesno{312},
\byear{2004},
p. \bfpage{169}
\end{bchapter}
\endbibitem

\bibitem[\protect\citeauthoryear{{Dai}}{2004}]{Dai04}
\begin{barticle}
\bauthor{\binits{Z.G.} \bsnm{{Dai}}},
\batitle{{Relativistic Wind Bubbles and Afterglow Signatures}}.
\bjtitle{\apj}
\bvolume{606},
\bfpage{1000}--\blpage{1005}
(\byear{2004}).
doi:\doiurl{10.1086/383019}
\end{barticle}
\endbibitem

\bibitem[\protect\citeauthoryear{{Daigne} and {Mochkovitch}}{1998}]{daigne:98}
\begin{barticle}
\bauthor{\binits{F.} \bsnm{{Daigne}}},
\bauthor{\binits{R.} \bsnm{{Mochkovitch}}},
\batitle{{Gamma-ray bursts from internal shocks in a relativistic wind:
  temporal and spectral properties}}.
\bjtitle{\mnras}
\bvolume{296},
\bfpage{275}--\blpage{286}
(\byear{1998}).
doi:\doiurl{10.1046/j.1365-8711.1998.01305.x}
\end{barticle}
\endbibitem

\bibitem[\protect\citeauthoryear{{Daigne} and {Mochkovitch}}{2000}]{daigne:00}
\begin{barticle}
\bauthor{\binits{F.} \bsnm{{Daigne}}},
\bauthor{\binits{R.} \bsnm{{Mochkovitch}}},
\batitle{{Gamma-ray bursts from internal shocks in a relativistic wind: a
  hydrodynamical study}}.
\bjtitle{\aap}
\bvolume{358},
\bfpage{1157}--\blpage{1166}
(\byear{2000})
\end{barticle}
\endbibitem

\bibitem[\protect\citeauthoryear{{Daigne} and {Mochkovitch}}{2002}]{daigne:02}
\begin{barticle}
\bauthor{\binits{F.} \bsnm{{Daigne}}},
\bauthor{\binits{R.} \bsnm{{Mochkovitch}}},
\batitle{{The expected thermal precursors of gamma-ray bursts in the internal
  shock model}}.
\bjtitle{\mnras}
\bvolume{336},
\bfpage{1271}--\blpage{1280}
(\byear{2002}).
doi:\doiurl{10.1046/j.1365-8711.2002.05875.x}
\end{barticle}
\endbibitem

\bibitem[\protect\citeauthoryear{{Daigne} and {Mochkovitch}}{2003}]{daigne:03}
\begin{barticle}
\bauthor{\binits{F.} \bsnm{{Daigne}}},
\bauthor{\binits{R.} \bsnm{{Mochkovitch}}},
\batitle{{The physics of pulses in gamma-ray bursts: emission processes,
  temporal profiles and time-lags}}.
\bjtitle{\mnras}
\bvolume{342},
\bfpage{587}--\blpage{592}
(\byear{2003}).
doi:\doiurl{10.1046/j.1365-8711.2003.06575.x}
\end{barticle}
\endbibitem

\bibitem[\protect\citeauthoryear{{Daigne} et~al.}{2011}]{daigne:11}
\begin{barticle}
\bauthor{\binits{F.} \bsnm{{Daigne}}},
\bauthor{\binits{{\v Z}.} \bsnm{{Bo{\v s}njak}}},
\bauthor{\binits{G.} \bsnm{{Dubus}}},
\batitle{{Reconciling observed gamma-ray burst prompt spectra with synchrotron
  radiation?}}
\bjtitle{\aap}
\bvolume{526},
\bfpage{110}
(\byear{2011}).
doi:\doiurl{10.1051/0004-6361/201015457}
\end{barticle}
\endbibitem

\bibitem[\protect\citeauthoryear{{Dall'Osso} et~al.}{2012}]{DallOsso12}
\begin{barticle}
\bauthor{\binits{S.} \bsnm{{Dall'Osso}}},
\bauthor{\binits{J.} \bsnm{{Granot}}},
\bauthor{\binits{T.} \bsnm{{Piran}}},
\batitle{{Magnetic field decay in neutron stars: from soft gamma repeaters to
  'weak-field magnetars'}}.
\bjtitle{\mnras}
\bvolume{422},
\bfpage{2878}--\blpage{2903}
(\byear{2012}).
doi:\doiurl{10.1111/j.1365-2966.2012.20612.x}
\end{barticle}
\endbibitem

\bibitem[\protect\citeauthoryear{{Dall'Osso} et~al.}{2009}]{DallOsso09}
\begin{barticle}
\bauthor{\binits{S.} \bsnm{{Dall'Osso}}},
\bauthor{\binits{S.N.} \bsnm{{Shore}}},
\bauthor{\binits{L.} \bsnm{{Stella}}},
\batitle{{Early evolution of newly born magnetars with a strong toroidal
  field}}.
\bjtitle{\mnras}
\bvolume{398},
\bfpage{1869}--\blpage{1885}
(\byear{2009}).
doi:\doiurl{10.1111/j.1365-2966.2008.14054.x}
\end{barticle}
\endbibitem

\bibitem[\protect\citeauthoryear{{Dall'Osso} et~al.}{2011}]{DallOsso11}
\begin{barticle}
\bauthor{\binits{S.} \bsnm{{Dall'Osso}}},
\bauthor{\binits{G.} \bsnm{{Stratta}}},
\bauthor{\binits{D.} \bsnm{{Guetta}}},
\bauthor{\binits{S.} \bsnm{{Covino}}},
\bauthor{\binits{G.} \bsnm{{De Cesare}}},
\bauthor{\binits{L.} \bsnm{{Stella}}},
\batitle{{Gamma-ray bursts afterglows with energy injection from a spinning
  down neutron star}}.
\bjtitle{\aap}
\bvolume{526},
\bfpage{121}
(\byear{2011}).
doi:\doiurl{10.1051/0004-6361/201014168}
\end{barticle}
\endbibitem

\bibitem[\protect\citeauthoryear{{Deng} and {Zhang}}{2014}]{deng:14}
\begin{barticle}
\bauthor{\binits{W.} \bsnm{{Deng}}},
\bauthor{\binits{B.} \bsnm{{Zhang}}},
\batitle{{Low Energy Spectral Index and E$_{p}$ Evolution of Quasi-thermal
  Photosphere Emission of Gamma-Ray Bursts}}.
\bjtitle{\apj}
\bvolume{785},
\bfpage{112}
(\byear{2014}).
doi:\doiurl{10.1088/0004-637X/785/2/112}
\end{barticle}
\endbibitem

\bibitem[\protect\citeauthoryear{{Derishev}}{2007}]{derishev:07}
\begin{barticle}
\bauthor{\binits{E.V.} \bsnm{{Derishev}}},
\batitle{{Synchrotron emission in the fast cooling regime: which spectra can be
  explained?}}
\bjtitle{\apss}
\bvolume{309},
\bfpage{157}--\blpage{161}
(\byear{2007}).
doi:\doiurl{10.1007/s10509-007-9421-z}
\end{barticle}
\endbibitem

\bibitem[\protect\citeauthoryear{{Derishev} et~al.}{2001}]{derishev:01}
\begin{barticle}
\bauthor{\binits{E.V.} \bsnm{{Derishev}}},
\bauthor{\binits{V.V.} \bsnm{{Kocharovsky}}},
\bauthor{\binits{V.V.} \bsnm{{Kocharovsky}}},
\batitle{{Physical parameters and emission mechanism in gamma-ray bursts}}.
\bjtitle{\aap}
\bvolume{372},
\bfpage{1071}--\blpage{1077}
(\byear{2001}).
doi:\doiurl{10.1051/0004-6361:20010586}
\end{barticle}
\endbibitem

\bibitem[\protect\citeauthoryear{{Drenkhahn} and {Spruit}}{2002}]{DS02}
\begin{barticle}
\bauthor{\binits{G.} \bsnm{{Drenkhahn}}},
\bauthor{\binits{H.C.} \bsnm{{Spruit}}},
\batitle{{Efficient acceleration and radiation in Poynting flux powered GRB
  outflows}}.
\bjtitle{\aap}
\bvolume{391},
\bfpage{1141}--\blpage{1153}
(\byear{2002}).
doi:\doiurl{10.1051/0004-6361:20020839}
\end{barticle}
\endbibitem

\bibitem[\protect\citeauthoryear{{Duncan} and {Thompson}}{1992}]{DT92}
\begin{barticle}
\bauthor{\binits{R.C.} \bsnm{{Duncan}}},
\bauthor{\binits{C.} \bsnm{{Thompson}}},
\batitle{{Formation of very strongly magnetized neutron stars - Implications
  for gamma-ray bursts}}.
\bjtitle{\apjl}
\bvolume{392},
\bfpage{9}--\blpage{13}
(\byear{1992}).
doi:\doiurl{10.1086/186413}
\end{barticle}
\endbibitem

\bibitem[\protect\citeauthoryear{{Eichler}}{1993}]{Eichler93}
\begin{barticle}
\bauthor{\binits{D.} \bsnm{{Eichler}}},
\batitle{{Magnetic Confinement of Jets}}.
\bjtitle{\apj}
\bvolume{419},
\bfpage{111}
(\byear{1993}).
doi:\doiurl{10.1086/173464}
\end{barticle}
\endbibitem

\bibitem[\protect\citeauthoryear{{Eichler} and {Granot}}{2006}]{EG06}
\begin{barticle}
\bauthor{\binits{D.} \bsnm{{Eichler}}},
\bauthor{\binits{J.} \bsnm{{Granot}}},
\batitle{{The Case for Anisotropic Afterglow Efficiency within Gamma-Ray Burst
  Jets}}.
\bjtitle{\apjl}
\bvolume{641},
\bfpage{5}--\blpage{8}
(\byear{2006}).
doi:\doiurl{10.1086/503667}
\end{barticle}
\endbibitem

\bibitem[\protect\citeauthoryear{{Eichler} and {Waxman}}{2005}]{EW05}
\begin{barticle}
\bauthor{\binits{D.} \bsnm{{Eichler}}},
\bauthor{\binits{E.} \bsnm{{Waxman}}},
\batitle{{The Efficiency of Electron Acceleration in Collisionless Shocks and
  Gamma-Ray Burst Energetics}}.
\bjtitle{\apj}
\bvolume{627},
\bfpage{861}--\blpage{867}
(\byear{2005}).
doi:\doiurl{10.1086/430596}
\end{barticle}
\endbibitem

\bibitem[\protect\citeauthoryear{{Eichler} et~al.}{1989}]{ELPS89}
\begin{barticle}
\bauthor{\binits{D.} \bsnm{{Eichler}}},
\bauthor{\binits{M.} \bsnm{{Livio}}},
\bauthor{\binits{T.} \bsnm{{Piran}}},
\bauthor{\binits{D.N.} \bsnm{{Schramm}}},
\batitle{{Nucleosynthesis, neutrino bursts and gamma-rays from coalescing
  neutron stars}}.
\bjtitle{\nat}
\bvolume{340},
\bfpage{126}--\blpage{128}
(\byear{1989}).
doi:\doiurl{10.1038/340126a0}
\end{barticle}
\endbibitem

\bibitem[\protect\citeauthoryear{{Fan}}{2010}]{Fan10}
\begin{barticle}
\bauthor{\binits{Y.-Z.} \bsnm{{Fan}}},
\batitle{{The spectrum of {$\gamma$}-ray burst: a clue}}.
\bjtitle{\mnras}
\bvolume{403},
\bfpage{483}--\blpage{490}
(\byear{2010}).
doi:\doiurl{10.1111/j.1365-2966.2009.16134.x}
\end{barticle}
\endbibitem

\bibitem[\protect\citeauthoryear{{Fan} et~al.}{2013a}]{Fan13b}
\begin{barticle}
\bauthor{\binits{Y.-Z.} \bsnm{{Fan}}}, \betal,
\batitle{{A Supramassive Magnetar Central Engine for GRB 130603B}}.
\bjtitle{\apjl}
\bvolume{779},
\bfpage{25}
(\byear{2013}a).
doi:\doiurl{10.1088/2041-8205/779/2/L25}
\end{barticle}
\endbibitem

\bibitem[\protect\citeauthoryear{{Fan} et~al.}{2013b}]{Fan13}
\begin{barticle}
\bauthor{\binits{Y.-Z.} \bsnm{{Fan}}}, \betal,
\batitle{{High-energy Emission of GRB 130427A: Evidence for Inverse Compton
  Radiation}}.
\bjtitle{\apj}
\bvolume{776},
\bfpage{95}
(\byear{2013}b).
doi:\doiurl{10.1088/0004-637X/776/2/95}
\end{barticle}
\endbibitem

\bibitem[\protect\citeauthoryear{{Fan} and {Xu}}{2006}]{FX06}
\begin{barticle}
\bauthor{\binits{Y.-Z.} \bsnm{{Fan}}},
\bauthor{\binits{D.} \bsnm{{Xu}}},
\batitle{{The X-ray afterglow flat segment in short GRB 051221A: Energy
  injection from a millisecond magnetar?}}
\bjtitle{\mnras}
\bvolume{372},
\bfpage{19}--\blpage{22}
(\byear{2006}).
doi:\doiurl{10.1111/j.1745-3933.2006.00217.x}
\end{barticle}
\endbibitem

\bibitem[\protect\citeauthoryear{{Frail} et~al.}{2000}]{Frail00}
\begin{barticle}
\bauthor{\binits{D.A.} \bsnm{{Frail}}}, \betal,
\batitle{{The Enigmatic Radio Afterglow of GRB 991216}}.
\bjtitle{\apjl}
\bvolume{538},
\bfpage{129}--\blpage{132}
(\byear{2000}).
doi:\doiurl{10.1086/312807}
\end{barticle}
\endbibitem

\bibitem[\protect\citeauthoryear{{Gao} et~al.}{2012}]{gao:12}
\begin{barticle}
\bauthor{\binits{H.} \bsnm{{Gao}}},
\bauthor{\binits{B.-B.} \bsnm{{Zhang}}},
\bauthor{\binits{B.} \bsnm{{Zhang}}},
\batitle{{Stepwise Filter Correlation Method and Evidence of Superposed
  Variability Components in Gamma-Ray Burst Prompt Emission Light Curves}}.
\bjtitle{\apj}
\bvolume{748},
\bfpage{134}
(\byear{2012}).
doi:\doiurl{10.1088/0004-637X/748/2/134}
\end{barticle}
\endbibitem

\bibitem[\protect\citeauthoryear{{Gendre} et~al.}{2010}]{Gendre10}
\begin{barticle}
\bauthor{\binits{B.} \bsnm{{Gendre}}}, \betal,
\batitle{{Testing gamma-ray burst models with the afterglow of GRB 090102}}.
\bjtitle{\mnras}
\bvolume{405},
\bfpage{2372}--\blpage{2380}
(\byear{2010}).
doi:\doiurl{10.1111/j.1365-2966.2010.16601.x}
\end{barticle}
\endbibitem

\bibitem[\protect\citeauthoryear{{Genet} and {Granot}}{2009}]{genet:09}
\begin{barticle}
\bauthor{\binits{F.} \bsnm{{Genet}}},
\bauthor{\binits{J.} \bsnm{{Granot}}},
\batitle{{Realistic analytic model for the prompt and high-latitude emission in
  GRBs}}.
\bjtitle{\mnras}
\bvolume{399},
\bfpage{1328}--\blpage{1346}
(\byear{2009}).
doi:\doiurl{10.1111/j.1365-2966.2009.15355.x}
\end{barticle}
\endbibitem

\bibitem[\protect\citeauthoryear{{Genet} et~al.}{2007}]{genet:07}
\begin{barticle}
\bauthor{\binits{F.} \bsnm{{Genet}}},
\bauthor{\binits{F.} \bsnm{{Daigne}}},
\bauthor{\binits{R.} \bsnm{{Mochkovitch}}},
\batitle{{Can the early X-ray afterglow of gamma-ray bursts be explained by a
  contribution from the reverse shock?}}
\bjtitle{\mnras}
\bvolume{381},
\bfpage{732}--\blpage{740}
(\byear{2007}).
doi:\doiurl{10.1111/j.1365-2966.2007.12243.x}
\end{barticle}
\endbibitem

\bibitem[\protect\citeauthoryear{{Ghisellini}}{2012}]{Ghisellini12}
\begin{barticle}
\bauthor{\binits{G.} \bsnm{{Ghisellini}}},
\batitle{{Jetted Active Galactic Nuclei}}.
\bjtitle{International Journal of Modern Physics Conference Series}
\bvolume{8},
\bfpage{1}--\blpage{12}
(\byear{2012}).
doi:\doiurl{10.1142/S2010194512004345}
\end{barticle}
\endbibitem

\bibitem[\protect\citeauthoryear{{Ghisellini} and
  {Celotti}}{1999}]{GhiselliniCelotti99}
\begin{barticle}
\bauthor{\binits{G.} \bsnm{{Ghisellini}}},
\bauthor{\binits{A.} \bsnm{{Celotti}}},
\batitle{{Quasi-thermal Comptonization and Gamma-Ray Bursts}}.
\bjtitle{\apjl}
\bvolume{511},
\bfpage{93}--\blpage{96}
(\byear{1999}).
doi:\doiurl{10.1086/311845}
\end{barticle}
\endbibitem

\bibitem[\protect\citeauthoryear{{Ghisellini} and {Lazzati}}{1999}]{GL99}
\begin{barticle}
\bauthor{\binits{G.} \bsnm{{Ghisellini}}},
\bauthor{\binits{D.} \bsnm{{Lazzati}}},
\batitle{{Polarization light curves and position angle variation of beamed
  gamma-ray bursts}}.
\bjtitle{\mnras}
\bvolume{309},
\bfpage{7}--\blpage{11}
(\byear{1999}).
doi:\doiurl{10.1046/j.1365-8711.1999.03025.x}
\end{barticle}
\endbibitem

\bibitem[\protect\citeauthoryear{{Ghisellini} et~al.}{2000}]{ghisellini:00}
\begin{barticle}
\bauthor{\binits{G.} \bsnm{{Ghisellini}}},
\bauthor{\binits{A.} \bsnm{{Celotti}}},
\bauthor{\binits{D.} \bsnm{{Lazzati}}},
\batitle{{Constraints on the emission mechanisms of gamma-ray bursts}}.
\bjtitle{\mnras}
\bvolume{313},
\bfpage{1}--\blpage{5}
(\byear{2000}).
doi:\doiurl{10.1046/j.1365-8711.2000.03354.x}
\end{barticle}
\endbibitem

\bibitem[\protect\citeauthoryear{{Giannios}}{2008}]{giannios:08}
\begin{barticle}
\bauthor{\binits{D.} \bsnm{{Giannios}}},
\batitle{{Prompt GRB emission from gradual energy dissipation}}.
\bjtitle{\aap}
\bvolume{480},
\bfpage{305}--\blpage{312}
(\byear{2008}).
doi:\doiurl{10.1051/0004-6361:20079085}
\end{barticle}
\endbibitem

\bibitem[\protect\citeauthoryear{{Giannios} and {Spruit}}{2006}]{GS06}
\begin{barticle}
\bauthor{\binits{D.} \bsnm{{Giannios}}},
\bauthor{\binits{H.C.} \bsnm{{Spruit}}},
\batitle{{The role of kink instability in Poynting-flux dominated jets}}.
\bjtitle{\aap}
\bvolume{450},
\bfpage{887}--\blpage{898}
(\byear{2006}).
doi:\doiurl{10.1051/0004-6361:20054107}
\end{barticle}
\endbibitem

\bibitem[\protect\citeauthoryear{{Giannios} and {Spruit}}{2007}]{giannios:07}
\begin{barticle}
\bauthor{\binits{D.} \bsnm{{Giannios}}},
\bauthor{\binits{H.C.} \bsnm{{Spruit}}},
\batitle{{Spectral and timing properties of a dissipative {$\gamma$}-ray burst
  photosphere}}.
\bjtitle{\aap}
\bvolume{469},
\bfpage{1}--\blpage{9}
(\byear{2007}).
doi:\doiurl{10.1051/0004-6361:20066739}
\end{barticle}
\endbibitem

\bibitem[\protect\citeauthoryear{{Goldreich} and {Julian}}{1970}]{GJ70}
\begin{barticle}
\bauthor{\binits{P.} \bsnm{{Goldreich}}},
\bauthor{\binits{W.H.} \bsnm{{Julian}}},
\batitle{{Stellar Winds}}.
\bjtitle{\apj}
\bvolume{160},
\bfpage{971}
(\byear{1970}).
doi:\doiurl{10.1086/150486}
\end{barticle}
\endbibitem

\bibitem[\protect\citeauthoryear{{Gompertz} et~al.}{2014}]{GOW14}
\begin{barticle}
\bauthor{\binits{B.P.} \bsnm{{Gompertz}}},
\bauthor{\binits{P.T.} \bsnm{{O'Brien}}},
\bauthor{\binits{G.A.} \bsnm{{Wynn}}},
\batitle{{Magnetar powered GRBs: explaining the extended emission and X-ray
  plateau of short GRB light curves}}.
\bjtitle{\mnras}
\bvolume{438},
\bfpage{240}--\blpage{250}
(\byear{2014}).
doi:\doiurl{10.1093/mnras/stt2165}
\end{barticle}
\endbibitem

\bibitem[\protect\citeauthoryear{{Goodman}}{1986}]{goodman:86}
\begin{barticle}
\bauthor{\binits{J.} \bsnm{{Goodman}}},
\batitle{{Are gamma-ray bursts optically thick?}}
\bjtitle{\apjl}
\bvolume{308},
\bfpage{47}--\blpage{50}
(\byear{1986}).
doi:\doiurl{10.1086/184741}
\end{barticle}
\endbibitem

\bibitem[\protect\citeauthoryear{{G{\"o}tz} et~al.}{2014}]{gotz:14}
\begin{barticle}
\bauthor{\binits{D.} \bsnm{{G{\"o}tz}}}, \betal,
\batitle{{GRB 140206A: the most distant polarized gamma-ray burst}}.
\bjtitle{\mnras}
\bvolume{444},
\bfpage{2776}--\blpage{2782}
(\byear{2014}).
doi:\doiurl{10.1093/mnras/stu1634}
\end{barticle}
\endbibitem

\bibitem[\protect\citeauthoryear{{G{\"o}tz} et~al.}{2009}]{gotz:09}
\begin{barticle}
\bauthor{\binits{D.} \bsnm{{G{\"o}tz}}},
\bauthor{\binits{P.} \bsnm{{Laurent}}},
\bauthor{\binits{F.} \bsnm{{Lebrun}}},
\bauthor{\binits{F.} \bsnm{{Daigne}}},
\bauthor{\binits{{\v Z}.} \bsnm{{Bo{\v s}njak}}},
\batitle{{Variable Polarization Measured in the Prompt Emission of GRB 041219A
  Using IBIS on Board INTEGRAL}}.
\bjtitle{\apjl}
\bvolume{695},
\bfpage{208}--\blpage{212}
(\byear{2009}).
doi:\doiurl{10.1088/0004-637X/695/2/L208}
\end{barticle}
\endbibitem

\bibitem[\protect\citeauthoryear{{Granot}}{2003}]{Granot03}
\begin{barticle}
\bauthor{\binits{J.} \bsnm{{Granot}}},
\batitle{{The Most Probable Cause for the High Gamma-Ray Polarization in GRB
  021206}}.
\bjtitle{\apjl}
\bvolume{596},
\bfpage{17}--\blpage{21}
(\byear{2003}).
doi:\doiurl{10.1086/379110}
\end{barticle}
\endbibitem

\bibitem[\protect\citeauthoryear{{Granot}}{2012a}]{Granot12a}
\begin{barticle}
\bauthor{\binits{J.} \bsnm{{Granot}}},
\batitle{{Interaction of a highly magnetized impulsive relativistic flow with
  an external medium}}.
\bjtitle{\mnras}
\bvolume{421},
\bfpage{2442}--\blpage{2466}
(\byear{2012}a).
doi:\doiurl{10.1111/j.1365-2966.2012.20473.x}
\end{barticle}
\endbibitem

\bibitem[\protect\citeauthoryear{{Granot}}{2012b}]{Granot12b}
\begin{barticle}
\bauthor{\binits{J.} \bsnm{{Granot}}},
\batitle{{The effects of sub-shells in highly magnetized relativistic flows}}.
\bjtitle{\mnras}
\bvolume{421},
\bfpage{2467}--\blpage{2477}
(\byear{2012}b).
doi:\doiurl{10.1111/j.1365-2966.2012.20474.x}
\end{barticle}
\endbibitem

\bibitem[\protect\citeauthoryear{{Granot} and {K{\"o}nigl}}{2003}]{GK03}
\begin{barticle}
\bauthor{\binits{J.} \bsnm{{Granot}}},
\bauthor{\binits{A.} \bsnm{{K{\"o}nigl}}},
\batitle{{Linear Polarization in Gamma-Ray Bursts: The Case for an Ordered
  Magnetic Field}}.
\bjtitle{\apjl}
\bvolume{594},
\bfpage{83}--\blpage{87}
(\byear{2003}).
doi:\doiurl{10.1086/378733}
\end{barticle}
\endbibitem

\bibitem[\protect\citeauthoryear{{Granot} and {Kumar}}{2006}]{GK06}
\begin{barticle}
\bauthor{\binits{J.} \bsnm{{Granot}}},
\bauthor{\binits{P.} \bsnm{{Kumar}}},
\batitle{{Distribution of gamma-ray burst ejecta energy with Lorentz factor}}.
\bjtitle{\mnras}
\bvolume{366},
\bfpage{13}--\blpage{16}
(\byear{2006}).
doi:\doiurl{10.1111/j.1745-3933.2005.00121.x}
\end{barticle}
\endbibitem

\bibitem[\protect\citeauthoryear{{Granot} and {Sari}}{2002}]{GS02}
\begin{barticle}
\bauthor{\binits{J.} \bsnm{{Granot}}},
\bauthor{\binits{R.} \bsnm{{Sari}}},
\batitle{{The Shape of Spectral Breaks in Gamma-Ray Burst Afterglows}}.
\bjtitle{\apj}
\bvolume{568},
\bfpage{820}--\blpage{829}
(\byear{2002}).
doi:\doiurl{10.1086/338966}
\end{barticle}
\endbibitem

\bibitem[\protect\citeauthoryear{{Granot} and {Taylor}}{2005}]{GT05}
\begin{barticle}
\bauthor{\binits{J.} \bsnm{{Granot}}},
\bauthor{\binits{G.B.} \bsnm{{Taylor}}},
\batitle{{Radio Flares and the Magnetic Field Structure in Gamma-Ray Burst
  Outflows}}.
\bjtitle{\apj}
\bvolume{625},
\bfpage{263}--\blpage{270}
(\byear{2005}).
doi:\doiurl{10.1086/429536}
\end{barticle}
\endbibitem

\bibitem[\protect\citeauthoryear{{Granot} et~al.}{2008}]{granot:08}
\begin{barticle}
\bauthor{\binits{J.} \bsnm{{Granot}}},
\bauthor{\binits{J.} \bsnm{{Cohen-Tanugi}}},
\bauthor{\binits{E.} \bsnm{{do Couto e Silva}}},
\batitle{{Opacity Buildup in Impulsive Relativistic Sources}}.
\bjtitle{\apj}
\bvolume{677},
\bfpage{92}--\blpage{126}
(\byear{2008}).
doi:\doiurl{10.1086/526414}
\end{barticle}
\endbibitem

\bibitem[\protect\citeauthoryear{{Granot} et~al.}{2011}]{GKS11}
\begin{barticle}
\bauthor{\binits{J.} \bsnm{{Granot}}},
\bauthor{\binits{S.S.} \bsnm{{Komissarov}}},
\bauthor{\binits{A.} \bsnm{{Spitkovsky}}},
\batitle{{Impulsive acceleration of strongly magnetized relativistic flows}}.
\bjtitle{\mnras}
\bvolume{411},
\bfpage{1323}--\blpage{1353}
(\byear{2011}).
doi:\doiurl{10.1111/j.1365-2966.2010.17770.x}
\end{barticle}
\endbibitem

\bibitem[\protect\citeauthoryear{{Granot} et~al.}{2006}]{GKP06}
\begin{barticle}
\bauthor{\binits{J.} \bsnm{{Granot}}},
\bauthor{\binits{A.} \bsnm{{K{\"o}nigl}}},
\bauthor{\binits{T.} \bsnm{{Piran}}},
\batitle{{Implications of the early X-ray afterglow light curves of Swift
  gamma-ray bursts}}.
\bjtitle{\mnras}
\bvolume{370},
\bfpage{1946}--\blpage{1960}
(\byear{2006}).
doi:\doiurl{10.1111/j.1365-2966.2006.10621.x}
\end{barticle}
\endbibitem

\bibitem[\protect\citeauthoryear{{Gruber} et~al.}{2014}]{gruber:14}
\begin{barticle}
\bauthor{\binits{D.} \bsnm{{Gruber}}}, \betal,
\batitle{{The Fermi GBM Gamma-Ray Burst Spectral Catalog: Four Years of Data}}.
\bjtitle{\apjs}
\bvolume{211},
\bfpage{12}
(\byear{2014}).
doi:\doiurl{10.1088/0067-0049/211/1/12}
\end{barticle}
\endbibitem

\bibitem[\protect\citeauthoryear{{Gruzinov} and {Waxman}}{1999}]{GW99}
\begin{barticle}
\bauthor{\binits{A.} \bsnm{{Gruzinov}}},
\bauthor{\binits{E.} \bsnm{{Waxman}}},
\batitle{{Gamma-Ray Burst Afterglow: Polarization and Analytic Light Curves}}.
\bjtitle{\apj}
\bvolume{511},
\bfpage{852}--\blpage{861}
(\byear{1999}).
doi:\doiurl{10.1086/306720}
\end{barticle}
\endbibitem

\bibitem[\protect\citeauthoryear{{Guetta} and {Granot}}{2003}]{GG03}
\begin{barticle}
\bauthor{\binits{D.} \bsnm{{Guetta}}},
\bauthor{\binits{J.} \bsnm{{Granot}}},
\batitle{{High-Energy Emission from the Prompt Gamma-Ray Burst}}.
\bjtitle{\apj}
\bvolume{585},
\bfpage{885}--\blpage{889}
(\byear{2003}).
doi:\doiurl{10.1086/346221}
\end{barticle}
\endbibitem

\bibitem[\protect\citeauthoryear{{Guidorzi} et~al.}{2012}]{guidorzi:12}
\begin{barticle}
\bauthor{\binits{C.} \bsnm{{Guidorzi}}}, \betal,
\batitle{{Average power density spectrum of Swift long gamma-ray bursts in the
  observer and in the source-rest frames}}.
\bjtitle{\mnras}
\bvolume{422},
\bfpage{1785}--\blpage{1803}
(\byear{2012}).
doi:\doiurl{10.1111/j.1365-2966.2012.20758.x}
\end{barticle}
\endbibitem

\bibitem[\protect\citeauthoryear{{Guiriec} et~al.}{2010}]{guiriec:10}
\begin{barticle}
\bauthor{\binits{S.} \bsnm{{Guiriec}}}, \betal,
\batitle{{Time-resolved Spectroscopy of the Three Brightest and Hardest Short
  Gamma-ray Bursts Observed with the Fermi Gamma-ray Burst Monitor}}.
\bjtitle{\apj}
\bvolume{725},
\bfpage{225}--\blpage{241}
(\byear{2010}).
doi:\doiurl{10.1088/0004-637X/725/1/225}
\end{barticle}
\endbibitem

\bibitem[\protect\citeauthoryear{{Guiriec} et~al.}{2011}]{Guiriec11}
\begin{barticle}
\bauthor{\binits{S.} \bsnm{{Guiriec}}}, \betal,
\batitle{{Detection of a Thermal Spectral Component in the Prompt Emission of
  GRB 100724B}}.
\bjtitle{\apjl}
\bvolume{727},
\bfpage{33}
(\byear{2011}).
doi:\doiurl{10.1088/2041-8205/727/2/L33}
\end{barticle}
\endbibitem

\bibitem[\protect\citeauthoryear{{Guiriec} et~al.}{2013}]{Guiriec13}
\begin{barticle}
\bauthor{\binits{S.} \bsnm{{Guiriec}}}, \betal,
\batitle{{Evidence for a Photospheric Component in the Prompt Emission of the
  Short GRB 120323A and Its Effects on the GRB Hardness-Luminosity Relation}}.
\bjtitle{\apj}
\bvolume{770},
\bfpage{32}
(\byear{2013}).
doi:\doiurl{10.1088/0004-637X/770/1/32}
\end{barticle}
\endbibitem

\bibitem[\protect\citeauthoryear{{Guiriec} et~al.}{2015}]{Guiriec15}
\begin{botherref}
\oauthor{\binits{S.} \bsnm{{Guiriec}}}, et al.,
{Towards a Better Understanding of the GRB Phenomenon: a New Model for GRB
  Prompt Emission and its effects on the New Non-Thermal
  $L_\mathrm{i}^\mathrm{NT}$-$E_\mathrm{peak,i}^\mathrm{rest,NT}$ relation}.
ArXiv e-prints
(2015)
\end{botherref}
\endbibitem

\bibitem[\protect\citeauthoryear{{Harrison} et~al.}{2001}]{Harrison01}
\begin{barticle}
\bauthor{\binits{F.A.} \bsnm{{Harrison}}}, \betal,
\batitle{{Broadband Observations of the Afterglow of GRB 000926: Observing the
  Effect of Inverse Compton Scattering}}.
\bjtitle{\apj}
\bvolume{559},
\bfpage{123}--\blpage{130}
(\byear{2001}).
doi:\doiurl{10.1086/322368}
\end{barticle}
\endbibitem

\bibitem[\protect\citeauthoryear{{Hasco{\"e}t} et~al.}{2012}]{hascoet:12b}
\begin{barticle}
\bauthor{\binits{R.} \bsnm{{Hasco{\"e}t}}},
\bauthor{\binits{F.} \bsnm{{Daigne}}},
\bauthor{\binits{R.} \bsnm{{Mochkovitch}}},
\batitle{{Accounting for the XRT early steep decay in models of the prompt
  gamma-ray burst emission}}.
\bjtitle{\aap}
\bvolume{542},
\bfpage{29}
(\byear{2012}).
doi:\doiurl{10.1051/0004-6361/201219339}
\end{barticle}
\endbibitem

\bibitem[\protect\citeauthoryear{{Hasco{\"e}t} et~al.}{2013}]{hascoet:13}
\begin{barticle}
\bauthor{\binits{R.} \bsnm{{Hasco{\"e}t}}},
\bauthor{\binits{F.} \bsnm{{Daigne}}},
\bauthor{\binits{R.} \bsnm{{Mochkovitch}}},
\batitle{{Prompt thermal emission in gamma-ray bursts}}.
\bjtitle{\aap}
\bvolume{551},
\bfpage{124}
(\byear{2013}).
doi:\doiurl{10.1051/0004-6361/201220023}
\end{barticle}
\endbibitem

\bibitem[\protect\citeauthoryear{{Hasco{\"e}t} et~al.}{2014}]{hascoet:14}
\begin{barticle}
\bauthor{\binits{R.} \bsnm{{Hasco{\"e}t}}},
\bauthor{\binits{F.} \bsnm{{Daigne}}},
\bauthor{\binits{R.} \bsnm{{Mochkovitch}}},
\batitle{{The prompt-early afterglow connection in gamma-ray bursts:
  implications for the early afterglow physics}}.
\bjtitle{\mnras}
\bvolume{442},
\bfpage{20}--\blpage{27}
(\byear{2014}).
doi:\doiurl{10.1093/mnras/stu750}
\end{barticle}
\endbibitem

\bibitem[\protect\citeauthoryear{{Hasco{\"e}t} et~al.}{2012}]{hascoet:12a}
\begin{barticle}
\bauthor{\binits{R.} \bsnm{{Hasco{\"e}t}}},
\bauthor{\binits{F.} \bsnm{{Daigne}}},
\bauthor{\binits{R.} \bsnm{{Mochkovitch}}},
\bauthor{\binits{V.} \bsnm{{Vennin}}},
\batitle{{Do Fermi Large Area Telescope observations imply very large Lorentz
  factors in gamma-ray burst outflows?}}
\bjtitle{\mnras}
\bvolume{421},
\bfpage{525}--\blpage{545}
(\byear{2012}).
doi:\doiurl{10.1111/j.1365-2966.2011.20332.x}
\end{barticle}
\endbibitem

\bibitem[\protect\citeauthoryear{{Hascoet} et~al.}{2015}]{hascoet:15}
\begin{botherref}
\oauthor{\binits{R.} \bsnm{{Hascoet}}},
\oauthor{\binits{A.M.} \bsnm{{Beloborodov}}},
\oauthor{\binits{F.} \bsnm{{Daigne}}},
\oauthor{\binits{R.} \bsnm{{Mochkovitch}}},
{X-ray flares from dense shells formed in gamma-ray burst explosions}.
ArXiv e-prints
(2015)
\end{botherref}
\endbibitem

\bibitem[\protect\citeauthoryear{{Heinz} and {Begelman}}{2000}]{HB00}
\begin{barticle}
\bauthor{\binits{S.} \bsnm{{Heinz}}},
\bauthor{\binits{M.C.} \bsnm{{Begelman}}},
\batitle{{Jet Acceleration by Tangled Magnetic Fields}}.
\bjtitle{\apj}
\bvolume{535},
\bfpage{104}--\blpage{117}
(\byear{2000}).
doi:\doiurl{10.1086/308820}
\end{barticle}
\endbibitem

\bibitem[\protect\citeauthoryear{{Kagan} et~al.}{2015}]{Kagan15}
\begin{barticle}
\bauthor{\binits{D.} \bsnm{{Kagan}}},
\bauthor{\binits{L.} \bsnm{{Sironi}}},
\bauthor{\binits{B.} \bsnm{{Cerutti}}},
\bauthor{\binits{D.} \bsnm{{Giannios}}},
\batitle{{Relativistic Magnetic Reconnection in Pair Plasmas and Its
  Astrophysical Applications}}.
\bjtitle{\ssr}
(\byear{2015}).
doi:\doiurl{10.1007/s11214-014-0132-9}
\end{barticle}
\endbibitem

\bibitem[\protect\citeauthoryear{{Kaneko} et~al.}{2006}]{kaneko:06}
\begin{barticle}
\bauthor{\binits{Y.} \bsnm{{Kaneko}}},
\bauthor{\binits{R.D.} \bsnm{{Preece}}},
\bauthor{\binits{M.S.} \bsnm{{Briggs}}},
\bauthor{\binits{W.S.} \bsnm{{Paciesas}}},
\bauthor{\binits{C.A.} \bsnm{{Meegan}}},
\bauthor{\binits{D.L.} \bsnm{{Band}}},
\batitle{{The Complete Spectral Catalog of Bright BATSE Gamma-Ray Bursts}}.
\bjtitle{\apjs}
\bvolume{166},
\bfpage{298}--\blpage{340}
(\byear{2006}).
doi:\doiurl{10.1086/505911}
\end{barticle}
\endbibitem

\bibitem[\protect\citeauthoryear{{Kawanaka} et~al.}{2013}]{KPK13}
\begin{barticle}
\bauthor{\binits{N.} \bsnm{{Kawanaka}}},
\bauthor{\binits{T.} \bsnm{{Piran}}},
\bauthor{\binits{J.H.} \bsnm{{Krolik}}},
\batitle{{Jet Luminosity from Neutrino-dominated Accretion Flows in Gamma-Ray
  Bursts}}.
\bjtitle{\apj}
\bvolume{766},
\bfpage{31}
(\byear{2013}).
doi:\doiurl{10.1088/0004-637X/766/1/31}
\end{barticle}
\endbibitem

\bibitem[\protect\citeauthoryear{{Keshet} et~al.}{2009}]{keshet:09}
\begin{barticle}
\bauthor{\binits{U.} \bsnm{{Keshet}}},
\bauthor{\binits{B.} \bsnm{{Katz}}},
\bauthor{\binits{A.} \bsnm{{Spitkovsky}}},
\bauthor{\binits{E.} \bsnm{{Waxman}}},
\batitle{{Magnetic Field Evolution in Relativistic Unmagnetized Collisionless
  Shocks}}.
\bjtitle{\apjl}
\bvolume{693},
\bfpage{127}--\blpage{130}
(\byear{2009}).
doi:\doiurl{10.1088/0004-637X/693/2/L127}
\end{barticle}
\endbibitem

\bibitem[\protect\citeauthoryear{{Klu{\'z}niak} and {Ruderman}}{1998}]{KR98}
\begin{barticle}
\bauthor{\binits{W.} \bsnm{{Klu{\'z}niak}}},
\bauthor{\binits{M.} \bsnm{{Ruderman}}},
\batitle{{The Central Engine of Gamma-Ray Bursters}}.
\bjtitle{\apjl}
\bvolume{505},
\bfpage{113}--\blpage{117}
(\byear{1998}).
doi:\doiurl{10.1086/311622}
\end{barticle}
\endbibitem

\bibitem[\protect\citeauthoryear{{Kobayashi} et~al.}{1997}]{kobayashi:97}
\begin{barticle}
\bauthor{\binits{S.} \bsnm{{Kobayashi}}},
\bauthor{\binits{T.} \bsnm{{Piran}}},
\bauthor{\binits{R.} \bsnm{{Sari}}},
\batitle{{Can Internal Shocks Produce the Variability in Gamma-Ray Bursts?}}
\bjtitle{\apj}
\bvolume{490},
\bfpage{92}
(\byear{1997}).
doi:\doiurl{10.1086/512791}
\end{barticle}
\endbibitem

\bibitem[\protect\citeauthoryear{{Komissarov}}{2011}]{Kom11}
\begin{barticle}
\bauthor{\binits{S.S.} \bsnm{{Komissarov}}},
\batitle{{Magnetic acceleration of relativistic jets.}}
\bjtitle{\memsai}
\bvolume{82},
\bfpage{95}
(\byear{2011})
\end{barticle}
\endbibitem

\bibitem[\protect\citeauthoryear{{Komissarov}}{2012}]{Kom12}
\begin{barticle}
\bauthor{\binits{S.S.} \bsnm{{Komissarov}}},
\batitle{{Shock dissipation in magnetically dominated impulsive flows}}.
\bjtitle{\mnras}
\bvolume{422},
\bfpage{326}--\blpage{346}
(\byear{2012}).
doi:\doiurl{10.1111/j.1365-2966.2012.20609.x}
\end{barticle}
\endbibitem

\bibitem[\protect\citeauthoryear{{Komissarov} et~al.}{2010}]{Kom10}
\begin{barticle}
\bauthor{\binits{S.S.} \bsnm{{Komissarov}}},
\bauthor{\binits{N.} \bsnm{{Vlahakis}}},
\bauthor{\binits{A.} \bsnm{{K{\"o}nigl}}},
\batitle{{Rarefaction acceleration of ultrarelativistic magnetized jets in
  gamma-ray burst sources}}.
\bjtitle{\mnras}
\bvolume{407},
\bfpage{17}--\blpage{28}
(\byear{2010}).
doi:\doiurl{10.1111/j.1365-2966.2010.16779.x}
\end{barticle}
\endbibitem

\bibitem[\protect\citeauthoryear{{Komissarov} et~al.}{2009}]{Kom09}
\begin{barticle}
\bauthor{\binits{S.S.} \bsnm{{Komissarov}}},
\bauthor{\binits{N.} \bsnm{{Vlahakis}}},
\bauthor{\binits{A.} \bsnm{{K{\"o}nigl}}},
\bauthor{\binits{M.V.} \bsnm{{Barkov}}},
\batitle{{Magnetic acceleration of ultrarelativistic jets in gamma-ray burst
  sources}}.
\bjtitle{\mnras}
\bvolume{394},
\bfpage{1182}--\blpage{1212}
(\byear{2009}).
doi:\doiurl{10.1111/j.1365-2966.2009.14410.x}
\end{barticle}
\endbibitem

\bibitem[\protect\citeauthoryear{{Kouveliotou} et~al.}{2013}]{Kouveliotou13}
\begin{barticle}
\bauthor{\binits{C.} \bsnm{{Kouveliotou}}}, \betal,
\batitle{{NuSTAR Observations of GRB 130427A Establish a Single Component
  Synchrotron Afterglow Origin for the Late Optical to Multi-GeV Emission}}.
\bjtitle{\apjl}
\bvolume{779},
\bfpage{1}
(\byear{2013}).
doi:\doiurl{10.1088/2041-8205/779/1/L1}
\end{barticle}
\endbibitem

\bibitem[\protect\citeauthoryear{{Kouveliotou} et~al.}{1993}]{Kouveliotou93}
\begin{barticle}
\bauthor{\binits{C.} \bsnm{{Kouveliotou}}},
\bauthor{\binits{C.A.} \bsnm{{Meegan}}},
\bauthor{\binits{G.J.} \bsnm{{Fishman}}},
\bauthor{\binits{N.P.} \bsnm{{Bhat}}},
\bauthor{\binits{M.S.} \bsnm{{Briggs}}},
\bauthor{\binits{T.M.} \bsnm{{Koshut}}},
\bauthor{\binits{W.S.} \bsnm{{Paciesas}}},
\bauthor{\binits{G.N.} \bsnm{{Pendleton}}},
\batitle{{Identification of two classes of gamma-ray bursts}}.
\bjtitle{\apjl}
\bvolume{413},
\bfpage{101}--\blpage{104}
(\byear{1993}).
doi:\doiurl{10.1086/186969}
\end{barticle}
\endbibitem

\bibitem[\protect\citeauthoryear{{Kulkarni} et~al.}{1999}]{Kulkarni99}
\begin{barticle}
\bauthor{\binits{S.R.} \bsnm{{Kulkarni}}}, \betal,
\batitle{{Discovery of a Radio Flare from GRB 990123}}.
\bjtitle{\apjl}
\bvolume{522},
\bfpage{97}--\blpage{100}
(\byear{1999}).
doi:\doiurl{10.1086/312227}
\end{barticle}
\endbibitem

\bibitem[\protect\citeauthoryear{{Kumar} and {McMahon}}{2008}]{KumarMcMahon08}
\begin{barticle}
\bauthor{\binits{P.} \bsnm{{Kumar}}},
\bauthor{\binits{E.} \bsnm{{McMahon}}},
\batitle{{A general scheme for modelling {$\gamma$}-ray burst prompt
  emission}}.
\bjtitle{\mnras}
\bvolume{384},
\bfpage{33}--\blpage{63}
(\byear{2008}).
doi:\doiurl{10.1111/j.1365-2966.2007.12621.x}
\end{barticle}
\endbibitem

\bibitem[\protect\citeauthoryear{{Kumar} and {Panaitescu}}{2000}]{kumar:00}
\begin{barticle}
\bauthor{\binits{P.} \bsnm{{Kumar}}},
\bauthor{\binits{A.} \bsnm{{Panaitescu}}},
\batitle{{Afterglow Emission from Naked Gamma-Ray Bursts}}.
\bjtitle{\apjl}
\bvolume{541},
\bfpage{51}--\blpage{54}
(\byear{2000}).
doi:\doiurl{10.1086/312905}
\end{barticle}
\endbibitem

\bibitem[\protect\citeauthoryear{{Kumar} and {Zhang}}{2015}]{KZ15}
\begin{barticle}
\bauthor{\binits{P.} \bsnm{{Kumar}}},
\bauthor{\binits{B.} \bsnm{{Zhang}}},
\batitle{{The physics of gamma-ray bursts \& relativistic jets}}.
\bjtitle{\physrep}
\bvolume{561},
\bfpage{1}--\blpage{109}
(\byear{2015}).
doi:\doiurl{10.1016/j.physrep.2014.09.008}
\end{barticle}
\endbibitem

\bibitem[\protect\citeauthoryear{{Kumar} et~al.}{2012}]{Kumar12}
\begin{barticle}
\bauthor{\binits{P.} \bsnm{{Kumar}}},
\bauthor{\binits{R.A.} \bsnm{{Hern{\'a}ndez}}},
\bauthor{\binits{{\v Z}.} \bsnm{{Bo{\v s}njak}}},
\bauthor{\binits{R.} \bsnm{{Barniol Duran}}},
\batitle{{Maximum synchrotron frequency for shock-accelerated particles}}.
\bjtitle{\mnras}
\bvolume{427},
\bfpage{40}--\blpage{44}
(\byear{2012}).
doi:\doiurl{10.1111/j.1745-3933.2012.01341.x}
\end{barticle}
\endbibitem

\bibitem[\protect\citeauthoryear{{Lazar} et~al.}{2009}]{lazar:09}
\begin{barticle}
\bauthor{\binits{A.} \bsnm{{Lazar}}},
\bauthor{\binits{E.} \bsnm{{Nakar}}},
\bauthor{\binits{T.} \bsnm{{Piran}}},
\batitle{{Gamma-Ray Burst Light Curves in the Relativistic Turbulence and
  Relativistic Subjet Models}}.
\bjtitle{\apjl}
\bvolume{695},
\bfpage{10}--\blpage{14}
(\byear{2009}).
doi:\doiurl{10.1088/0004-637X/695/1/L10}
\end{barticle}
\endbibitem

\bibitem[\protect\citeauthoryear{{Lemoine}}{2013}]{Lemoine13}
\begin{barticle}
\bauthor{\binits{M.} \bsnm{{Lemoine}}},
\batitle{{Synchrotron signature of a relativistic blast wave with decaying
  microturbulence}}.
\bjtitle{\mnras}
\bvolume{428},
\bfpage{845}--\blpage{866}
(\byear{2013}).
doi:\doiurl{10.1093/mnras/sts081}
\end{barticle}
\endbibitem

\bibitem[\protect\citeauthoryear{{Lemoine} et~al.}{2013}]{LLW13}
\begin{barticle}
\bauthor{\binits{M.} \bsnm{{Lemoine}}},
\bauthor{\binits{Z.} \bsnm{{Li}}},
\bauthor{\binits{X.-Y.} \bsnm{{Wang}}},
\batitle{{On the magnetization of gamma-ray burst blast waves}}.
\bjtitle{\mnras}
\bvolume{435},
\bfpage{3009}--\blpage{3016}
(\byear{2013}).
doi:\doiurl{10.1093/mnras/stt1494}
\end{barticle}
\endbibitem

\bibitem[\protect\citeauthoryear{{Levinson}}{2010}]{Levinson10}
\begin{barticle}
\bauthor{\binits{A.} \bsnm{{Levinson}}},
\batitle{{Interaction of a Magnetized Shell with an Ambient Medium: Limits on
  Impulsive Magnetic Acceleration}}.
\bjtitle{\apj}
\bvolume{720},
\bfpage{1490}--\blpage{1499}
(\byear{2010}).
doi:\doiurl{10.1088/0004-637X/720/2/1490}
\end{barticle}
\endbibitem

\bibitem[\protect\citeauthoryear{{Levinson} and {Eichler}}{1993}]{LE93}
\begin{barticle}
\bauthor{\binits{A.} \bsnm{{Levinson}}},
\bauthor{\binits{D.} \bsnm{{Eichler}}},
\batitle{{Baryon Purity in Cosmological Gamma-Ray Bursts as a Manifestation of
  Event Horizons}}.
\bjtitle{\apj}
\bvolume{418},
\bfpage{386}
(\byear{1993}).
doi:\doiurl{10.1086/173397}
\end{barticle}
\endbibitem

\bibitem[\protect\citeauthoryear{{Levinson} and {Globus}}{2013}]{LG13}
\begin{barticle}
\bauthor{\binits{A.} \bsnm{{Levinson}}},
\bauthor{\binits{N.} \bsnm{{Globus}}},
\batitle{{Ultra-relativistic, Neutrino-driven Flows in Gamma-Ray Bursts: A
  Double Transonic Flow Solution in a Schwarzschild Spacetime}}.
\bjtitle{\apj}
\bvolume{770},
\bfpage{159}
(\byear{2013}).
doi:\doiurl{10.1088/0004-637X/770/2/159}
\end{barticle}
\endbibitem

\bibitem[\protect\citeauthoryear{{Liang} et~al.}{2006}]{liang:06}
\begin{barticle}
\bauthor{\binits{E.W.} \bsnm{{Liang}}}, \betal,
\batitle{{Testing the Curvature Effect and Internal Origin of Gamma-Ray Burst
  Prompt Emissions and X-Ray Flares with Swift Data}}.
\bjtitle{\apj}
\bvolume{646},
\bfpage{351}--\blpage{357}
(\byear{2006}).
doi:\doiurl{10.1086/504684}
\end{barticle}
\endbibitem

\bibitem[\protect\citeauthoryear{{Liang} et~al.}{2007}]{Liang07}
\begin{barticle}
\bauthor{\binits{E.} \bsnm{{Liang}}},
\bauthor{\binits{B.} \bsnm{{Zhang}}},
\bauthor{\binits{F.} \bsnm{{Virgili}}},
\bauthor{\binits{Z.G.} \bsnm{{Dai}}},
\batitle{{Low-Luminosity Gamma-Ray Bursts as a Unique Population: Luminosity
  Function, Local Rate, and Beaming Factor}}.
\bjtitle{\apj}
\bvolume{662},
\bfpage{1111}--\blpage{1118}
(\byear{2007}).
doi:\doiurl{10.1086/517959}
\end{barticle}
\endbibitem

\bibitem[\protect\citeauthoryear{{Lindner} et~al.}{2010}]{Lindner10}
\begin{barticle}
\bauthor{\binits{C.C.} \bsnm{{Lindner}}},
\bauthor{\binits{M.} \bsnm{{Milosavljevi{\'c}}}},
\bauthor{\binits{S.M.} \bsnm{{Couch}}},
\bauthor{\binits{P.} \bsnm{{Kumar}}},
\batitle{{Collapsar Accretion and the Gamma-Ray Burst X-Ray Light Curve}}.
\bjtitle{\apj}
\bvolume{713},
\bfpage{800}--\blpage{815}
(\byear{2010}).
doi:\doiurl{10.1088/0004-637X/713/2/800}
\end{barticle}
\endbibitem

\bibitem[\protect\citeauthoryear{{Lithwick} and {Sari}}{2001}]{lithwick:01}
\begin{barticle}
\bauthor{\binits{Y.} \bsnm{{Lithwick}}},
\bauthor{\binits{R.} \bsnm{{Sari}}},
\batitle{{Lower Limits on Lorentz Factors in Gamma-Ray Bursts}}.
\bjtitle{\apj}
\bvolume{555},
\bfpage{540}--\blpage{545}
(\byear{2001}).
doi:\doiurl{10.1086/321455}
\end{barticle}
\endbibitem

\bibitem[\protect\citeauthoryear{{Liu} et~al.}{2013}]{Liu13}
\begin{barticle}
\bauthor{\binits{R.-Y.} \bsnm{{Liu}}},
\bauthor{\binits{X.-Y.} \bsnm{{Wang}}},
\bauthor{\binits{X.-F.} \bsnm{{Wu}}},
\batitle{{Interpretation of the Unprecedentedly Long-lived High-energy Emission
  of GRB 130427A}}.
\bjtitle{\apjl}
\bvolume{773},
\bfpage{20}
(\byear{2013}).
doi:\doiurl{10.1088/2041-8205/773/2/L20}
\end{barticle}
\endbibitem

\bibitem[\protect\citeauthoryear{{Loeb} and {Perna}}{1998}]{LP98}
\begin{barticle}
\bauthor{\binits{A.} \bsnm{{Loeb}}},
\bauthor{\binits{R.} \bsnm{{Perna}}},
\batitle{{Microlensing of Gamma-Ray Burst Afterglows}}.
\bjtitle{\apj}
\bvolume{495},
\bfpage{597}--\blpage{603}
(\byear{1998}).
doi:\doiurl{10.1086/305337}
\end{barticle}
\endbibitem

\bibitem[\protect\citeauthoryear{{Lu} et~al.}{2012}]{lu:12}
\begin{barticle}
\bauthor{\binits{R.-J.} \bsnm{{Lu}}}, \betal,
\batitle{{A Comprehensive Analysis of Fermi Gamma-Ray Burst Data. II. E $_{p}$
  Evolution Patterns and Implications for the Observed Spectrum-Luminosity
  Relations}}.
\bjtitle{\apj}
\bvolume{756},
\bfpage{112}
(\byear{2012}).
doi:\doiurl{10.1088/0004-637X/756/2/112}
\end{barticle}
\endbibitem

\bibitem[\protect\citeauthoryear{{Lundman} et~al.}{2013}]{lundman:13}
\begin{barticle}
\bauthor{\binits{C.} \bsnm{{Lundman}}},
\bauthor{\binits{A.} \bsnm{{Pe'er}}},
\bauthor{\binits{F.} \bsnm{{Ryde}}},
\batitle{{A theory of photospheric emission from relativistic, collimated
  outflows}}.
\bjtitle{\mnras}
\bvolume{428},
\bfpage{2430}--\blpage{2442}
(\byear{2013}).
doi:\doiurl{10.1093/mnras/sts219}
\end{barticle}
\endbibitem

\bibitem[\protect\citeauthoryear{{Lyubarskii}}{1999}]{Lyub99}
\begin{barticle}
\bauthor{\binits{Y.E.} \bsnm{{Lyubarskii}}},
\batitle{{Kink instability of relativistic force-free jets}}.
\bjtitle{\mnras}
\bvolume{308},
\bfpage{1006}--\blpage{1010}
(\byear{1999}).
doi:\doiurl{10.1046/j.1365-8711.1999.02763.x}
\end{barticle}
\endbibitem

\bibitem[\protect\citeauthoryear{{Lyubarskij}}{1992}]{Lyub92}
\begin{barticle}
\bauthor{\binits{Y.E.} \bsnm{{Lyubarskij}}},
\batitle{{Energy release in strongly magnetized relativistic winds}}.
\bjtitle{Soviet Astronomy Letters}
\bvolume{18},
\bfpage{356}
(\byear{1992})
\end{barticle}
\endbibitem

\bibitem[\protect\citeauthoryear{{Lyubarsky}}{2009}]{Lyub09}
\begin{barticle}
\bauthor{\binits{Y.} \bsnm{{Lyubarsky}}},
\batitle{{Asymptotic Structure of Poynting-Dominated Jets}}.
\bjtitle{\apj}
\bvolume{698},
\bfpage{1570}--\blpage{1589}
(\byear{2009}).
doi:\doiurl{10.1088/0004-637X/698/2/1570}
\end{barticle}
\endbibitem

\bibitem[\protect\citeauthoryear{{Lyubarsky} and {Kirk}}{2001}]{LK01}
\begin{barticle}
\bauthor{\binits{Y.} \bsnm{{Lyubarsky}}},
\bauthor{\binits{J.G.} \bsnm{{Kirk}}},
\batitle{{Reconnection in a Striped Pulsar Wind}}.
\bjtitle{\apj}
\bvolume{547},
\bfpage{437}--\blpage{448}
(\byear{2001}).
doi:\doiurl{10.1086/318354}
\end{barticle}
\endbibitem

\bibitem[\protect\citeauthoryear{{Lyubarsky}}{2003}]{Lyub03}
\begin{barticle}
\bauthor{\binits{Y.E.} \bsnm{{Lyubarsky}}},
\batitle{{The termination shock in a striped pulsar wind}}.
\bjtitle{\mnras}
\bvolume{345},
\bfpage{153}--\blpage{160}
(\byear{2003}).
doi:\doiurl{10.1046/j.1365-8711.2003.06927.x}
\end{barticle}
\endbibitem

\bibitem[\protect\citeauthoryear{{Lyubarsky}}{2010a}]{Lyub10b}
\begin{barticle}
\bauthor{\binits{Y.E.} \bsnm{{Lyubarsky}}},
\batitle{{A New Mechanism for Dissipation of Alternating Fields in
  Poynting-dominated Outflows}}.
\bjtitle{\apjl}
\bvolume{725},
\bfpage{234}--\blpage{238}
(\byear{2010}a).
doi:\doiurl{10.1088/2041-8205/725/2/L234}
\end{barticle}
\endbibitem

\bibitem[\protect\citeauthoryear{{Lyubarsky}}{2010b}]{Lyub10a}
\begin{barticle}
\bauthor{\binits{Y.E.} \bsnm{{Lyubarsky}}},
\batitle{{Transformation of the Poynting flux into kinetic energy in
  relativistic jets}}.
\bjtitle{\mnras}
\bvolume{402},
\bfpage{353}--\blpage{361}
(\byear{2010}b).
doi:\doiurl{10.1111/j.1365-2966.2009.15877.x}
\end{barticle}
\endbibitem

\bibitem[\protect\citeauthoryear{{Lyutikov}}{2006}]{Lyut06}
\begin{barticle}
\bauthor{\binits{M.} \bsnm{{Lyutikov}}},
\batitle{{The electromagnetic model of gamma-ray bursts}}.
\bjtitle{New Journal of Physics}
\bvolume{8},
\bfpage{119}
(\byear{2006}).
doi:\doiurl{10.1088/1367-2630/8/7/119}
\end{barticle}
\endbibitem

\bibitem[\protect\citeauthoryear{{Lyutikov}}{2010}]{Lyut10}
\begin{barticle}
\bauthor{\binits{M.} \bsnm{{Lyutikov}}},
\batitle{{A high-sigma model of pulsar wind nebulae}}.
\bjtitle{\mnras}
\bvolume{405},
\bfpage{1809}--\blpage{1815}
(\byear{2010}).
doi:\doiurl{10.1111/j.1365-2966.2010.16553.x}
\end{barticle}
\endbibitem

\bibitem[\protect\citeauthoryear{{Lyutikov}}{2011}]{Lyut11}
\begin{barticle}
\bauthor{\binits{M.} \bsnm{{Lyutikov}}},
\batitle{{Dynamics of strongly magnetized ejecta in gamma-ray bursts}}.
\bjtitle{\mnras}
\bvolume{411},
\bfpage{422}--\blpage{426}
(\byear{2011}).
doi:\doiurl{10.1111/j.1365-2966.2010.17696.x}
\end{barticle}
\endbibitem

\bibitem[\protect\citeauthoryear{{Lyutikov} and {Blandford}}{2003}]{LB03}
\begin{botherref}
\oauthor{\binits{M.} \bsnm{{Lyutikov}}},
\oauthor{\binits{R.} \bsnm{{Blandford}}},
{Gamma Ray Bursts as Electromagnetic Outflows}.
ArXiv Astrophysics e-prints
(2003)
\end{botherref}
\endbibitem

\bibitem[\protect\citeauthoryear{{Lyutikov} et~al.}{2003}]{LPB03}
\begin{barticle}
\bauthor{\binits{M.} \bsnm{{Lyutikov}}},
\bauthor{\binits{V.I.} \bsnm{{Pariev}}},
\bauthor{\binits{R.D.} \bsnm{{Blandford}}},
\batitle{{Polarization of Prompt Gamma-Ray Burst Emission: Evidence for
  Electromagnetically Dominated Outflow}}.
\bjtitle{\apj}
\bvolume{597},
\bfpage{998}--\blpage{1009}
(\byear{2003}).
doi:\doiurl{10.1086/378497}
\end{barticle}
\endbibitem

\bibitem[\protect\citeauthoryear{{MacFadyen} and {Woosley}}{1999}]{MW99}
\begin{barticle}
\bauthor{\binits{A.I.} \bsnm{{MacFadyen}}},
\bauthor{\binits{S.E.} \bsnm{{Woosley}}},
\batitle{{Collapsars: Gamma-Ray Bursts and Explosions in ``Failed
  Supernovae''}}.
\bjtitle{\apj}
\bvolume{524},
\bfpage{262}--\blpage{289}
(\byear{1999}).
doi:\doiurl{10.1086/307790}
\end{barticle}
\endbibitem

\bibitem[\protect\citeauthoryear{{Margalit} et~al.}{2015}]{Margalit15}
\begin{botherref}
\oauthor{\binits{B.} \bsnm{{Margalit}}},
\oauthor{\binits{B.D.} \bsnm{{Metzger}}},
\oauthor{\binits{A.M.} \bsnm{{Beloborodov}}},
{Does the Collapse of a Supramassive Neutron Star Leave a Debris Disk?}
ArXiv e-prints
(2015)
\end{botherref}
\endbibitem

\bibitem[\protect\citeauthoryear{{Maselli} et~al.}{2014}]{Maselli14}
\begin{barticle}
\bauthor{\binits{A.} \bsnm{{Maselli}}}, \betal,
\batitle{{GRB 130427A: A Nearby Ordinary Monster}}.
\bjtitle{Science}
\bvolume{343},
\bfpage{48}--\blpage{51}
(\byear{2014}).
doi:\doiurl{10.1126/science.1242279}
\end{barticle}
\endbibitem

\bibitem[\protect\citeauthoryear{{Matzner}}{2003}]{Matzner03}
\begin{barticle}
\bauthor{\binits{C.D.} \bsnm{{Matzner}}},
\batitle{{Supernova hosts for gamma-ray burst jets: dynamical constraints}}.
\bjtitle{\mnras}
\bvolume{345},
\bfpage{575}--\blpage{589}
(\byear{2003}).
doi:\doiurl{10.1046/j.1365-8711.2003.06969.x}
\end{barticle}
\endbibitem

\bibitem[\protect\citeauthoryear{{McGlynn} et~al.}{2007}]{mcglynn:07}
\begin{barticle}
\bauthor{\binits{S.} \bsnm{{McGlynn}}}, \betal,
\batitle{{Polarisation studies of the prompt gamma-ray emission from GRB
  041219a using the spectrometer aboard INTEGRAL}}.
\bjtitle{\aap}
\bvolume{466},
\bfpage{895}--\blpage{904}
(\byear{2007}).
doi:\doiurl{10.1051/0004-6361:20066179}
\end{barticle}
\endbibitem

\bibitem[\protect\citeauthoryear{{McKinney} and {Uzdensky}}{2012}]{mckinney:12}
\begin{barticle}
\bauthor{\binits{J.C.} \bsnm{{McKinney}}},
\bauthor{\binits{D.A.} \bsnm{{Uzdensky}}},
\batitle{{A reconnection switch to trigger gamma-ray burst jet dissipation}}.
\bjtitle{\mnras}
\bvolume{419},
\bfpage{573}--\blpage{607}
(\byear{2012}).
doi:\doiurl{10.1111/j.1365-2966.2011.19721.x}
\end{barticle}
\endbibitem

\bibitem[\protect\citeauthoryear{{Medvedev} and {Loeb}}{1999}]{ML99}
\begin{barticle}
\bauthor{\binits{M.V.} \bsnm{{Medvedev}}},
\bauthor{\binits{A.} \bsnm{{Loeb}}},
\batitle{{Generation of Magnetic Fields in the Relativistic Shock of Gamma-Ray
  Burst Sources}}.
\bjtitle{\apj}
\bvolume{526},
\bfpage{697}--\blpage{706}
(\byear{1999}).
doi:\doiurl{10.1086/308038}
\end{barticle}
\endbibitem

\bibitem[\protect\citeauthoryear{{M{\'e}sz{\'a}ros} and
  {Rees}}{2000}]{meszaros:00}
\begin{barticle}
\bauthor{\binits{P.} \bsnm{{M{\'e}sz{\'a}ros}}},
\bauthor{\binits{M.J.} \bsnm{{Rees}}},
\batitle{{Steep Slopes and Preferred Breaks in Gamma-Ray Burst Spectra: The
  Role of Photospheres and Comptonization}}.
\bjtitle{\apj}
\bvolume{530},
\bfpage{292}--\blpage{298}
(\byear{2000}).
doi:\doiurl{10.1086/308371}
\end{barticle}
\endbibitem

\bibitem[\protect\citeauthoryear{{Meszaros} et~al.}{1993}]{meszaros:93}
\begin{barticle}
\bauthor{\binits{P.} \bsnm{{Meszaros}}},
\bauthor{\binits{P.} \bsnm{{Laguna}}},
\bauthor{\binits{M.J.} \bsnm{{Rees}}},
\batitle{{Gasdynamics of relativistically expanding gamma-ray burst sources -
  Kinematics, energetics, magnetic fields, and efficiency}}.
\bjtitle{\apj}
\bvolume{415},
\bfpage{181}--\blpage{190}
(\byear{1993}).
doi:\doiurl{10.1086/173154}
\end{barticle}
\endbibitem

\bibitem[\protect\citeauthoryear{{M{\'e}sz{\'a}ros} et~al.}{2002}]{meszaros:02}
\begin{barticle}
\bauthor{\binits{P.} \bsnm{{M{\'e}sz{\'a}ros}}},
\bauthor{\binits{E.} \bsnm{{Ramirez-Ruiz}}},
\bauthor{\binits{M.J.} \bsnm{{Rees}}},
\bauthor{\binits{B.} \bsnm{{Zhang}}},
\batitle{{X-Ray-rich Gamma-Ray Bursts, Photospheres, and Variability}}.
\bjtitle{\apj}
\bvolume{578},
\bfpage{812}--\blpage{817}
(\byear{2002}).
doi:\doiurl{10.1086/342611}
\end{barticle}
\endbibitem

\bibitem[\protect\citeauthoryear{{Metzger} et~al.}{2008}]{Metzger08}
\begin{barticle}
\bauthor{\binits{B.D.} \bsnm{{Metzger}}},
\bauthor{\binits{E.} \bsnm{{Quataert}}},
\bauthor{\binits{T.A.} \bsnm{{Thompson}}},
\batitle{{Short-duration gamma-ray bursts with extended emission from
  protomagnetar spin-down}}.
\bjtitle{\mnras}
\bvolume{385},
\bfpage{1455}--\blpage{1460}
(\byear{2008}).
doi:\doiurl{10.1111/j.1365-2966.2008.12923.x}
\end{barticle}
\endbibitem

\bibitem[\protect\citeauthoryear{{Metzger} et~al.}{2011}]{Metzger11}
\begin{barticle}
\bauthor{\binits{B.D.} \bsnm{{Metzger}}},
\bauthor{\binits{D.} \bsnm{{Giannios}}},
\bauthor{\binits{T.A.} \bsnm{{Thompson}}},
\bauthor{\binits{N.} \bsnm{{Bucciantini}}},
\bauthor{\binits{E.} \bsnm{{Quataert}}},
\batitle{{The protomagnetar model for gamma-ray bursts}}.
\bjtitle{\mnras}
\bvolume{413},
\bfpage{2031}--\blpage{2056}
(\byear{2011}).
doi:\doiurl{10.1111/j.1365-2966.2011.18280.x}
\end{barticle}
\endbibitem

\bibitem[\protect\citeauthoryear{{Mimica} and {Aloy}}{2010}]{mimica:10}
\begin{barticle}
\bauthor{\binits{P.} \bsnm{{Mimica}}},
\bauthor{\binits{M.A.} \bsnm{{Aloy}}},
\batitle{{On the dynamic efficiency of internal shocks in magnetized
  relativistic outflows}}.
\bjtitle{\mnras}
\bvolume{401},
\bfpage{525}--\blpage{532}
(\byear{2010}).
doi:\doiurl{10.1111/j.1365-2966.2009.15669.x}
\end{barticle}
\endbibitem

\bibitem[\protect\citeauthoryear{{Molinari} et~al.}{2007}]{Molinari07}
\begin{barticle}
\bauthor{\binits{E.} \bsnm{{Molinari}}}, \betal,
\batitle{{REM observations of GRB 060418 and GRB 060607A: the onset of the
  afterglow and the initial fireball Lorentz factor determination}}.
\bjtitle{\aap}
\bvolume{469},
\bfpage{13}--\blpage{16}
(\byear{2007}).
doi:\doiurl{10.1051/0004-6361:20077388}
\end{barticle}
\endbibitem

\bibitem[\protect\citeauthoryear{{Mundell} et~al.}{2007a}]{Mundell07}
\begin{barticle}
\bauthor{\binits{C.G.} \bsnm{{Mundell}}}, \betal,
\batitle{{Early Optical Polarization of a Gamma-Ray Burst Afterglow}}.
\bjtitle{Science}
\bvolume{315},
\bfpage{1822}
(\byear{2007}a).
doi:\doiurl{10.1126/science.1138484}
\end{barticle}
\endbibitem

\bibitem[\protect\citeauthoryear{{Mundell} et~al.}{2007b}]{Mundell07b}
\begin{barticle}
\bauthor{\binits{C.G.} \bsnm{{Mundell}}}, \betal,
\batitle{{The Remarkable Afterglow of GRB 061007: Implications for Optical
  Flashes and GRB Fireballs}}.
\bjtitle{\apj}
\bvolume{660},
\bfpage{489}--\blpage{495}
(\byear{2007}b).
doi:\doiurl{10.1086/512605}
\end{barticle}
\endbibitem

\bibitem[\protect\citeauthoryear{{Mundell} et~al.}{2013}]{Mundell13}
\begin{barticle}
\bauthor{\binits{C.G.} \bsnm{{Mundell}}}, \betal,
\batitle{{Highly polarized light from stable ordered magnetic fields in GRB
  120308A}}.
\bjtitle{\nat}
\bvolume{504},
\bfpage{119}--\blpage{121}
(\byear{2013}).
doi:\doiurl{10.1038/nature12814}
\end{barticle}
\endbibitem

\bibitem[\protect\citeauthoryear{{Nakar} and {Oren}}{2004}]{NO04}
\begin{barticle}
\bauthor{\binits{E.} \bsnm{{Nakar}}},
\bauthor{\binits{Y.} \bsnm{{Oren}}},
\batitle{{Polarization and Light-Curve Variability: The ``Patchy-Shell''
  Model}}.
\bjtitle{\apjl}
\bvolume{602},
\bfpage{97}--\blpage{100}
(\byear{2004}).
doi:\doiurl{10.1086/382729}
\end{barticle}
\endbibitem

\bibitem[\protect\citeauthoryear{{Nakar} and {Sari}}{2012}]{NS12}
\begin{barticle}
\bauthor{\binits{E.} \bsnm{{Nakar}}},
\bauthor{\binits{R.} \bsnm{{Sari}}},
\batitle{{Relativistic Shock Breakouts -- A Variety of Gamma-Ray Flares: From
  Low-luminosity Gamma-Ray Bursts to Type Ia Supernovae}}.
\bjtitle{\apj}
\bvolume{747},
\bfpage{88}
(\byear{2012}).
doi:\doiurl{10.1088/0004-637X/747/2/88}
\end{barticle}
\endbibitem

\bibitem[\protect\citeauthoryear{{Nakar} et~al.}{2009}]{nakar:09}
\begin{barticle}
\bauthor{\binits{E.} \bsnm{{Nakar}}},
\bauthor{\binits{S.} \bsnm{{Ando}}},
\bauthor{\binits{R.} \bsnm{{Sari}}},
\batitle{{Klein-Nishina Effects on Optically Thin Synchrotron and Synchrotron
  Self-Compton Spectrum}}.
\bjtitle{\apj}
\bvolume{703},
\bfpage{675}--\blpage{691}
(\byear{2009}).
doi:\doiurl{10.1088/0004-637X/703/1/675}
\end{barticle}
\endbibitem

\bibitem[\protect\citeauthoryear{{Nakar} et~al.}{2003}]{NPW03}
\begin{barticle}
\bauthor{\binits{E.} \bsnm{{Nakar}}},
\bauthor{\binits{T.} \bsnm{{Piran}}},
\bauthor{\binits{E.} \bsnm{{Waxman}}},
\batitle{{Implications of the bold gamma-ray polarization of GRB 021206}}.
\bjtitle{\jcap}
\bvolume{10},
\bfpage{5}
(\byear{2003}).
doi:\doiurl{10.1088/1475-7516/2003/10/005}
\end{barticle}
\endbibitem

\bibitem[\protect\citeauthoryear{{Narayan} et~al.}{2011}]{narayan:11}
\begin{barticle}
\bauthor{\binits{R.} \bsnm{{Narayan}}},
\bauthor{\binits{P.} \bsnm{{Kumar}}},
\bauthor{\binits{A.} \bsnm{{Tchekhovskoy}}},
\batitle{{Constraints on cold magnetized shocks in gamma-ray bursts}}.
\bjtitle{\mnras}
\bvolume{416},
\bfpage{2193}--\blpage{2201}
(\byear{2011}).
doi:\doiurl{10.1111/j.1365-2966.2011.19197.x}
\end{barticle}
\endbibitem

\bibitem[\protect\citeauthoryear{{Nava} et~al.}{2015}]{NNP15}
\begin{botherref}
\oauthor{\binits{L.} \bsnm{{Nava}}},
\oauthor{\binits{E.} \bsnm{{Nakar}}},
\oauthor{\binits{T.} \bsnm{{Piran}}},
{Linear and circular polarization in GRB afterglows}.
ArXiv e-prints
(2015)
\end{botherref}
\endbibitem

\bibitem[\protect\citeauthoryear{{Nava} et~al.}{2011}]{nava:11}
\begin{barticle}
\bauthor{\binits{L.} \bsnm{{Nava}}},
\bauthor{\binits{G.} \bsnm{{Ghirlanda}}},
\bauthor{\binits{G.} \bsnm{{Ghisellini}}},
\bauthor{\binits{A.} \bsnm{{Celotti}}},
\batitle{{Spectral properties of 438 GRBs detected by Fermi/GBM}}.
\bjtitle{\aap}
\bvolume{530},
\bfpage{21}
(\byear{2011}).
doi:\doiurl{10.1051/0004-6361/201016270}
\end{barticle}
\endbibitem

\bibitem[\protect\citeauthoryear{{Nousek} et~al.}{2006}]{nousek06}
\begin{barticle}
\bauthor{\binits{J.A.} \bsnm{{Nousek}}},
\bauthor{\binits{C.} \bsnm{{Kouveliotou}}},
\bauthor{\binits{D.} \bsnm{{Grupe}}},
\bauthor{\binits{K.L.} \bsnm{{Page}}},
\bauthor{\binits{J.} \bsnm{{Granot}}},
\bauthor{\binits{E.} \bsnm{{Ramirez-Ruiz}}}, \betal,
\batitle{{Evidence for a Canonical Gamma-Ray Burst Afterglow Light Curve in the
  Swift XRT Data}}.
\bjtitle{\apj}
\bvolume{642},
\bfpage{389}--\blpage{400}
(\byear{2006}).
doi:\doiurl{10.1086/500724}
\end{barticle}
\endbibitem

\bibitem[\protect\citeauthoryear{{O'Brien} et~al.}{2006}]{obrien06}
\begin{barticle}
\bauthor{\binits{P.T.} \bsnm{{O'Brien}}}, \betal,
\batitle{{The Early X-Ray Emission from GRBs}}.
\bjtitle{\apj}
\bvolume{647},
\bfpage{1213}--\blpage{1237}
(\byear{2006}).
doi:\doiurl{10.1086/505457}
\end{barticle}
\endbibitem

\bibitem[\protect\citeauthoryear{{Paczynski}}{1986}]{paczynski:86}
\begin{barticle}
\bauthor{\binits{B.} \bsnm{{Paczynski}}},
\batitle{{Gamma-ray bursters at cosmological distances}}.
\bjtitle{\apjl}
\bvolume{308},
\bfpage{43}--\blpage{46}
(\byear{1986}).
doi:\doiurl{10.1086/184740}
\end{barticle}
\endbibitem

\bibitem[\protect\citeauthoryear{{Pe'er}}{2008}]{peer:08}
\begin{barticle}
\bauthor{\binits{A.} \bsnm{{Pe'er}}},
\batitle{{Temporal Evolution of Thermal Emission from Relativistically
  Expanding Plasma}}.
\bjtitle{\apj}
\bvolume{682},
\bfpage{463}--\blpage{473}
(\byear{2008}).
doi:\doiurl{10.1086/588136}
\end{barticle}
\endbibitem

\bibitem[\protect\citeauthoryear{{Pe'er} et~al.}{2006}]{peer:06}
\begin{barticle}
\bauthor{\binits{A.} \bsnm{{Pe'er}}},
\bauthor{\binits{P.} \bsnm{{M{\'e}sz{\'a}ros}}},
\bauthor{\binits{M.J.} \bsnm{{Rees}}},
\batitle{{The Observable Effects of a Photospheric Component on GRB and XRF
  Prompt Emission Spectrum}}.
\bjtitle{\apj}
\bvolume{642},
\bfpage{995}--\blpage{1003}
(\byear{2006}).
doi:\doiurl{10.1086/501424}
\end{barticle}
\endbibitem

\bibitem[\protect\citeauthoryear{{Peng} et~al.}{2005}]{PKG05}
\begin{barticle}
\bauthor{\binits{F.} \bsnm{{Peng}}},
\bauthor{\binits{A.} \bsnm{{K{\"o}nigl}}},
\bauthor{\binits{J.} \bsnm{{Granot}}},
\batitle{{Two-Component Jet Models of Gamma-Ray Burst Sources}}.
\bjtitle{\apj}
\bvolume{626},
\bfpage{966}--\blpage{977}
(\byear{2005}).
doi:\doiurl{10.1086/430045}
\end{barticle}
\endbibitem

\bibitem[\protect\citeauthoryear{{Perley} et~al.}{2014}]{Perley14}
\begin{barticle}
\bauthor{\binits{D.A.} \bsnm{{Perley}}}, \betal,
\batitle{{The Afterglow of GRB 130427A from 1 to 10$^{16}$ GHz}}.
\bjtitle{\apj}
\bvolume{781},
\bfpage{37}
(\byear{2014}).
doi:\doiurl{10.1088/0004-637X/781/1/37}
\end{barticle}
\endbibitem

\bibitem[\protect\citeauthoryear{{Piran}}{1999}]{Piran99}
\begin{barticle}
\bauthor{\binits{T.} \bsnm{{Piran}}},
\batitle{{Gamma-ray bursts and the fireball model}}.
\bjtitle{\physrep}
\bvolume{314},
\bfpage{575}--\blpage{667}
(\byear{1999}).
doi:\doiurl{10.1016/S0370-1573(98)00127-6}
\end{barticle}
\endbibitem

\bibitem[\protect\citeauthoryear{{Piran}}{2004}]{Piran04}
\begin{barticle}
\bauthor{\binits{T.} \bsnm{{Piran}}},
\batitle{{The physics of gamma-ray bursts}}.
\bjtitle{Reviews of Modern Physics}
\bvolume{76},
\bfpage{1143}--\blpage{1210}
(\byear{2004}).
doi:\doiurl{10.1103/RevModPhys.76.1143}
\end{barticle}
\endbibitem

\bibitem[\protect\citeauthoryear{{Piran}}{2005}]{Piran05m}
\begin{bchapter}
\bauthor{\binits{T.} \bsnm{{Piran}}},
\bctitle{{Magnetic Fields in Gamma-Ray Bursts: A Short Overview}},
in \bbtitle{Magnetic Fields in the Universe: From Laboratory and Stars to
  Primordial Structures.},
ed. by \beditor{\binits{E.M.} \bsnm{{de Gouveia dal Pino}}},
\beditor{\binits{G.} \bsnm{{Lugones}}},
\beditor{\binits{A.} \bsnm{{Lazarian}}}
\bsertitle{American Institute of Physics Conference Series},
vol. \bseriesno{784},
\byear{2005},
pp. \bfpage{164}--\blpage{174}.
doi:\doiurl{10.1063/1.2077181}
\end{bchapter}
\endbibitem

\bibitem[\protect\citeauthoryear{{Piran} and {Nakar}}{2010}]{PN10}
\begin{barticle}
\bauthor{\binits{T.} \bsnm{{Piran}}},
\bauthor{\binits{E.} \bsnm{{Nakar}}},
\batitle{{On the External Shock Synchrotron Model for Gamma-ray Bursts' GeV
  Emission}}.
\bjtitle{\apjl}
\bvolume{718},
\bfpage{63}--\blpage{67}
(\byear{2010}).
doi:\doiurl{10.1088/2041-8205/718/2/L63}
\end{barticle}
\endbibitem

\bibitem[\protect\citeauthoryear{{Piran} et~al.}{2009}]{piran:09}
\begin{barticle}
\bauthor{\binits{T.} \bsnm{{Piran}}},
\bauthor{\binits{R.} \bsnm{{Sari}}},
\bauthor{\binits{Y.-C.} \bsnm{{Zou}}},
\batitle{{Observational limits on inverse Compton processes in gamma-ray
  bursts}}.
\bjtitle{\mnras}
\bvolume{393},
\bfpage{1107}--\blpage{1113}
(\byear{2009}).
doi:\doiurl{10.1111/j.1365-2966.2008.14198.x}
\end{barticle}
\endbibitem

\bibitem[\protect\citeauthoryear{{Preece} et~al.}{2014}]{preece:14}
\begin{barticle}
\bauthor{\binits{R.} \bsnm{{Preece}}}, \betal,
\batitle{{The First Pulse of the Extremely Bright GRB 130427A: A Test Lab for
  Synchrotron Shocks}}.
\bjtitle{Science}
\bvolume{343},
\bfpage{51}--\blpage{54}
(\byear{2014}).
doi:\doiurl{10.1126/science.1242302}
\end{barticle}
\endbibitem

\bibitem[\protect\citeauthoryear{{Preece} et~al.}{1998}]{Preece98}
\begin{barticle}
\bauthor{\binits{R.D.} \bsnm{{Preece}}},
\bauthor{\binits{M.S.} \bsnm{{Briggs}}},
\bauthor{\binits{R.S.} \bsnm{{Mallozzi}}},
\bauthor{\binits{G.N.} \bsnm{{Pendleton}}},
\bauthor{\binits{W.S.} \bsnm{{Paciesas}}},
\bauthor{\binits{D.L.} \bsnm{{Band}}},
\batitle{{The Synchrotron Shock Model Confronts a ``Line of Death'' in the
  BATSE Gamma-Ray Burst Data}}.
\bjtitle{\apjl}
\bvolume{506},
\bfpage{23}--\blpage{26}
(\byear{1998}).
doi:\doiurl{10.1086/311644}
\end{barticle}
\endbibitem

\bibitem[\protect\citeauthoryear{{Preece} et~al.}{2000}]{preece:00}
\begin{barticle}
\bauthor{\binits{R.D.} \bsnm{{Preece}}},
\bauthor{\binits{M.S.} \bsnm{{Briggs}}},
\bauthor{\binits{R.S.} \bsnm{{Mallozzi}}},
\bauthor{\binits{G.N.} \bsnm{{Pendleton}}},
\bauthor{\binits{W.S.} \bsnm{{Paciesas}}},
\bauthor{\binits{D.L.} \bsnm{{Band}}},
\batitle{{The BATSE Gamma-Ray Burst Spectral Catalog. I. High Time Resolution
  Spectroscopy of Bright Bursts Using High Energy Resolution Data}}.
\bjtitle{\apjs}
\bvolume{126},
\bfpage{19}--\blpage{36}
(\byear{2000}).
doi:\doiurl{10.1086/313289}
\end{barticle}
\endbibitem

\bibitem[\protect\citeauthoryear{{Press} et~al.}{1992}]{Press92}
\begin{bbook}
\bauthor{\binits{W.H.} \bsnm{{Press}}},
\bauthor{\binits{S.A.} \bsnm{{Teukolsky}}},
\bauthor{\binits{W.T.} \bsnm{{Vetterling}}},
\bauthor{\binits{B.P.} \bsnm{{Flannery}}},
\bbtitle{{Numerical recipes in C. The art of scientific computing}}
\byear{1992}
\end{bbook}
\endbibitem

\bibitem[\protect\citeauthoryear{{Racusin} et~al.}{2008}]{Racusin08}
\begin{barticle}
\bauthor{\binits{J.L.} \bsnm{{Racusin}}},
\bauthor{\binits{S.V.} \bsnm{{Karpov}}},
\bauthor{\binits{M.} \bsnm{{Sokolowski}}},
\bauthor{\binits{J.} \bsnm{{Granot}}}, \betal,
\batitle{{Broadband observations of the naked-eye {$\gamma$}-ray burst
  GRB080319B}}.
\bjtitle{\nat}
\bvolume{455},
\bfpage{183}--\blpage{188}
(\byear{2008}).
doi:\doiurl{10.1038/nature07270}
\end{barticle}
\endbibitem

\bibitem[\protect\citeauthoryear{{Rees} and {Meszaros}}{1994}]{rees:94}
\begin{barticle}
\bauthor{\binits{M.J.} \bsnm{{Rees}}},
\bauthor{\binits{P.} \bsnm{{Meszaros}}},
\batitle{{Unsteady outflow models for cosmological gamma-ray bursts}}.
\bjtitle{\apjl}
\bvolume{430},
\bfpage{93}--\blpage{96}
(\byear{1994}).
doi:\doiurl{10.1086/187446}
\end{barticle}
\endbibitem

\bibitem[\protect\citeauthoryear{{Rees} and {M{\'e}sz{\'a}ros}}{2005}]{rees:05}
\begin{barticle}
\bauthor{\binits{M.J.} \bsnm{{Rees}}},
\bauthor{\binits{P.} \bsnm{{M{\'e}sz{\'a}ros}}},
\batitle{{Dissipative Photosphere Models of Gamma-Ray Bursts and X-Ray
  Flashes}}.
\bjtitle{\apj}
\bvolume{628},
\bfpage{847}--\blpage{852}
(\byear{2005}).
doi:\doiurl{10.1086/430818}
\end{barticle}
\endbibitem

\bibitem[\protect\citeauthoryear{{Rezzolla} and {Kumar}}{2015}]{Rezzolla15}
\begin{barticle}
\bauthor{\binits{L.} \bsnm{{Rezzolla}}},
\bauthor{\binits{P.} \bsnm{{Kumar}}},
\batitle{{A Novel Paradigm for Short Gamma-Ray Bursts With Extended X-Ray
  Emission}}.
\bjtitle{\apj}
\bvolume{802},
\bfpage{95}
(\byear{2015}).
doi:\doiurl{10.1088/0004-637X/802/2/95}
\end{barticle}
\endbibitem

\bibitem[\protect\citeauthoryear{{Rosswog}}{2007}]{Rosswog07}
\begin{bchapter}
\bauthor{\binits{S.} \bsnm{{Rosswog}}},
\bctitle{{Last Moments in the Life of a Compact Binary System: Gravitational
  Waves, Gamma-Ray Bursts and Magnetar Formation}},
in \bbtitle{Revista Mexicana de Astronomia y Astrofisica, vol. 27}.
\bsertitle{Revista Mexicana de Astronomia y Astrofisica, vol. 27},
vol. \bseriesno{27},
\byear{2007},
pp. \bfpage{57}--\blpage{79}
\end{bchapter}
\endbibitem

\bibitem[\protect\citeauthoryear{{Rowlinson} et~al.}{2010}]{Rowlinson10}
\begin{barticle}
\bauthor{\binits{A.} \bsnm{{Rowlinson}}}, \betal,
\batitle{{The unusual X-ray emission of the short Swift GRB 090515: evidence
  for the formation of a magnetar?}}
\bjtitle{\mnras}
\bvolume{409},
\bfpage{531}--\blpage{540}
(\byear{2010}).
doi:\doiurl{10.1111/j.1365-2966.2010.17354.x}
\end{barticle}
\endbibitem

\bibitem[\protect\citeauthoryear{{Rowlinson} et~al.}{2013}]{Rowlinson13}
\begin{barticle}
\bauthor{\binits{A.} \bsnm{{Rowlinson}}},
\bauthor{\binits{P.T.} \bsnm{{O'Brien}}},
\bauthor{\binits{B.D.} \bsnm{{Metzger}}},
\bauthor{\binits{N.R.} \bsnm{{Tanvir}}},
\bauthor{\binits{A.J.} \bsnm{{Levan}}},
\batitle{{Signatures of magnetar central engines in short GRB light curves}}.
\bjtitle{\mnras}
\bvolume{430},
\bfpage{1061}--\blpage{1087}
(\byear{2013}).
doi:\doiurl{10.1093/mnras/sts683}
\end{barticle}
\endbibitem

\bibitem[\protect\citeauthoryear{{Rowlinson} et~al.}{2014}]{Rowlinson14}
\begin{barticle}
\bauthor{\binits{A.} \bsnm{{Rowlinson}}},
\bauthor{\binits{B.P.} \bsnm{{Gompertz}}},
\bauthor{\binits{M.} \bsnm{{Dainotti}}},
\bauthor{\binits{P.T.} \bsnm{{O'Brien}}},
\bauthor{\binits{R.A.M.J.} \bsnm{{Wijers}}},
\bauthor{\binits{A.J.} \bsnm{{van der Horst}}},
\batitle{{Constraining properties of GRB magnetar central engines using the
  observed plateau luminosity and duration correlation}}.
\bjtitle{\mnras}
\bvolume{443},
\bfpage{1779}--\blpage{1787}
(\byear{2014}).
doi:\doiurl{10.1093/mnras/stu1277}
\end{barticle}
\endbibitem

\bibitem[\protect\citeauthoryear{{Rutledge} and {Fox}}{2004}]{RutledgeFox04}
\begin{barticle}
\bauthor{\binits{R.E.} \bsnm{{Rutledge}}},
\bauthor{\binits{D.B.} \bsnm{{Fox}}},
\batitle{{Re-analysis of polarization in the {$\gamma$}-ray flux of GRB
  021206}}.
\bjtitle{\mnras}
\bvolume{350},
\bfpage{1288}--\blpage{1300}
(\byear{2004}).
doi:\doiurl{10.1111/j.1365-2966.2004.07665.x}
\end{barticle}
\endbibitem

\bibitem[\protect\citeauthoryear{{Ryde} et~al.}{2010}]{Ryde10}
\begin{barticle}
\bauthor{\binits{F.} \bsnm{{Ryde}}}, \betal,
\batitle{{Identification and Properties of the Photospheric Emission in
  GRB090902B}}.
\bjtitle{\apjl}
\bvolume{709},
\bfpage{172}--\blpage{177}
(\byear{2010}).
doi:\doiurl{10.1088/2041-8205/709/2/L172}
\end{barticle}
\endbibitem

\bibitem[\protect\citeauthoryear{{Ryde} et~al.}{2011}]{Ryde11}
\begin{barticle}
\bauthor{\binits{F.} \bsnm{{Ryde}}}, \betal,
\batitle{{Observational evidence of dissipative photospheres in gamma-ray
  bursts}}.
\bjtitle{\mnras}
\bvolume{415},
\bfpage{3693}--\blpage{3705}
(\byear{2011}).
doi:\doiurl{10.1111/j.1365-2966.2011.18985.x}
\end{barticle}
\endbibitem

\bibitem[\protect\citeauthoryear{{Sakamoto} et~al.}{2005}]{sakamoto:05}
\begin{barticle}
\bauthor{\binits{T.} \bsnm{{Sakamoto}}}, \betal,
\batitle{{Global Characteristics of X-Ray Flashes and X-Ray-Rich Gamma-Ray
  Bursts Observed by HETE-2}}.
\bjtitle{\apj}
\bvolume{629},
\bfpage{311}--\blpage{327}
(\byear{2005}).
doi:\doiurl{10.1086/431235}
\end{barticle}
\endbibitem

\bibitem[\protect\citeauthoryear{{Sari}}{1999}]{Sari99}
\begin{barticle}
\bauthor{\binits{R.} \bsnm{{Sari}}},
\batitle{{Linear Polarization and Proper Motion in the Afterglow of Beamed
  Gamma-Ray Bursts}}.
\bjtitle{\apjl}
\bvolume{524},
\bfpage{43}--\blpage{46}
(\byear{1999}).
doi:\doiurl{10.1086/312294}
\end{barticle}
\endbibitem

\bibitem[\protect\citeauthoryear{{Sari} and {Esin}}{2001}]{SE01}
\begin{barticle}
\bauthor{\binits{R.} \bsnm{{Sari}}},
\bauthor{\binits{A.A.} \bsnm{{Esin}}},
\batitle{{On the Synchrotron Self-Compton Emission from Relativistic Shocks and
  Its Implications for Gamma-Ray Burst Afterglows}}.
\bjtitle{\apj}
\bvolume{548},
\bfpage{787}--\blpage{799}
(\byear{2001}).
doi:\doiurl{10.1086/319003}
\end{barticle}
\endbibitem

\bibitem[\protect\citeauthoryear{{Sari} and {Piran}}{1995}]{SP95}
\begin{barticle}
\bauthor{\binits{R.} \bsnm{{Sari}}},
\bauthor{\binits{T.} \bsnm{{Piran}}},
\batitle{{Hydrodynamic Timescales and Temporal Structure of Gamma-Ray Bursts}}.
\bjtitle{\apjl}
\bvolume{455},
\bfpage{143}
(\byear{1995}).
doi:\doiurl{10.1086/309835}
\end{barticle}
\endbibitem

\bibitem[\protect\citeauthoryear{{Sari} and {Piran}}{1997}]{SP97}
\begin{barticle}
\bauthor{\binits{R.} \bsnm{{Sari}}},
\bauthor{\binits{T.} \bsnm{{Piran}}},
\batitle{{Variability in Gamma-Ray Bursts: A Clue}}.
\bjtitle{\apj}
\bvolume{485},
\bfpage{270}
(\byear{1997}).
doi:\doiurl{10.1086/304428}
\end{barticle}
\endbibitem

\bibitem[\protect\citeauthoryear{{Sari} and {Piran}}{1999}]{SP99}
\begin{barticle}
\bauthor{\binits{R.} \bsnm{{Sari}}},
\bauthor{\binits{T.} \bsnm{{Piran}}},
\batitle{{GRB 990123: The Optical Flash and the Fireball Model}}.
\bjtitle{\apjl}
\bvolume{517},
\bfpage{109}--\blpage{112}
(\byear{1999}).
doi:\doiurl{10.1086/312039}
\end{barticle}
\endbibitem

\bibitem[\protect\citeauthoryear{{Sari} et~al.}{1998}]{SPN98}
\begin{barticle}
\bauthor{\binits{R.} \bsnm{{Sari}}},
\bauthor{\binits{T.} \bsnm{{Piran}}},
\bauthor{\binits{R.} \bsnm{{Narayan}}},
\batitle{{Spectra and Light Curves of Gamma-Ray Burst Afterglows}}.
\bjtitle{\apjl}
\bvolume{497},
\bfpage{17}--\blpage{20}
(\byear{1998}).
doi:\doiurl{10.1086/311269}
\end{barticle}
\endbibitem

\bibitem[\protect\citeauthoryear{{Shemi} and {Piran}}{1990}]{shemi:90}
\begin{barticle}
\bauthor{\binits{A.} \bsnm{{Shemi}}},
\bauthor{\binits{T.} \bsnm{{Piran}}},
\batitle{{The appearance of cosmic fireballs}}.
\bjtitle{\apjl}
\bvolume{365},
\bfpage{55}--\blpage{58}
(\byear{1990}).
doi:\doiurl{10.1086/185887}
\end{barticle}
\endbibitem

\bibitem[\protect\citeauthoryear{{Sironi} and {Spitkovsky}}{2014}]{sironi:14}
\begin{barticle}
\bauthor{\binits{L.} \bsnm{{Sironi}}},
\bauthor{\binits{A.} \bsnm{{Spitkovsky}}},
\batitle{{Relativistic Reconnection: An Efficient Source of Non-thermal
  Particles}}.
\bjtitle{\apjl}
\bvolume{783},
\bfpage{21}
(\byear{2014}).
doi:\doiurl{10.1088/2041-8205/783/1/L21}
\end{barticle}
\endbibitem

\bibitem[\protect\citeauthoryear{{Sironi} et~al.}{2015}]{SKL15}
\begin{barticle}
\bauthor{\binits{L.} \bsnm{{Sironi}}},
\bauthor{\binits{U.} \bsnm{{Keshet}}},
\bauthor{\binits{M.} \bsnm{{Lemoine}}},
\batitle{{Relativistic Shocks: Particle Acceleration and Magnetization}}.
\bjtitle{\ssr}
(\byear{2015}).
doi:\doiurl{10.1007/s11214-015-0102-2}
\end{barticle}
\endbibitem

\bibitem[\protect\citeauthoryear{{Soderberg} et~al.}{2006}]{Soderberg06}
\begin{barticle}
\bauthor{\binits{A.M.} \bsnm{{Soderberg}}}, \betal,
\batitle{{Relativistic ejecta from X-ray flash XRF 060218 and the rate of
  cosmic explosions}}.
\bjtitle{\nat}
\bvolume{442},
\bfpage{1014}--\blpage{1017}
(\byear{2006}).
doi:\doiurl{10.1038/nature05087}
\end{barticle}
\endbibitem

\bibitem[\protect\citeauthoryear{{Spitkovsky}}{2006}]{Spitkovsky06}
\begin{barticle}
\bauthor{\binits{A.} \bsnm{{Spitkovsky}}},
\batitle{{Time-dependent Force-free Pulsar Magnetospheres: Axisymmetric and
  Oblique Rotators}}.
\bjtitle{\apjl}
\bvolume{648},
\bfpage{51}--\blpage{54}
(\byear{2006}).
doi:\doiurl{10.1086/507518}
\end{barticle}
\endbibitem

\bibitem[\protect\citeauthoryear{{Spruit} et~al.}{2001}]{spruit:01}
\begin{barticle}
\bauthor{\binits{H.C.} \bsnm{{Spruit}}},
\bauthor{\binits{F.} \bsnm{{Daigne}}},
\bauthor{\binits{G.} \bsnm{{Drenkhahn}}},
\batitle{{Large scale magnetic fields and their dissipation in GRB fireballs}}.
\bjtitle{\aap}
\bvolume{369},
\bfpage{694}--\blpage{705}
(\byear{2001}).
doi:\doiurl{10.1051/0004-6361:20010131}
\end{barticle}
\endbibitem

\bibitem[\protect\citeauthoryear{{Spruit} et~al.}{1997}]{Spruit97}
\begin{barticle}
\bauthor{\binits{H.C.} \bsnm{{Spruit}}},
\bauthor{\binits{T.} \bsnm{{Foglizzo}}},
\bauthor{\binits{R.} \bsnm{{Stehle}}},
\batitle{{Collimation of magnetically driven jets from accretion discs}}.
\bjtitle{\mnras}
\bvolume{288},
\bfpage{333}--\blpage{342}
(\byear{1997})
\end{barticle}
\endbibitem

\bibitem[\protect\citeauthoryear{{Steele} et~al.}{2009}]{Steele09}
\begin{barticle}
\bauthor{\binits{I.A.} \bsnm{{Steele}}},
\bauthor{\binits{C.G.} \bsnm{{Mundell}}},
\bauthor{\binits{R.J.} \bsnm{{Smith}}},
\bauthor{\binits{S.} \bsnm{{Kobayashi}}},
\bauthor{\binits{C.} \bsnm{{Guidorzi}}},
\batitle{{Ten per cent polarized optical emission from GRB090102}}.
\bjtitle{\nat}
\bvolume{462},
\bfpage{767}--\blpage{769}
(\byear{2009}).
doi:\doiurl{10.1038/nature08590}
\end{barticle}
\endbibitem

\bibitem[\protect\citeauthoryear{{Tam} et~al.}{2013}]{Tam13}
\begin{barticle}
\bauthor{\binits{P.-H.T.} \bsnm{{Tam}}},
\bauthor{\binits{Q.-W.} \bsnm{{Tang}}},
\bauthor{\binits{S.-J.} \bsnm{{Hou}}},
\bauthor{\binits{R.-Y.} \bsnm{{Liu}}},
\bauthor{\binits{X.-Y.} \bsnm{{Wang}}},
\batitle{{Discovery of an Extra Hard Spectral Component in the High-energy
  Afterglow Emission of GRB 130427A}}.
\bjtitle{\apjl}
\bvolume{771},
\bfpage{13}
(\byear{2013}).
doi:\doiurl{10.1088/2041-8205/771/1/L13}
\end{barticle}
\endbibitem

\bibitem[\protect\citeauthoryear{{Tchekhovskoy} et~al.}{2010}]{Tchek10}
\begin{barticle}
\bauthor{\binits{A.} \bsnm{{Tchekhovskoy}}},
\bauthor{\binits{R.} \bsnm{{Narayan}}},
\bauthor{\binits{J.C.} \bsnm{{McKinney}}},
\batitle{{Magnetohydrodynamic simulations of gamma-ray burst jets: Beyond the
  progenitor star}}.
\bjtitle{\na}
\bvolume{15},
\bfpage{749}--\blpage{754}
(\byear{2010}).
doi:\doiurl{10.1016/j.newast.2010.03.001}
\end{barticle}
\endbibitem

\bibitem[\protect\citeauthoryear{{Thompson}}{1994}]{thompson:94}
\begin{barticle}
\bauthor{\binits{C.} \bsnm{{Thompson}}},
\batitle{{A Model of Gamma-Ray Bursts}}.
\bjtitle{\mnras}
\bvolume{270},
\bfpage{480}
(\byear{1994})
\end{barticle}
\endbibitem

\bibitem[\protect\citeauthoryear{{Troja} et~al.}{2007}]{troja07}
\begin{barticle}
\bauthor{\binits{E.} \bsnm{{Troja}}}, \betal,
\batitle{{Swift Observations of GRB 070110: An Extraordinary X-Ray Afterglow
  Powered by the Central Engine}}.
\bjtitle{\apj}
\bvolume{665},
\bfpage{599}--\blpage{607}
(\byear{2007}).
doi:\doiurl{10.1086/519450}
\end{barticle}
\endbibitem

\bibitem[\protect\citeauthoryear{{Uhm} and {Beloborodov}}{2007}]{uhm:07}
\begin{barticle}
\bauthor{\binits{Z.L.} \bsnm{{Uhm}}},
\bauthor{\binits{A.M.} \bsnm{{Beloborodov}}},
\batitle{{On the Mechanism of Gamma-Ray Burst Afterglows}}.
\bjtitle{\apjl}
\bvolume{665},
\bfpage{93}--\blpage{96}
(\byear{2007}).
doi:\doiurl{10.1086/519837}
\end{barticle}
\endbibitem

\bibitem[\protect\citeauthoryear{{Uhm} and {Zhang}}{2014}]{uhm:14}
\begin{barticle}
\bauthor{\binits{Z.L.} \bsnm{{Uhm}}},
\bauthor{\binits{B.} \bsnm{{Zhang}}},
\batitle{{Fast-cooling synchrotron radiation in a decaying magnetic field and
  {$\gamma$}-ray burst emission mechanism}}.
\bjtitle{Nature Physics}
\bvolume{10},
\bfpage{351}--\blpage{356}
(\byear{2014}).
doi:\doiurl{10.1038/nphys2932}
\end{barticle}
\endbibitem

\bibitem[\protect\citeauthoryear{{Uhm} et~al.}{2012}]{uhm:12}
\begin{barticle}
\bauthor{\binits{Z.L.} \bsnm{{Uhm}}},
\bauthor{\binits{B.} \bsnm{{Zhang}}},
\bauthor{\binits{R.} \bsnm{{Hasco{\"e}t}}},
\bauthor{\binits{F.} \bsnm{{Daigne}}},
\bauthor{\binits{R.} \bsnm{{Mochkovitch}}},
\bauthor{\binits{I.H.} \bsnm{{Park}}},
\batitle{{Dynamics and Afterglow Light Curves of Gamma-Ray Burst Blast Waves
  with a Long-lived Reverse Shock}}.
\bjtitle{\apj}
\bvolume{761},
\bfpage{147}
(\byear{2012}).
doi:\doiurl{10.1088/0004-637X/761/2/147}
\end{barticle}
\endbibitem

\bibitem[\protect\citeauthoryear{{Usov}}{1992}]{Usov92}
\begin{barticle}
\bauthor{\binits{V.V.} \bsnm{{Usov}}},
\batitle{{Millisecond pulsars with extremely strong magnetic fields as a
  cosmological source of gamma-ray bursts}}.
\bjtitle{\nat}
\bvolume{357},
\bfpage{472}--\blpage{474}
(\byear{1992}).
doi:\doiurl{10.1038/357472a0}
\end{barticle}
\endbibitem

\bibitem[\protect\citeauthoryear{{Vestrand} et~al.}{2014}]{Vestrand14}
\begin{barticle}
\bauthor{\binits{W.T.} \bsnm{{Vestrand}}}, \betal,
\batitle{{The Bright Optical Flash and Afterglow from the Gamma-Ray Burst GRB
  130427A}}.
\bjtitle{Science}
\bvolume{343},
\bfpage{38}--\blpage{41}
(\byear{2014}).
doi:\doiurl{10.1126/science.1242316}
\end{barticle}
\endbibitem

\bibitem[\protect\citeauthoryear{{Vietri} and {Stella}}{1998}]{VS98}
\begin{barticle}
\bauthor{\binits{M.} \bsnm{{Vietri}}},
\bauthor{\binits{L.} \bsnm{{Stella}}},
\batitle{{A Gamma-Ray Burst Model with Small Baryon Contamination}}.
\bjtitle{\apjl}
\bvolume{507},
\bfpage{45}--\blpage{48}
(\byear{1998}).
doi:\doiurl{10.1086/311674}
\end{barticle}
\endbibitem

\bibitem[\protect\citeauthoryear{{Vietri} and {Stella}}{1999}]{VS99}
\begin{barticle}
\bauthor{\binits{M.} \bsnm{{Vietri}}},
\bauthor{\binits{L.} \bsnm{{Stella}}},
\batitle{{Supranova Events from Spun-up Neutron Stars: An Explosion in Search
  of an Observation}}.
\bjtitle{\apjl}
\bvolume{527},
\bfpage{43}--\blpage{46}
(\byear{1999}).
doi:\doiurl{10.1086/312386}
\end{barticle}
\endbibitem

\bibitem[\protect\citeauthoryear{{Virgili} et~al.}{2009}]{Virgili09}
\begin{barticle}
\bauthor{\binits{F.J.} \bsnm{{Virgili}}},
\bauthor{\binits{E.-W.} \bsnm{{Liang}}},
\bauthor{\binits{B.} \bsnm{{Zhang}}},
\batitle{{Low-luminosity gamma-ray bursts as a distinct GRB population: a
  firmer case from multiple criteria constraints}}.
\bjtitle{\mnras}
\bvolume{392},
\bfpage{91}--\blpage{103}
(\byear{2009}).
doi:\doiurl{10.1111/j.1365-2966.2008.14063.x}
\end{barticle}
\endbibitem

\bibitem[\protect\citeauthoryear{{Vurm} et~al.}{2011}]{vurm:11}
\begin{barticle}
\bauthor{\binits{I.} \bsnm{{Vurm}}},
\bauthor{\binits{A.M.} \bsnm{{Beloborodov}}},
\bauthor{\binits{J.} \bsnm{{Poutanen}}},
\batitle{{Gamma-Ray Bursts from Magnetized Collisionally Heated Jets}}.
\bjtitle{\apj}
\bvolume{738},
\bfpage{77}
(\byear{2011}).
doi:\doiurl{10.1088/0004-637X/738/1/77}
\end{barticle}
\endbibitem

\bibitem[\protect\citeauthoryear{{Vurm} et~al.}{2013}]{vurm:13}
\begin{barticle}
\bauthor{\binits{I.} \bsnm{{Vurm}}},
\bauthor{\binits{Y.} \bsnm{{Lyubarsky}}},
\bauthor{\binits{T.} \bsnm{{Piran}}},
\batitle{{On Thermalization in Gamma-Ray Burst Jets and the Peak Energies of
  Photospheric Spectra}}.
\bjtitle{\apj}
\bvolume{764},
\bfpage{143}
(\byear{2013}).
doi:\doiurl{10.1088/0004-637X/764/2/143}
\end{barticle}
\endbibitem

\bibitem[\protect\citeauthoryear{{Wang} et~al.}{2009}]{wang:09}
\begin{barticle}
\bauthor{\binits{X.-Y.} \bsnm{{Wang}}},
\bauthor{\binits{Z.} \bsnm{{Li}}},
\bauthor{\binits{Z.-G.} \bsnm{{Dai}}},
\bauthor{\binits{P.} \bsnm{{M{\'e}sz{\'a}ros}}},
\batitle{{GRB 080916C: On the Radiation Origin of the Prompt Emission from
  keV/MeV TO GeV}}.
\bjtitle{\apjl}
\bvolume{698},
\bfpage{98}--\blpage{102}
(\byear{2009}).
doi:\doiurl{10.1088/0004-637X/698/2/L98}
\end{barticle}
\endbibitem

\bibitem[\protect\citeauthoryear{{Wiersema} et~al.}{2014}]{Wiersema14}
\begin{barticle}
\bauthor{\binits{K.} \bsnm{{Wiersema}}}, \betal,
\batitle{{Circular polarization in the optical afterglow of GRB 121024A}}.
\bjtitle{\nat}
\bvolume{509},
\bfpage{201}--\blpage{204}
(\byear{2014}).
doi:\doiurl{10.1038/nature13237}
\end{barticle}
\endbibitem

\bibitem[\protect\citeauthoryear{{Wigger} et~al.}{2004}]{Wigger04}
\begin{barticle}
\bauthor{\binits{C.} \bsnm{{Wigger}}},
\bauthor{\binits{W.} \bsnm{{Hajdas}}},
\bauthor{\binits{K.} \bsnm{{Arzner}}},
\bauthor{\binits{M.} \bsnm{{G{\"u}del}}},
\bauthor{\binits{A.} \bsnm{{Zehnder}}},
\batitle{{Gamma-Ray Burst Polarization: Limits from RHESSI Measurements}}.
\bjtitle{\apj}
\bvolume{613},
\bfpage{1088}--\blpage{1100}
(\byear{2004}).
doi:\doiurl{10.1086/423163}
\end{barticle}
\endbibitem

\bibitem[\protect\citeauthoryear{{Willingale} et~al.}{2010}]{willingale:10}
\begin{barticle}
\bauthor{\binits{R.} \bsnm{{Willingale}}},
\bauthor{\binits{F.} \bsnm{{Genet}}},
\bauthor{\binits{J.} \bsnm{{Granot}}},
\bauthor{\binits{P.T.} \bsnm{{O'Brien}}},
\batitle{{The spectral-temporal properties of the prompt pulses and rapid decay
  phase of gamma-ray bursts}}.
\bjtitle{\mnras}
\bvolume{403},
\bfpage{1296}--\blpage{1316}
(\byear{2010}).
doi:\doiurl{10.1111/j.1365-2966.2009.16187.x}
\end{barticle}
\endbibitem

\bibitem[\protect\citeauthoryear{{Woosley}}{1993}]{Woosley93}
\begin{barticle}
\bauthor{\binits{S.E.} \bsnm{{Woosley}}},
\batitle{{Gamma-ray bursts from stellar mass accretion disks around black
  holes}}.
\bjtitle{\apj}
\bvolume{405},
\bfpage{273}--\blpage{277}
(\byear{1993}).
doi:\doiurl{10.1086/172359}
\end{barticle}
\endbibitem

\bibitem[\protect\citeauthoryear{{Woosley} and {Bloom}}{2006}]{WB06}
\begin{barticle}
\bauthor{\binits{S.E.} \bsnm{{Woosley}}},
\bauthor{\binits{J.S.} \bsnm{{Bloom}}},
\batitle{{The Supernova Gamma-Ray Burst Connection}}.
\bjtitle{\araa}
\bvolume{44},
\bfpage{507}--\blpage{556}
(\byear{2006}).
doi:\doiurl{10.1146/annurev.astro.43.072103.150558}
\end{barticle}
\endbibitem

\bibitem[\protect\citeauthoryear{{Yonetoku} et~al.}{2011}]{yonetoku:11}
\begin{barticle}
\bauthor{\binits{D.} \bsnm{{Yonetoku}}}, \betal,
\batitle{{Detection of Gamma-Ray Polarization in Prompt Emission of GRB
  100826A}}.
\bjtitle{\apjl}
\bvolume{743},
\bfpage{30}
(\byear{2011}).
doi:\doiurl{10.1088/2041-8205/743/2/L30}
\end{barticle}
\endbibitem

\bibitem[\protect\citeauthoryear{{Yonetoku} et~al.}{2012}]{yonetoku:12}
\begin{barticle}
\bauthor{\binits{D.} \bsnm{{Yonetoku}}}, \betal,
\batitle{{Magnetic Structures in Gamma-Ray Burst Jets Probed by Gamma-Ray
  Polarization}}.
\bjtitle{\apjl}
\bvolume{758},
\bfpage{1}
(\byear{2012}).
doi:\doiurl{10.1088/2041-8205/758/1/L1}
\end{barticle}
\endbibitem

\bibitem[\protect\citeauthoryear{{Zhang} et~al.}{2006}]{zhang06}
\begin{barticle}
\bauthor{\binits{B.} \bsnm{{Zhang}}}, \betal,
\batitle{{Physical Processes Shaping Gamma-Ray Burst X-Ray Afterglow Light
  Curves: Theoretical Implications from the Swift X-Ray Telescope
  Observations}}.
\bjtitle{\apj}
\bvolume{642},
\bfpage{354}--\blpage{370}
(\byear{2006}).
doi:\doiurl{10.1086/500723}
\end{barticle}
\endbibitem

\bibitem[\protect\citeauthoryear{{Zhang} and {Yan}}{2011}]{ZhangYan11}
\begin{barticle}
\bauthor{\binits{B.} \bsnm{{Zhang}}},
\bauthor{\binits{H.} \bsnm{{Yan}}},
\batitle{{The Internal-collision-induced Magnetic Reconnection and Turbulence
  (ICMART) Model of Gamma-ray Bursts}}.
\bjtitle{\apj}
\bvolume{726},
\bfpage{90}
(\byear{2011}).
doi:\doiurl{10.1088/0004-637X/726/2/90}
\end{barticle}
\endbibitem

\bibitem[\protect\citeauthoryear{{Zhang} and {Zhang}}{2014}]{zhang:14}
\begin{barticle}
\bauthor{\binits{B.} \bsnm{{Zhang}}},
\bauthor{\binits{B.} \bsnm{{Zhang}}},
\batitle{{Gamma-Ray Burst Prompt Emission Light Curves and Power Density
  Spectra in the ICMART Model}}.
\bjtitle{\apj}
\bvolume{782},
\bfpage{92}
(\byear{2014}).
doi:\doiurl{10.1088/0004-637X/782/2/92}
\end{barticle}
\endbibitem

\bibitem[\protect\citeauthoryear{{Zhang} et~al.}{2003}]{ZWM03}
\begin{barticle}
\bauthor{\binits{W.} \bsnm{{Zhang}}},
\bauthor{\binits{S.E.} \bsnm{{Woosley}}},
\bauthor{\binits{A.I.} \bsnm{{MacFadyen}}},
\batitle{{Relativistic Jets in Collapsars}}.
\bjtitle{\apj}
\bvolume{586},
\bfpage{356}--\blpage{371}
(\byear{2003}).
doi:\doiurl{10.1086/367609}
\end{barticle}
\endbibitem

\bibitem[\protect\citeauthoryear{{Zhao} et~al.}{2014}]{zhao:14}
\begin{barticle}
\bauthor{\binits{X.} \bsnm{{Zhao}}},
\bauthor{\binits{Z.} \bsnm{{Li}}},
\bauthor{\binits{X.} \bsnm{{Liu}}},
\bauthor{\binits{B.-b.} \bsnm{{Zhang}}},
\bauthor{\binits{J.} \bsnm{{Bai}}},
\bauthor{\binits{P.} \bsnm{{M{\'e}sz{\'a}ros}}},
\batitle{{Gamma-Ray Burst Spectrum with Decaying Magnetic Field}}.
\bjtitle{\apj}
\bvolume{780},
\bfpage{12}
(\byear{2014}).
doi:\doiurl{10.1088/0004-637X/780/1/12}
\end{barticle}
\endbibitem

\end{thebibliography}
\nocite{*}

\end{document}